\pdfoutput=1
\documentclass[11pt]{article}

\usepackage{geometry}
\usepackage{amsmath, amssymb, amsthm, mathtools}
\usepackage{bm}
\usepackage{mathrsfs}

\usepackage{booktabs}
\usepackage{tabularx}
\usepackage{multirow}

\usepackage{pdflscape}
\usepackage{rotating}
\usepackage{setspace}
\usepackage{enumitem}
\usepackage{titlesec}
\usepackage{indentfirst}
\usepackage{color}
\usepackage[round]{natbib}

\usepackage[pdftex,breaklinks,colorlinks,citecolor=blue,urlcolor=blue]{hyperref}
\usepackage{dcolumn}
\usepackage{bookmark}
\usepackage{cleveref}

\newcommand{\Title}{Polynomial-Preserving Processes and Volatility Modelling in Discrete Time}

\newcommand{\Authors}{Maciej Augustyniak\footnote{Quantact Actuarial and Financial Mathematics Laboratory and D\'epartement de math\'ematiques et de statistique, Universit\'e de Montr\'eal, PO Box 6128, Station Centre-ville, Montr\'eal, QC, Canada, H3C 3J7; email: \href{mailto:maciej.augustyniak@umontreal.ca}{\nolinkurl{maciej.augustyniak@umontreal.ca}}} \quad
Alexandru Badescu\footnote{Corresponding author. Department of Mathematics and Statistics, University of Calgary, 2500 University Drive NW, Calgary, AB, Canada, T2N 1N4; email: \href{mailto:abadescu@ucalgary.ca}{\nolinkurl{abadescu@ucalgary.ca}}} \quad
Zekun Zhang\footnote{Department of Mathematics and Statistics, University of Calgary, 2500 University Drive NW, Calgary, AB, Canada, T2N 1N4; email: \href{mailto:zekun.zhang@ucalgary.ca}{\nolinkurl{zekun.zhang@ucalgary.ca}}}
}

\newcommand{\Keywords}{Polynomial-preserving processes; hybrid GARCH–stochastic volatility models; weak limits; option pricing}

\newcommand{\JEL}{C58, G12, G13}

\newcolumntype{d}[1]{D{.}{.}{#1}}
\newcolumntype{Y}{>{\raggedleft\arraybackslash}X}
\newcolumntype{C}{>{\centering\arraybackslash}X}

\newcommand{\Var}{{\rm Var}}
\newcommand{\Cov}{{\rm Cov}}
\newcommand{\Corr}{{\rm Corr}}
\newcommand{\E}{{\rm E}}

\newcommand{\BigO}[1]{\operatorname{O}\left(#1\right)}

\DeclareMathOperator{\Pol}{Pol}

\newcommand{\Pm}{\mathbb{P}}
\newcommand{\Qm}{\mathbb{Q}}

\newcommand{\F}{\mathcal{F}}
\newcommand{\Ft}{\mathbb{F}}
\newcommand{\Rplus}{\mathbb{R}_{+}}
\newcommand{\HT}{\mathsf{H}}
\newcommand{\trans}{\mathcal{G}}

\newtheorem{definition}{Definition}
\newtheorem{lemma}{Lemma}
\newtheorem{remark}{Remark}
\newtheorem{proposition}{Proposition}
\newtheorem{corollary}{Corollary}
\newtheorem{example}{Example}
\newtheorem{assumption}{Assumption}

\newtheorem*{examplecontinued}{Example~\ref{ex:bivariate_polynomial_moment} (continued)}

\titleformat{\section}{\normalfont\bfseries}{\thesection}{1em}{}
\titleformat{\subsection}{\normalfont\bfseries}{\thesubsection}{1em}{}
\titleformat{\subsubsection}{\normalfont\bfseries}{\thesubsubsection}{1em}{}

\titlespacing*{\section}{0pt}{12pt}{6pt}
\titlespacing*{\subsection}{0pt}{12pt}{6pt}
\titlespacing*{\subsubsection}{0pt}{12pt}{6pt}

\hypersetup{hypertexnames=false,
  pdftitle={Polynomial-Preserving Processes and Volatility Modelling in Discrete Time},
  pdfauthor={Maciej Augustyniak, Alexandru Badescu, Zekun Zhang}}
\makeatletter
\newcommand{\ArxivBibScope}[1]{%
  \def\@extra@b@citeb{#1}%
  \if@filesw\immediate\write\@auxout{\string\gdef\string\@extra@binfo{#1}}\fi}
\makeatother
\begin{document}
\ArxivBibScope{}
\bookmark[level=0,page=1]{Manuscript}
\makeatletter
\let\ArxivMakeTitle\maketitle
\let\ArxivTitleBlock\@maketitle
\let\ArxivTitle\title
\let\ArxivAuthor\author
\let\ArxivDate\date
\let\ArxivThanks\thanks
\makeatother
\allowdisplaybreaks
\singlespacing

\newif\ifblindedfirstpage
\blindedfirstpagefalse

\ifblindedfirstpage
	\title{\Title\bigskip}
	\author{\vspace{-0.5cm}}
\else
	\title{\Title
	\bigskip}
	\author{\Authors}
\fi

\date{\vspace{-0.5cm}}

\maketitle

\begin{abstract}
\noindent  This paper develops a discrete-time framework for volatility modelling and derivative pricing based on polynomial preservation.  The framework provides exact moments of arbitrary order and serial cross-moments of variance and return processes over any finite horizon through finite-dimensional matrix representations. This analytical tractability is maintained while accommodating flexible leverage and variance-of-variance dynamics and non-Gaussian innovations. We introduce two novel classes of stochastic volatility models that nest widely used discrete-time specifications and study their statistical properties. Under two distinct scaling regimes, we establish weak convergence to polynomial diffusions and jump-diffusions and convergence of polynomial moments, linking discrete-time moment recursions to continuous-time matrix-exponential formulas. An empirical analysis of S\&P~500 returns and a large panel of options spanning 1996--2023 shows that the proposed specifications improve return fit and reduce mean weekly implied volatility root mean squared errors by approximately 8--41\% relative to their nested benchmarks, depending on the variance dynamics and innovation distribution.
\bigskip


\bigskip
\noindent {\it JEL Classification:} \JEL .
\end{abstract}

\newpage
\bigskip

\onehalfspacing

\section{Introduction}\label{sec:introduction}

Although volatility is a central determinant of the dynamics of asset returns and option prices, the leverage effect and the volatility of the conditional variance also play an important role in shaping the asymmetry and heavy-tailedness of conditional return distributions, as well as volatility smiles. Models with similar expected variance paths can therefore imply markedly different return tails and smiles because they impose different responses to return shocks and different uncertainty about future variance.\footnote{See, e.g., \citet{hesto93,jones03,christoffersen2009shape,ChristoffersenJacobsMimouni2010,GruberTebaldiTrojani2021,Bollerslev2023,ChongTodorov2024} in continuous time, and \citet{christoffersen2004garch,christoffersen2006option,babaouglu2018option,Han2020,Chorro22} in discrete time.}  Capturing these features, however, involves a well-known trade-off between flexibility and tractability: specifications rich enough to reproduce them often lack closed-form expressions for key quantities, whereas tractable specifications typically restrict how return shocks and variance uncertainty can interact. We address this challenge by developing a general discrete-time polynomial-preserving framework in which conditional moments can be computed exactly at any horizon, and we use it to construct new stochastic volatility (SV) models, establish their weak limits, and characterize polynomial preservation under the pricing measure.  Exact moments are valuable because they make the link between model parameters and the term structures of variance, skewness, and kurtosis explicit, and they provide multi-horizon forecasts and inputs for derivative valuation without simulation error or the associated computational burden. Relative to their nested benchmarks, the proposed models improve the fit to returns and a large panel of equity index options.

Polynomial preservation means that the conditional expectation of any polynomial in the future state is a polynomial of no higher degree in the current state. Consequently, conditional moments of any given order can be computed through a finite-dimensional system involving only moments of that order or lower.  Polynomial processes have been extensively studied in continuous time. Their foundations are developed by \citet{cuchiero2012polynomial,filipovic2016,FilipovicLarsson2020}, with conditional polynomial moments obtained from finite-dimensional matrix exponentials. Financial and econometric applications include density approximations for likelihood inference and option pricing \citep[see, e.g.,][]{FilipovicMayerhoferSchneider2013}, variance-swap term structures \citep[see, e.g.,][]{FilipovicGourierMancini2016}, option-price expansions \citep[see, e.g.,][]{Ackerer18,AckererFilipovic2020}, and moment-based simulation \citep[see, e.g.,][]{Brignone2026}. %
Only recently have \citet{KallsenRichert25,KallsenRichertEstimation2026} developed a general discrete-time polynomial state-space framework for filtering and estimation. Their framework includes the standard symmetric GARCH(1,1) model as a special case, but their primary focus is on inference rather than on characterizing the dynamics that preserve polynomiality in discrete-time SV models with leverage effects. We address this important modelling question and its implications for derivative pricing and continuous-time limits.

First, following the operator approach of \citet{filipovic2016}, we characterize polynomial preservation for a class of discrete-time Markov chains and use the resulting restrictions to construct two new classes of volatility models. We consider hybrid GARCH--SV models whose variance dynamics involve the first two nonconstant Hermite polynomials of the innovation, with state-dependent coefficient functions (henceforth innovation loadings).\footnote{\citet{Meddahi2001Eigenfunction} also uses Hermite polynomials for volatility modelling. In his Hermite SV framework, variance is specified as a linear combination of these polynomials evaluated at a latent Gaussian AR(1) factor. Our formulation instead applies Hermite polynomials to innovations in the variance equation, and our objective is to characterize the innovation loadings that preserve polynomiality.}  For jointly symmetric innovations with finite moments of all orders, we identify the restrictions on conditional means and innovation loadings that preserve polynomial moments of variance alone or jointly with returns. The framework accommodates non-Gaussian innovations and distinct, correlated return and variance shocks. In the Gaussian case, the first class, variance-polynomial hybrid GARCH--SV (VPGARCH-SV), extends the NGARCH dynamics of \citet{engle1993measuring}, used for option valuation, for example, by \citet{duan1995garch}, while the second, jointly polynomial hybrid GARCH--SV (JPGARCH-SV), extends the affine HNGARCH model of \citet{heston2000closed} and its SV counterpart studied by \citet{Jacobs24}.  We establish necessary and sufficient conditions for nonnegative conditional variance within the maintained parameter domains. Both specifications assume affine conditional mean variance dynamics while allowing richer variance-of-variance functions and state-dependent leverage effects. Both specifications yield exact conditional moments and serial cross-moments of variance, cumulative variance, and squared conditionally demeaned returns, while JPGARCH-SV also provides cumulative-return moments.\footnote{The latter property also holds for VPGARCH-SV with independent innovations. Related analytical results include GARCH moment conditions \citep[see, e.g.,][]{HeTerasvirta99,LingMcAleer02,LingMcAleer02a}, volatility moment restrictions \citep[see, e.g.,][]{Meddahi04}, and cumulative-return moments \citep[see, e.g.,][]{wongso2003,AlexanderLazarStanescu21}.} Such higher-order and cross-date moments can be cumbersome to derive even for standard GARCH specifications. For the proposed classes, the model-specific task is to construct the one-step transition matrices from the model parameters and innovation moments; the same matrix formulas then apply across specifications and forecast horizons. At any fixed moment order, the matrix dimensions do not grow with the horizon.

Second, we connect our proposed discrete-time models to polynomial diffusions and jump-diffusions, with convergence of both the dynamics and their polynomial moments. Building on the weak-limit literature \citep[see, e.g.,][]{nelson1990arch,corradi2000reconsidering,Duan97,Duan06,badescu2024discrete}, we derive two limiting regimes under the physical measure. Both regimes preserve polynomial variance dynamics and yield joint polynomiality in log price and variance for JPGARCH-SV. Its diffusion limit combines quadratic variance of variance with linear instantaneous return--variance covariance on an unbounded variance state space, distinct from the bounded-state Jacobi model of \citet{Ackerer18}. Scaling with standardized L\'evy increments yields members of the polynomial jump-diffusion class of \citet{FilipovicLarsson2020}.  This construction includes a COGARCH variance specification \citep[see, e.g.,][]{KluppelbergLindnerMaller2004,Kallsen2009} and allows Brownian and jump components in both state coordinates. We give sufficient conditions for nonnegative approximations and global weak convergence. For the corresponding polynomial states, moment recursions converge to continuous-time matrix-exponential formulas. These results provide a new discrete-time construction of continuous-time stochastic volatility models with polynomial variance dynamics.

Third, we characterize polynomial preservation under the pricing measure. We adopt variance-dependent pricing kernels, which are widely used in discrete-time option valuation and allow separate equity and variance risk-preference parameters \citep[see, e.g.,][]{christoffersen2013capturing,corsi13,majew15,bades17}. Specifically, we derive the transformed dynamics under the required cumulant-generating-function conditions and characterize simultaneous polynomial preservation under jointly Gaussian innovations. Preservation imposes additional restrictions  on risk preferences and innovation loadings and can fail even under a monotonic pricing kernel. The results identify when exact risk-neutral moment recursions are available and clarify why polynomial physical dynamics can remain useful for derivative valuation when the pricing dynamics are nonpolynomial.  Corresponding local risk-neutral diffusion limits are established under additional regularity conditions.

Finally, we assess the empirical relevance of the additional flexibility of the proposed models using daily S\&P~500 returns and 664,592 option observations across 1,443 weekly dates spanning 1996--2023. Return estimation by maximum likelihood and weekly option calibration use the GARCH special cases under Gaussian and symmetric normal inverse Gaussian (NIG) innovations. Both proposed specifications substantially improve  the fit relative to their nested benchmarks. JPGARCH reduces the mean weekly implied volatility root mean squared error (IVRMSE) relative to HNGARCH by approximately 41\% under Gaussian innovations and 8\% under NIG innovations; the corresponding gains of VPGARCH over NGARCH are 11\% and 14\%, respectively. VPGARCH achieves the lowest mean weekly pricing error under both distributions. NIG innovations improve option fit much more for HNGARCH and JPGARCH than for NGARCH and VPGARCH, suggesting that the value of heavier innovation tails depends on the variance dynamics. Exact physical moments further reveal differences in the persistence of variance uncertainty that expected-variance term structures alone do not capture.

Section~\ref{sec:polynomial_garch} presents the moment framework, characterizes polynomial-preserving hybrid GARCH--SV dynamics, and derives their moment implications. Section~\ref{sec:polynomial_diffusion_limits} develops the diffusion and jump limits. Section~\ref{sec:empirical_analysis} presents the empirical illustration, and Section~\ref{sec:conclusions} concludes. Appendix~\ref{sec:risk_neutral_dynamics} studies the pricing kernel and risk-neutral preservation.

\section{Polynomial-Preserving Hybrid GARCH--SV Models}\label{sec:polynomial_garch}

This section characterizes hybrid GARCH--SV dynamics with analytically tractable moments under the physical measure. After presenting the general moment framework, we identify the innovation loading specifications that ensure polynomial preservation and construct two canonical model classes. We then establish positivity and moment conditions and derive moment formulas for variance and return-based quantities, including cumulative variance and returns.

\subsection{Discrete-time polynomial-preserving Markov chains}\label{subsec:general_framework}

We formulate polynomial preservation for time-homogeneous Markov chains using transition operators, with the same degree restriction on conditional moments as in \citet{KallsenRichert25}. Let $\tau=1/n$, where $n\geq1$ is an integer, and fix an integer horizon $L\geq1$. On a filtered probability space $\left(\Omega^{(n)},\F^{(n)},\Ft^{(n)},\Pm^{(n)}\right)$, with $\Ft^{(n)}=\{\F^{(n)}_{k\tau}\}_{k=0}^{nL}$ and physical measure $\Pm^{(n)}$, consider a time-homogeneous Markov chain $\bm{X}^{(n)}=\{\bm{X}^{(n)}_{k\tau}\}_{k=0}^{nL}$ on a Borel state space $E\subseteq\mathbb{R}^d$, adapted and Markov with respect to $\Ft^{(n)}$. Its one-step transition kernel $\Pi^{(n)}\left(\bm{x},\cdot\right)$ gives the conditional distribution of the next state given the current state $\bm{x}\in E$. Throughout this subsection, $n$ is fixed; its indexing is retained for the weak-limit analysis in Section~\ref{sec:polynomial_diffusion_limits}. Let $\Pol_m\left(E\right)$ denote the restrictions to $E$ of real polynomials on $\mathbb{R}^d$ of total degree at most $m$, write $\Pol\left(E\right)=\bigcup_{m\geq0}\Pol_m\left(E\right)$, and set $N_m=\dim\Pol_m\left(E\right)$.\footnote{The degree of a polynomial on $E$ is the smallest degree of a polynomial on $\mathbb{R}^d$ restricting to it. When $E$ has nonempty interior, as in the applications below, the restriction map is one-to-one and $N_m=\binom{m+d}{d}$. See \citet{filipovic2016} for polynomial spaces on semialgebraic state spaces.} Throughout, $\Rplus:=\left[0,\infty\right)$. For $p\in\Pol\left(E\right)$ satisfying the one-step integrability condition
\begin{equation}\label{eq:polynomial_integrability}
\int_E\left|p\left(\bm{z}\right)\right|\Pi^{(n)}\left(\bm{x},d\bm{z}\right)<\infty,\qquad \bm{x}\in E,
\end{equation}
define the one-step transition operator by
\begin{equation}\label{eq:transition_operator}\trans^{(n)}p\left(\bm{x}\right):=\int_E p\left(\bm{z}\right)\Pi^{(n)}\left(\bm{x},d\bm{z}\right)=\E^{\Pm^{(n)}}\left[p\left(\bm{X}^{(n)}_{(k+1)\tau}\right)\,\middle|\,\bm{X}^{(n)}_{k\tau}=\bm{x}\right],\qquad \bm{x}\in E.\end{equation}
The following definition requires this conditional expectation to remain polynomial without increasing degree.
\begin{definition}\label{def:pol_preserving_chain}
Let $m\geq0$ be an integer. The process $\bm{X}^{(n)}$ is \emph{$m$-polynomial preserving} if Condition~\eqref{eq:polynomial_integrability} holds for every $p\in\Pol_m\left(E\right)$ and
\begin{equation*}\trans^{(n)}\Pol_r\left(E\right)\subseteq\Pol_r\left(E\right),\qquad r=0,\ldots,m.\end{equation*}
It is \emph{polynomial preserving} if it is $m$-polynomial preserving for every integer $m\geq0$.
\end{definition}

The following illustration shows what polynomial preservation means for a two-dimensional state.

\begin{example}\label{ex:bivariate_polynomial_moment}
Take $d=2$ and $m=2$, with $E=\mathbb{R}^2$. The polynomial $p(\bm{x})=x_1x_2$ has total degree two, so a polynomial-preserving chain satisfies
\begin{equation*}
\E^{\Pm^{(n)}}\left[X^{(n)}_{1,(k+1)\tau}X^{(n)}_{2,(k+1)\tau}\,\middle|\,\bm{X}^{(n)}_{k\tau}=\bm{x}\right]=c_0+c_1x_1+c_2x_2+c_3x_1^2+c_4x_1x_2+c_5x_2^2,
\end{equation*}
for constants $c_0,\ldots,c_5$. For example, in a jointly polynomial SV model where $X_1$ represents the return and $X_2$ the variance state, the same polynomial form gives their one-step conditional cross-moment.
\end{example}

The degree restriction closes the moment system under repeated conditioning; the proofs below use full polynomial preservation to establish conditional integrability over multiple steps and for products at different dates. Conditional expectations of signed random variables are understood in the generalized sense whenever the conditional expectation of their absolute value is finite almost surely.

For each $m\geq0$, choose a basis $\bm{\HT}_m=\left(\HT_{m,1},\ldots,\HT_{m,N_m}\right)^\top$ of $\Pol_m\left(E\right)$, with $\bm{\HT}_0\equiv1$, such that the first $N_r$ entries of $\bm{\HT}_m$ coincide with $\bm{\HT}_r$ for every $0\leq r\leq m$. Each $p\in\Pol_m\left(E\right)$ has a unique coordinate vector $\bm{p}\in\mathbb{R}^{N_m}$ satisfying
\begin{equation}\label{eq:coordinate_representation}p\left(\bm{x}\right)=\bm{\HT}_m\left(\bm{x}\right)^\top\bm{p}.\end{equation}
If $\bm{X}^{(n)}$ is $m$-polynomial preserving, the restriction of $\trans^{(n)}$ to $\Pol_m\left(E\right)$ has a unique coordinate matrix $\bm{Q}^{(n)}_m$, characterized by
\begin{equation}\label{eq:basis_vector_transition}\trans^{(n)}\bm{\HT}_m\left(\bm{x}\right)=\left(\bm{Q}^{(n)}_m\right)^\top\bm{\HT}_m\left(\bm{x}\right),\end{equation}
where $\trans^{(n)}$ acts componentwise. Degree preservation makes $\bm{Q}^{(n)}_m$ block upper triangular; the transpose reflects the convention that $\bm{Q}^{(n)}_m$ acts on polynomial coordinate vectors. In a monomial basis, its columns contain the coefficients of the one-step conditional moments. The following proposition records the corresponding matrix-power moment formula in our notation; see \citet{KallsenRichert25}.

\begin{proposition}\label{prop:finite_dimensional_representation}
Suppose that $\bm{X}^{(n)}$ is polynomial preserving. Let $p\in\Pol_m\left(E\right)$ have coordinate vector $\bm{p}$ as in Equation~\eqref{eq:coordinate_representation}. Then, for all integers $k\geq0$ and $s\geq0$ such that $k+s\leq nL$,
\begin{equation}\label{eq:polynomial_expectation}\E^{\Pm^{(n)}}\left[p\left(\bm{X}^{(n)}_{(k+s)\tau}\right)\,\middle|\,\F^{(n)}_{k\tau}\right]=\bm{\HT}_m\left(\bm{X}^{(n)}_{k\tau}\right)^\top\left(\bm{Q}^{(n)}_m\right)^s\bm{p},\end{equation}
with the convention $\left(\bm{Q}^{(n)}_m\right)^0=\bm{I}_{N_m}$.
\end{proposition}

\begin{proof}
See Supplementary Material~\ref{sm:proof_finite_dimensional_representation}.
\end{proof}

\begin{examplecontinued}
The space $\Pol_2(E)$ has dimension $N_2=1+2+3=6$. Choose the degree-ordered monomial basis
\begin{equation*}
\bm{\HT}_2(\bm{x})=\left(\underbrace{1}_{\text{degree }0},\ \underbrace{x_1,x_2}_{\text{degree }1},\ \underbrace{x_1^2,x_1x_2,x_2^2}_{\text{degree }2}\right)^\top.
\end{equation*}
The corresponding one-step transition matrix has the block form
\begin{equation*}
\bm Q^{(n)}_2=
\begin{pmatrix}
1 & \bm Q^{(n)}_{01} & \bm Q^{(n)}_{02}\\
\bm 0 & \bm Q^{(n)}_{11} & \bm Q^{(n)}_{12}\\
\bm 0 & \bm 0 & \bm Q^{(n)}_{22}
\end{pmatrix}
\in\mathbb{R}^{6\times6}.
\end{equation*}
The row and column blocks correspond to degrees $0,1,2$ and have sizes $1,2,3$, respectively. For example, $\bm Q^{(n)}_{12}$ is $2\times3$ and $\bm Q^{(n)}_{22}$ is $3\times3$.  The third column gives the coefficients of the conditional mean of the second coordinate, while the fifth column is $(c_0,\ldots,c_5)^\top$, corresponding to the mixed second moment above. By Proposition~\ref{prop:finite_dimensional_representation}, the corresponding columns of $\left(\bm Q^{(n)}_2\right)^s$ give the coefficients of these conditional moments $s$ steps ahead.
\end{examplecontinued}

Equation~\eqref{eq:polynomial_expectation} computes conditional polynomial moments $s$ time steps ahead from a single matrix $\bm{Q}^{(n)}_m$ whose dimension $N_m$ does not grow with $s$. The $s$-step transition operator is $\trans_s^{(n)}=\left(\trans^{(n)}\right)^s$, with $\trans_0^{(n)}$ the identity, so the chain sampled every $s$ time steps, $s\geq1$, is also polynomial preserving. When the initial state is deterministic, setting $k=0$ gives the corresponding finite-horizon moments.

The nested bases yield a recursion in which moments of degree $r$ evolve through their own diagonal block and moments of lower degree. Following the nested-matrix approach of \citet{DawPender2023}, Supplementary Material~\ref{sm:matryoshka_details} implements this recursion using block matrix--vector products, avoiding full matrix powers for consecutive horizons from a given current state and reusing lower-degree calculations when extending to higher orders.

The nonconstant degree blocks also determine long-horizon moment stability. For this purpose, extend the chain to all $k\geq0$ using the same transition kernel. Writing $\bm{Q}^{(n)}_{rr}$ for the degree-$r$ diagonal block and $\rho\left(\cdot\right)$ for the spectral radius, the condition
\begin{equation}\label{eq:spectral_stationarity_condition}\rho\left(\bm{Q}^{(n)}_{rr}\right)<1,\qquad r=1,\ldots,m,\end{equation}
ensures that the conditional moments in Equation~\eqref{eq:polynomial_expectation} converge as $s\to\infty$ to finite limits independent of the current state. Proposition~\ref{sm:prop:matryoshka_stationary_moments} computes these limits recursively; if an invariant distribution with finite moments through degree $m$ exists, they coincide with its moments.

Cross-date moments are also needed to compute serial covariances and moments of cumulative quantities. These follow by combining the transition matrices with polynomial multiplication. For $p\in\Pol_a\left(E\right)$ and an integer $b\geq0$, the map $\mathcal{M}_p:\Pol_b\left(E\right)\to\Pol_{a+b}\left(E\right)$, $f\mapsto pf$, has coordinate matrix $\bm{M}_p\in\mathbb{R}^{N_{a+b}\times N_b}$ determined by
\begin{equation}\label{eq:polynomial_multiplication_matrix}p\left(\bm{x}\right)\bm{\HT}_b\left(\bm{x}\right)^\top=\bm{\HT}_{a+b}\left(\bm{x}\right)^\top\bm{M}_p.\end{equation}
The dependence of $\bm{M}_p$ on $b$ is suppressed when clear from the context. Repeated conditioning then gives the following backward recursion, a discrete-time counterpart of the correlator formulas developed by \citet{BenthLavagnini2021} for one-dimensional continuous-time polynomial processes.

\begin{corollary}\label{cor:polynomial_cross_moments}
Suppose that $\bm{X}^{(n)}$ is polynomial preserving, and let $J\geq1$ be an integer. For $j=1,\ldots,J$, let $p_j\in\Pol_{m_j}\left(E\right)$ have coordinate vector $\bm{p}_j$, and let $k,s_1,\ldots,s_J\geq0$ be integers such that $k+s_1+\cdots+s_J\leq nL$. Set $\mu_j=m_j+\cdots+m_J$, let $\bm{v}_J=\bm{p}_J$, and define recursively, for $j=J-1,\ldots,1$,
\begin{equation*}\bm{v}_j=\bm{M}_{p_j}\left(\bm{Q}^{(n)}_{\mu_{j+1}}\right)^{s_{j+1}}\bm{v}_{j+1}\in\mathbb{R}^{N_{\mu_j}},\end{equation*}
where $\bm{M}_{p_j}$ is the multiplication matrix of Equation~\eqref{eq:polynomial_multiplication_matrix} acting from $\Pol_{\mu_{j+1}}\left(E\right)$ to $\Pol_{\mu_j}\left(E\right)$. Then
\begin{equation}\label{eq:conditional_polynomial_product}\E^{\Pm^{(n)}}\left[\prod_{j=1}^{J}p_j\left(\bm{X}^{(n)}_{\left(k+s_1+\cdots+s_j\right)\tau}\right)\,\middle|\,\F^{(n)}_{k\tau}\right]=\bm{\HT}_{\mu_1}\left(\bm{X}^{(n)}_{k\tau}\right)^\top\left(\bm{Q}^{(n)}_{\mu_1}\right)^{s_1}\bm{v}_1.\end{equation}
\end{corollary}

\begin{proof}
See Supplementary Material~\ref{sm:proof_polynomial_cross_moments}.
\end{proof}

The two-date case illustrates the calculation. For $p\in\Pol_{m_1}\left(E\right)$ and $q\in\Pol_{m_2}\left(E\right)$ with coordinate vector $\bm{q}$, and nonnegative integers $k,s,\ell$ such that $k+s+\ell\leq nL$, Equation~\eqref{eq:conditional_polynomial_product} gives
\begin{equation*}\E^{\Pm^{(n)}}\left[p\left(\bm{X}^{(n)}_{(k+s)\tau}\right)q\left(\bm{X}^{(n)}_{(k+s+\ell)\tau}\right)\mid\F^{(n)}_{k\tau}\right]=\bm{\HT}_{m_1+m_2}\left(\bm{X}^{(n)}_{k\tau}\right)^\top\left(\bm{Q}^{(n)}_{m_1+m_2}\right)^s\bm{M}_p\left(\bm{Q}^{(n)}_{m_2}\right)^\ell\bm{q}.\end{equation*}
Subtracting the product of the two single-date forecasts gives the conditional covariance.

For $p,q\in\Pol_m\left(E\right)$ and an integer $s\geq0$, define the forecast-error covariance polynomial
\begin{equation*}V^{(n)}_{p,q,s}\left(\bm x\right):=\trans_s^{(n)}\left(pq\right)\left(\bm x\right)-\left(\trans_s^{(n)}p\right)\left(\bm x\right)\left(\trans_s^{(n)}q\right)\left(\bm x\right).\end{equation*}
For $k+s\leq nL$, $V^{(n)}_{p,q,s}\left(\bm X^{(n)}_{k\tau}\right)$ is the conditional covariance of the $s$-step forecast errors for $p$ and $q$; in particular, $V^{(n)}_{p,p,s}$ gives the conditional mean-square forecast error. Each $V^{(n)}_{p,q,s}$ belongs to $\Pol_{2m}\left(E\right)$ and is computable from transition matrices through degree $2m$. Its scaled one-step limit is studied in Section~\ref{subsec:general_diffusion_framework}; its variance-chain version, evaluated after the initial shock, also gives the variance-of-variance news impact curves in Section~\ref{subsec:return_estimation}.

Expanding powers of sums into cross-date products also gives moments of cumulative quantities. For $p\in\Pol_a\left(E\right)$, transition matrices through degree $ar$ suffice for the $r$th moment of a sum of its evaluations, independently of the number of dates, although the number of terms may grow with the horizon. When $a\leq1$, appending the running sum to the state preserves polynomiality and gives a fixed-size matrix-power representation. Section~\ref{subsec:temporal_aggregates} uses this approach for cumulative variance and returns. %

\subsection{Polynomial-preserving hybrid GARCH--SV dynamics under the physical measure}
\label{subsec:polynomial_garch_models_p}

We consider a discrete-time financial market consisting of a risk-free bond and a risky asset with price process $S^{(n)}=\left\{S^{(n)}_{k\tau}\right\}_{k=0}^{nL}$. Let $Y^{(n)}=\left\{Y^{(n)}_{k\tau}\right\}_{k=0}^{nL}$ denote its logarithm, so that $Y^{(n)}_{k\tau}=\log S^{(n)}_{k\tau}$. The difference $y^{(n)}_{(k+1)\tau}:=Y^{(n)}_{(k+1)\tau}-Y^{(n)}_{k\tau}$ is the risky asset's one-step log return (henceforth return), and $h^{(n)}_{k\tau}$ denotes its conditional variance rate, so that $\Var^{\Pm^{(n)}}\!\left[y^{(n)}_{(k+1)\tau}\mid\F^{(n)}_{k\tau}\right]=\tau h^{(n)}_{k\tau}$.

In the framework of Section~\ref{subsec:general_framework}, variance preservation corresponds to the one-dimensional state $\bm{X}^{(n)}=h^{(n)}$ with $d=1$, while joint return--variance preservation corresponds to the two-dimensional state $\bm{X}^{(n)}=\left(y^{(n)},h^{(n)}\right)^\top$ with $d=2$.  Characterizing polynomial preservation for these states in terms of the conditional means and innovation loadings leads to two canonical model classes. We work with jointly symmetric standardized innovations, allowing non-Gaussian conditional kurtosis. We incorporate stochastic volatility by allowing the variance innovation to differ from, and remain correlated with, the return innovation, in the spirit of the affine Gaussian HNGARCH-SV model studied by \citet{Jacobs24}.\footnote{A related approach adds an independent variance innovation to GARCH dynamics; see \citet{EscobarStentoftYe2023}.} Furthermore, under Gaussian innovations, the resulting classes extend the NGARCH model of \citet{engle1993measuring}, used for option valuation by \citet{duan1995garch}, and the affine HNGARCH model of \citet{heston2000closed}.\footnote{The characterization concerns the innovation representation below, not all GARCH models. Other specifications, including some within the augmented GARCH framework of \citet{Duan97}, may require different innovation functions or transformed states.} We assume the following $n$-indexed dynamics under $\Pm^{(n)}$:
\begin{align}
y^{(n)}_{(k+1)\tau} &= \psi_0\!\left(h^{(n)}_{k\tau};\tau\right)+\sqrt{\tau h^{(n)}_{k\tau}}\,\epsilon^{(n)}_{1,(k+1)\tau}, \label{eq:general_return_dynamics_p}\\
h^{(n)}_{(k+1)\tau} &= \psi_1\!\left(h^{(n)}_{k\tau};\tau\right)+\psi_2\!\left(h^{(n)}_{k\tau};\tau\right)\epsilon^{(n)}_{2,(k+1)\tau}+\psi_3\!\left(h^{(n)}_{k\tau};\tau\right)\left[\left(\epsilon^{(n)}_{2,(k+1)\tau}\right)^2-1\right]. \label{eq:general_variance_dynamics_p}
\end{align}
Here, the innovation pairs are i.i.d. under $\Pm^{(n)}$, with a joint distribution common to all $n$ satisfying
\begin{equation*}\left(\epsilon_1,\epsilon_2\right)\overset{\mathcal{D}}{=}\left(-\epsilon_1,-\epsilon_2\right), \qquad \epsilon_i\overset{\Pm^{(n)}}{\sim} D_i\left(0,1\right),\quad i=1,2, \qquad \Corr^{\Pm^{(n)}}\left[\epsilon_1,\epsilon_2\right]=\E^{\Pm^{(n)}}\left[\epsilon_1\epsilon_2\right]=\varrho\in\left[0,1\right].\end{equation*}
The marginal laws $D_i\left(0,1\right)$ are standardized and have finite moments of all orders, but need not coincide. Write $m_{4,i}:=\E^{\Pm^{(n)}}\left[\epsilon_i^4\right]$, $i=1,2$, and $m_4:=m_{4,2}>1$ for the fourth moment of the variance innovation. The restriction $\varrho\geq0$ is a sign convention for this class: replacing $\epsilon_2$ by $-\epsilon_2$ and $\psi_2$ by $-\psi_2$, while reflecting the joint law accordingly, leaves the dynamics unchanged and reverses the innovation correlation.

We assume $h^{(n)}_0=h_0\in\Rplus$ to be deterministic and $\Ft^{(n)}$ to be the augmented natural filtration of the innovation pairs, with $\F^{(n)}_{k\tau}=\sigma\!\left(\epsilon^{(n)}_{1,s\tau},\epsilon^{(n)}_{2,s\tau}\mid1\leq s\leq k\right)$ up to completion. In addition, without loss of generality we set $y^{(n)}_0:=0$. All conditional moments below refer to this full-information filtration, which may contain more information than the observed return history. The functions $\psi_0$ and $\psi_1$ specify the conditional means of the return and next-period variance rate, while $\psi_2$ and $\psi_3$ are the loadings on the linear and centred quadratic variance-innovation terms. Their parameters may depend on $\tau$, and we write this dependence as $\psi_j\left(\cdot;\tau\right)$. We require the variance dynamics to preserve $\Rplus$ and explicit restrictions are given below for the two specifications.

Henceforth, we append the suffix -SV to denote the hybrid versions of the GARCH specifications and retain the original names for their $\varrho=1$ special cases. When $\varrho=1$, standardization implies $\epsilon_1=\epsilon_2$ almost surely, and Equations~\eqref{eq:general_return_dynamics_p}--\eqref{eq:general_variance_dynamics_p} recover the corresponding GARCH dynamics. For $\varrho<1$, the framework accommodates variance innovations that are not determined by the return innovation.\footnote{In the jointly Gaussian case, the representation $\epsilon_2=\varrho\epsilon_1+\sqrt{1-\varrho^2}\,u$, with $u\sim N\left(0,1\right)$ independent of $\epsilon_1$, makes the additional source of variance uncertainty explicit.} At a current variance state $x>0$, joint symmetry gives
\begin{align*}\Cov^{\Pm^{(n)}}\!\left[y^{(n)}_{(k+1)\tau},h^{(n)}_{(k+1)\tau}\mid h^{(n)}_{k\tau}=x\right]&=\varrho\sqrt{\tau x}\,\psi_2\left(x;\tau\right),\\ \Var^{\Pm^{(n)}}\!\left[h^{(n)}_{(k+1)\tau}\mid h^{(n)}_{k\tau}=x\right]&=\psi_2\left(x;\tau\right)^2+\left(m_4-1\right)\psi_3\left(x;\tau\right)^2.\end{align*}
This implies that at a given variance level, changing $\psi_3$ while holding $\psi_2$ and $\varrho$ fixed can change the variability of future variance without affecting its covariance with returns. Moreover, the relative magnitudes of $\psi_2$ and $\psi_3$ can vary with the current variance, allowing the conditional return--variance correlation to be state-dependent even when $\varrho$ is constant.\footnote{Note that in both canonical specifications introduced below we assume $\psi_2$ is a nonnegative function multiplied by a signed parameter $\eta\left(\tau\right)$, which governs the sign of nonzero return--variance covariance when $\varrho>0$. The parameter $\eta\left(\tau\right)$ also affects the marginal variance dynamics, whereas changing $\varrho$ with the marginal laws fixed leaves those dynamics unchanged.}

Under these dynamics, $h^{(n)}$ is a time-homogeneous Markov chain on $\Rplus$, while $\left(y^{(n)},h^{(n)}\right)$ is a time-homogeneous Markov chain on $\mathbb{R}\times\Rplus$, both with respect to $\Ft^{(n)}$. For each fixed $\tau$, we suppress the second argument in polynomial-space statements and write $\psi_j$ for the function $x\mapsto\psi_j\left(x;\tau\right)$. Under an affine restriction on $\psi_3$, the following characterization holds for every joint innovation distribution satisfying the stated symmetry and moment assumptions.\footnote{When the variance innovation is Gaussian, polynomial preservation itself implies that $\psi_3$ is affine. Supplementary Material~\ref{sm:sec:higher_order_hermite} gives a higher-order Hermite extension under jointly Gaussian innovations.}
\begin{proposition}
\label{prop:garch_polynomial_characterization}
Fix $\tau$ and $\varrho\in\left[0,1\right]$, and suppose that $\psi_3\in\Pol_1\left(\Rplus\right)$, while $\psi_0,\psi_1,\psi_2:\Rplus\to\mathbb{R}$ are Borel measurable.

\begin{enumerate}[label=(\roman*)]
\item The process $h^{(n)}$ is polynomial preserving on $\Rplus$ if and only if
\begin{equation*}\psi_1\in\Pol_1\left(\Rplus\right), \qquad \psi_2^2\in\Pol_2\left(\Rplus\right).\end{equation*}

\item The process $\left(y^{(n)},h^{(n)}\right)$ is polynomial preserving on $\mathbb{R}\times\Rplus$ if
\begin{equation*}\psi_0,\psi_1\in\Pol_1\left(\Rplus\right), \qquad \psi_2\in\operatorname{span}\left\{x\mapsto\sqrt{x}\right\}.\end{equation*}
These conditions are also necessary when $\varrho>0$. For independent innovations, joint preservation is equivalent to variance preservation together with $\psi_0\in\Pol_1\left(\Rplus\right)$.
\end{enumerate}
\end{proposition}

\begin{proof}
See Appendix~\ref{app:proof_garch_polynomial_characterization}.
\end{proof}
For $\varrho>0$, the square-root restriction on $\psi_2$ arises from mixed return--variance moments. Joint polynomial preservation enables the cumulative-return moment calculations in Section~\ref{subsec:analytic_moments}.

For both  specifications, we set\footnote{An affine $\psi_0$ is needed for joint preservation, but not for variance preservation. We use the same conditional mean return specification in both models to isolate the effects of their variance dynamics.}
\begin{equation}
\label{eq:common_mean_parametrizations} \psi_0\left(x;\tau\right)=r\tau+\lambda\left(\tau\right)x, \qquad \psi_1\left(x;\tau\right)=\omega\left(\tau\right)+\phi\left(\tau\right)x, \qquad \psi_3\left(x;\tau\right)=b_0\left(\tau\right)+b_1\left(\tau\right)x,
\end{equation}
where $\omega\left(\tau\right)=\left[1-\phi\left(\tau\right)\right]\sigma^2\left(\tau\right)$. Here, $r$ is the continuously compounded risk-free  rate, $\lambda\left(\tau\right)$ governs the conditional expected excess log return, and $\phi\left(\tau\right)$ controls variance persistence. Under $0\leq\phi\left(\tau\right)<1$, the conditional mean of the variance converges geometrically to $\sigma^2\left(\tau\right)$.\footnote{For stationary versions of both specifications with finite mean variance, the conditionally demeaned returns $y^{(n)}_{(k+1)\tau}-\psi_0\left(h^{(n)}_{k\tau};\tau\right)$ belong to the SR-SARV$\left(1\right)$ class of \citet{Meddahi04}, since their conditional variance $\tau h^{(n)}_{k\tau}$ has affine conditional mean dynamics. However, our additional restrictions yield polynomial conditional moments of variance at every degree and, under the joint restrictions, polynomial mixed return--variance moments; these properties are not implied by the SR-SARV$\left(1\right)$ structure.} The two models differ in the remaining loading $\psi_2$, specified next.

\subsubsection{The variance-polynomial hybrid GARCH--SV model (VPGARCH-SV)}
\label{subsec:variance_polynomial_dynamics}

For variance preservation, part~\textnormal{(i)} of Proposition~\ref{prop:garch_polynomial_characterization} permits a quadratic squared loading. Within the nondegenerate quadratic class, we use the following scale-normalized specification
\begin{equation}
\psi_2\left(x;\tau\right)=\eta\left(\tau\right)\sqrt{a_0\left(\tau\right)+a_1\left(\tau\right)x+x^2}. \label{eq:variance_polynomial_loading_functions}
\end{equation}
For $\eta\left(\tau\right)\neq0$, normalizing the coefficient on $x^2$ removes scale redundancy between $\eta\left(\tau\right)$ and the polynomial coefficients, while leaving $\eta\left(\tau\right)$ unrestricted in sign. Together with Equation~\eqref{eq:common_mean_parametrizations}, this loading defines the variance-polynomial hybrid GARCH--SV model, henceforth denoted by VPGARCH-SV, through Equations~\eqref{eq:general_return_dynamics_p}--\eqref{eq:general_variance_dynamics_p}. Its variance process is polynomial preserving by Proposition~\ref{prop:garch_polynomial_characterization}. The NGARCH-SV specification and its NGARCH special case are nested as follows.

\begin{remark}
\label{rmk:ngarch_special_case}
The NGARCH-SV specification is obtained by setting
\begin{equation*}
a_0\left(\tau\right)=a_1\left(\tau\right)=b_0\left(\tau\right)=0, \qquad b_1\left(\tau\right)=\alpha\left(\tau\right), \qquad \eta\left(\tau\right)=-2\alpha\left(\tau\right)\gamma\left(\tau\right),
\end{equation*}
where $\alpha\left(\tau\right)$ and $\gamma\left(\tau\right)$ denote the conventional NGARCH scale and leverage parameters, respectively. Under these restrictions, Equation~\eqref{eq:general_variance_dynamics_p} becomes
\begin{equation*}
h^{(n)}_{(k+1)\tau}=\omega\left(\tau\right)+\phi\left(\tau\right)h^{(n)}_{k\tau}+\alpha\left(\tau\right)h^{(n)}_{k\tau}\left[\left(\epsilon^{(n)}_{2,(k+1)\tau}\right)^2-2\gamma\left(\tau\right)\epsilon^{(n)}_{2,(k+1)\tau}-1\right].
\end{equation*}
At $\varrho=1$, this recovers the NGARCH variance dynamics of \citet{engle1993measuring}. Equivalently, if $\beta\left(\tau\right)$ denotes the conventional NGARCH persistence parameter, then $\phi\left(\tau\right)=\beta\left(\tau\right)+\alpha\left(\tau\right)\left[1+\gamma\left(\tau\right)^2\right]$.\footnote{At $\varrho=1$, the same variance dynamics admit the bilinear GARCH$\left(1,1\right)$ representation of \citet{StortiVitale2003}: its asymmetric term is proportional to the conditionally demeaned return times its conditional standard deviation.} Setting $\gamma\left(\tau\right)=0$ and $\varrho=1$ gives the standard GARCH$\left(1,1\right)$ variance equation.
\end{remark}

Polynomial preservation does not by itself ensure a nonnegative variance. The following restrictions guarantee nonnegativity for every current state and every real value of $\epsilon_2$. They do not depend on $\varrho$, since the variance dynamics involves only $\epsilon_2$, irrespective of its dependence on $\epsilon_1$.

\begin{proposition}
\label{prop:variance_polynomial_positivity}
Suppose that $0\leq\phi\left(\tau\right)<1$ and $\sigma^2\left(\tau\right) >0$. Under the VPGARCH-SV dynamics from Equations~\eqref{eq:general_return_dynamics_p}--\eqref{eq:general_variance_dynamics_p}, \eqref{eq:common_mean_parametrizations}, and~\eqref{eq:variance_polynomial_loading_functions}, define
\begin{align*}
\delta_2\left(\tau\right)&=\eta\left(\tau\right)^2-4b_1\left(\tau\right)\left[\phi\left(\tau\right)-b_1\left(\tau\right)\right], \\
\delta_1\left(\tau\right)&=\eta\left(\tau\right)^2a_1\left(\tau\right)-4\left(b_0\left(\tau\right)\phi\left(\tau\right)+b_1\left(\tau\right)\omega\left(\tau\right)-2b_0\left(\tau\right)b_1\left(\tau\right)\right), \\
\delta_0\left(\tau\right)&=\eta\left(\tau\right)^2a_0\left(\tau\right)-4b_0\left(\tau\right)\left(\omega\left(\tau\right)-b_0\left(\tau\right)\right).
\end{align*}
Under the parameter restrictions
\begin{equation*}
a_0\left(\tau\right)\geq0, \qquad a_1\left(\tau\right)\geq-2\sqrt{a_0\left(\tau\right)}, \qquad b_0\left(\tau\right)\geq0, \qquad b_1\left(\tau\right)>0,
\end{equation*}
necessary and sufficient conditions for the conditional variance to be non-negative are
\begin{equation}
\delta_2\left(\tau\right)\leq0, \qquad \delta_0\left(\tau\right)\leq0, \qquad \delta_1\left(\tau\right)\leq2\sqrt{\delta_2\left(\tau\right)\delta_0\left(\tau\right)}.
\label{eq:positivity_delta_restrictions}
\end{equation}
\end{proposition}

\begin{proof}
See Supplementary Material~\ref{sm:proof_variance_polynomial_positivity}.
\end{proof}

The parameters $a_0\left(\tau\right)$ and $a_1\left(\tau\right)$ enrich return--variance dependence relative to NGARCH-SV. The one-step conditional covariance is
\begin{equation*}
\Cov^{\Pm^{(n)}}\left[y^{(n)}_{(k+1)\tau},h^{(n)}_{(k+1)\tau}\mid\F^{(n)}_{k\tau}\right]=\varrho\eta\left(\tau\right)\sqrt{\tau h^{(n)}_{k\tau}\left[a_0\left(\tau\right)+a_1\left(\tau\right)h^{(n)}_{k\tau}+\left(h^{(n)}_{k\tau}\right)^2\right]}.
\end{equation*}
In the NGARCH-SV case, the covariance is proportional to $\left(h^{(n)}_{k\tau}\right)^{3/2}$. For $\varrho>0$ and $\eta\left(\tau\right)\neq0$, the covariance above is not polynomial in the variance state, so VPGARCH-SV does not preserve polynomials jointly in returns and variance. Under independent innovations, it is jointly polynomial by part~\textnormal{(ii)} of Proposition~\ref{prop:garch_polynomial_characterization}.

The one-step conditional variance of $h^{(n)}$ is then
\begin{equation*}
\Var^{\Pm^{(n)}}\left[h^{(n)}_{(k+1)\tau}\mid\F^{(n)}_{k\tau}\right]=v_0\left(\tau\right)+v_1\left(\tau\right)h^{(n)}_{k\tau}+v_2\left(\tau\right)\left(h^{(n)}_{k\tau}\right)^2,
\end{equation*}
where the coefficients $v_0, v_1$ and $v_2$ are given by:
\begin{equation*}\begin{aligned}v_0\left(\tau\right)&=\eta\left(\tau\right)^2a_0\left(\tau\right)+\left(m_4-1\right)b_0\left(\tau\right)^2, \qquad v_1\left(\tau\right)=\eta\left(\tau\right)^2a_1\left(\tau\right)+2\left(m_4-1\right)b_0\left(\tau\right)b_1\left(\tau\right),\\ v_2\left(\tau\right)&=\eta\left(\tau\right)^2+\left(m_4-1\right)b_1\left(\tau\right)^2.\end{aligned}\end{equation*}
Thus, VPGARCH-SV adds constant and linear terms to the quadratic conditional variance-of-variance relation of NGARCH-SV while retaining closed recursions for variance moments. The conditional return--variance correlation is constant in the nondegenerate NGARCH-SV case, whereas the additional loadings generally make it state dependent under VPGARCH-SV.

\subsubsection{The jointly polynomial hybrid GARCH--SV model (JPGARCH-SV)}
\label{subsec:joint_polynomial_dynamics}

To obtain joint return--variance preservation for every $\varrho\in\left[0,1\right]$, we use the loading in part~\textnormal{(ii)} of Proposition~\ref{prop:garch_polynomial_characterization},
\begin{equation}
\psi_2\left(x;\tau\right)=\eta\left(\tau\right)\sqrt{x}. \label{eq:joint_polynomial_loading_functions}
\end{equation}
Together with Equation~\eqref{eq:common_mean_parametrizations}, this loading defines the jointly polynomial hybrid GARCH--SV model, denoted by JPGARCH-SV, through Equations~\eqref{eq:general_return_dynamics_p}--\eqref{eq:general_variance_dynamics_p}. Its joint return--variance state is polynomial preserving, which will also permit exact cumulative-return moment calculations.\footnote{Although joint preservation implies variance preservation, the two canonical parametrizations are not nested: for nonzero $\eta\left(\tau\right)$, the squared loading has degree two under VPGARCH-SV and degree one under JPGARCH-SV. They coincide only when $\eta\left(\tau\right)=0$.}

At $\varrho=1$, JPGARCH extends the HNGARCH variance formulation by allowing the centred-quadratic loading to depend on the current variance through $b_1\left(\tau\right)$. With jointly Gaussian innovations and $b_1\left(\tau\right)=0$, JPGARCH-SV contains the HNGARCH-SV specification of \citet{Jacobs24}.

\begin{remark}
\label{rmk:hngarch_special_case}
The HNGARCH-SV specification is obtained by setting
\begin{equation}
b_0\left(\tau\right)=\alpha\left(\tau\right), \qquad b_1\left(\tau\right)=0, \qquad \eta\left(\tau\right)=-2\alpha\left(\tau\right)\gamma\left(\tau\right), \label{eq:hngarch_parameter_mapping}
\end{equation}
where $\alpha\left(\tau\right)$ and $\gamma\left(\tau\right)$ are the conventional HNGARCH scale and leverage parameters, respectively. Under these restrictions, Equation~\eqref{eq:general_variance_dynamics_p} becomes
\begin{equation*}
h^{(n)}_{(k+1)\tau}=\omega\left(\tau\right)+\phi\left(\tau\right)h^{(n)}_{k\tau}+\alpha\left(\tau\right)\left[\left(\epsilon^{(n)}_{2,(k+1)\tau}\right)^2-2\gamma\left(\tau\right)\sqrt{h^{(n)}_{k\tau}}\,\epsilon^{(n)}_{2,(k+1)\tau}-1\right].
\end{equation*}
At $\varrho=1$, this recovers the HNGARCH variance specification of \citet{heston2000closed}. Equivalently, if $\omega_{\mathrm{HN}}\left(\tau\right)$ and $\beta\left(\tau\right)$ denote the conventional HNGARCH intercept and persistence parameters, then $\omega\left(\tau\right)=\omega_{\mathrm{HN}}\left(\tau\right)+\alpha\left(\tau\right)$ and $\phi\left(\tau\right)=\beta\left(\tau\right)+\alpha\left(\tau\right)\gamma\left(\tau\right)^2$.\footnote{With $\varrho=1$, the restriction $b_0\left(\tau\right)=0$ gives the quadratic GARCH$\left(1,1\right)$ (QGARCH) variance dynamics associated with \citet{Engle1990,Sentana1995}, termed GQARCH$\left(1,1\right)$ in the latter paper. Indeed, for $b_1\left(\tau\right)>0$, completing the square gives
\begin{equation*}\begin{aligned} h^{(n)}_{(k+1)\tau}={}&\left[\omega\left(\tau\right)-\frac{\eta\left(\tau\right)^2}{4b_1\left(\tau\right)}\right]+\left[\phi\left(\tau\right)-b_1\left(\tau\right)\right] h^{(n)}_{k\tau} +\frac{b_1\left(\tau\right)}{\tau}\left[y^{(n)}_{(k+1)\tau}-\psi_0\left(h^{(n)}_{k\tau};\tau\right)+\frac{\eta\left(\tau\right)\sqrt{\tau}}{2b_1\left(\tau\right)}\right]^2.\end{aligned}\end{equation*}
Multiplying by $\tau$ gives the conventional equation for the one-step return variance. Proposition~\ref{prop:joint_polynomial_positivity} ensures nonnegative completed-square intercept and persistence coefficient. This correspondence concerns the one-lag specification.}
\end{remark}

The following restrictions ensure nonnegative variance under JPGARCH-SV. As in the variance-polynomial case, they hold for every real value of $\epsilon_2$ and do not depend on $\varrho$.

\begin{proposition}
\label{prop:joint_polynomial_positivity}
Suppose that $0\leq\phi\left(\tau\right)<1$, $\sigma^2\left(\tau\right)>0$, and $b_0\left(\tau\right)+b_1\left(\tau\right)>0$. Necessary and sufficient conditions for the conditional variance to be non-negative under the JPGARCH-SV dynamics are
\begin{equation}
b_0\left(\tau\right)\geq0, \qquad b_1\left(\tau\right)\geq0, \qquad b_0\left(\tau\right)\leq\omega\left(\tau\right), \qquad b_1\left(\tau\right)\leq\phi\left(\tau\right), \label{eq:joint_positivity_loading_restrictions}
\end{equation}
together with
\begin{equation}
\eta\left(\tau\right)^2\leq4\left(\sqrt{b_0\left(\tau\right)\left[\phi\left(\tau\right)-b_1\left(\tau\right)\right]}+\sqrt{b_1\left(\tau\right)\left[\omega\left(\tau\right)-b_0\left(\tau\right)\right]}\right)^2. \label{eq:joint_positivity_leverage_restriction}
\end{equation}
\end{proposition}

\begin{proof}
See Supplementary Material~\ref{sm:proof_joint_polynomial_positivity}.
\end{proof}

The one-step conditional return--variance covariance is
\begin{equation*}
\Cov^{\Pm^{(n)}}\left[y^{(n)}_{(k+1)\tau},h^{(n)}_{(k+1)\tau}\mid\F^{(n)}_{k\tau}\right]=\varrho\eta\left(\tau\right)\sqrt{\tau}\,h^{(n)}_{k\tau}.
\end{equation*}
This covariance is linear in the variance state.

The one-step conditional variance of $h^{(n)}$ is
\begin{equation*}
\Var^{\Pm^{(n)}}\left[h^{(n)}_{(k+1)\tau}\mid\F^{(n)}_{k\tau}\right]=w_0\left(\tau\right)+w_1\left(\tau\right)h^{(n)}_{k\tau}+w_2\left(\tau\right)\left(h^{(n)}_{k\tau}\right)^2,
\end{equation*}
where the coefficients $w_0, w_1$ and $w_2$ are given by:
\begin{equation*}\begin{aligned}w_0\left(\tau\right)&=\left(m_4-1\right)b_0\left(\tau\right)^2, \qquad w_1\left(\tau\right)=\eta\left(\tau\right)^2+2\left(m_4-1\right)b_0\left(\tau\right)b_1\left(\tau\right),\\ w_2\left(\tau\right)&=\left(m_4-1\right)b_1\left(\tau\right)^2.\end{aligned}\end{equation*}
The state-dependent quadratic-innovation loading therefore extends the affine conditional variance-of-variance relation of HNGARCH-SV to a quadratic one without losing joint polynomial preservation.\footnote{The symmetric square-root GARCH specification of \citet{ishida2002modeling} instead has affine $\psi_1$, $\psi_2\left(x;\tau\right)=0$, and $\psi_3\left(x;\tau\right)=\alpha\left(\tau\right)\sqrt{x}$, with $\alpha\left(\tau\right)>0$. It is outside the two  specifications. Under Gaussian innovations, its third conditional moment of the variance contains $8\alpha\left(\tau\right)^3x^{3/2}$, so its linear variance-of-variance function does not yield polynomial preservation at every order.} Under the HNGARCH-SV parameter mapping, the expression reduces to $\left(m_4-1\right)\alpha\left(\tau\right)^2+4\alpha\left(\tau\right)^2\gamma\left(\tau\right)^2h^{(n)}_{k\tau}$, the Gaussian formula following from $m_4=3$.

For $\varrho>0$, $b_0\left(\tau\right),b_1\left(\tau\right)>0$, and $\eta\left(\tau\right)\ne0$, absolute conditional return--variance correlation increases up to $x=b_0\left(\tau\right)/b_1\left(\tau\right)$ and decreases thereafter. In the HNGARCH-SV case with $\varrho>0$, $b_0\left(\tau\right)>0$, and $\eta\left(\tau\right)\ne0$, it instead increases with $x$. JPGARCH-SV thus permits the strongest return--variance dependence at an intermediate variance level.\footnote{Supplementary Material~\ref{sm:sec:two_component_extension} develops a two-component hybrid-SV structure extending the variance specifications of \citet{christoffersen2008option,chris10b}.}

The next subsection examines how the different loading specifications affect moment stability and moment calculations for return-based and cumulative quantities.

\subsection{Analytic moments}
\label{subsec:analytic_moments}

We now apply the framework of Section~\ref{subsec:general_framework} to identify the model-specific coefficients and parameter restrictions governing variance moments, and to determine which return-based and cumulative quantities retain polynomial moment closure.

\subsubsection{Multi-period and stationary moments of the variance process}
\label{subsec:moment_implications}

For long-horizon analysis, we extend the VPGARCH-SV and JPGARCH-SV variance dynamics to all $k\geq0$. Under the corresponding positivity restrictions, the variance process of each model admits an invariant probability distribution $\pi^{(n)}$, obtained as a weak subsequential limit of the Ces\`aro averages of its marginal laws.\footnote{Both variance processes are weak Feller on $\Rplus$. Their affine conditional means, with $0\leq\phi\left(\tau\right)<1$, give uniformly bounded first moments from the deterministic initial state. The Ces\`aro averages of their marginal laws are therefore tight, and any weak subsequential limit is invariant.} The conditions below ensure convergence to finite stationary moments; no uniqueness of the invariant distribution is required.

Let $\bm Q^{(n)}_m$ denote the matrix representation of the transition operator of $h^{(n)}$ on $\Pol_m\left(\Rplus\right)$ for either model, and let $\bm p^{(m)}_{h^i}$ denote the coordinate vector of $x^i$ with respect to the degree-ordered basis $\bm{\HT}_m$. Then, for $i=1,\ldots,m$ and all integers $k,s\geq0$,
\begin{equation*}\E^{\Pm^{(n)}}\left[\left(h^{(n)}_{(k+s)\tau}\right)^i\mid\F^{(n)}_{k\tau}\right]=\bm{\HT}_m\left(h^{(n)}_{k\tau}\right)^\top\left(\bm Q^{(n)}_m\right)^s\bm p^{(m)}_{h^i}.\end{equation*}
For either model, the one-step expansion determines the entries of $\bm Q^{(n)}_m$ from the parameters and the even moments of $\epsilon_2$ up to order $2m$.

To illustrate the construction of the transition matrix, consider $m=2$. Let $d_j\left(\tau\right)=v_j\left(\tau\right)$ under VPGARCH-SV and $d_j\left(\tau\right)=w_j\left(\tau\right)$ under JPGARCH-SV, $j=0,1,2$, where these coefficients are defined in Sections~\ref{subsec:variance_polynomial_dynamics} and~\ref{subsec:joint_polynomial_dynamics}. At a current variance state $h^{(n)}_{k\tau}=x$, the conditional mean and variance give
\begin{align*}
\E^{\Pm^{(n)}}\left[h^{(n)}_{(k+1)\tau}\mid h^{(n)}_{k\tau}=x\right]&=\omega\left(\tau\right)+\phi\left(\tau\right)x,\\
\E^{\Pm^{(n)}}\left[\left(h^{(n)}_{(k+1)\tau}\right)^2\mid h^{(n)}_{k\tau}=x\right]&=\left[\omega\left(\tau\right)+\phi\left(\tau\right)x\right]^2+d_0\left(\tau\right)+d_1\left(\tau\right)x+d_2\left(\tau\right)x^2.
\end{align*}
With respect to the monomial basis $\bm{\HT}_2\left(x\right)=\left(1,x,x^2\right)^\top$, collecting these polynomial coefficients in the corresponding columns gives
\begin{equation*}
\bm Q^{(n)}_2=
\begin{pmatrix}
1&\omega\left(\tau\right)&\omega\left(\tau\right)^2+d_0\left(\tau\right)\\
0&\phi\left(\tau\right)&2\omega\left(\tau\right)\phi\left(\tau\right)+d_1\left(\tau\right)\\
0&0&\phi\left(\tau\right)^2+d_2\left(\tau\right)
\end{pmatrix}.
\end{equation*}
For a fixed variance-innovation marginal law $D_2$, the variance transition matrix and its moment-stability conditions do not depend on $\varrho$. The model-specific task is to identify the diagonal entries and thereby express the general stability condition in Equation~\eqref{eq:spectral_stationarity_condition} directly in terms of the model parameters. For fixed $\tau$, evaluate Equation~\eqref{eq:general_variance_dynamics_p} at the current variance $h^{(n)}_{k\tau}=x>0$, using the respective VPGARCH-SV and JPGARCH-SV loadings. Dividing the resulting next-period variance by $x$ and letting $x\to\infty$ yields the growth factors
\begin{equation*}A_{\mathsf{VP}}=\phi\left(\tau\right)+\eta\left(\tau\right)\epsilon^{(n)}_{2,(k+1)\tau}+b_1\left(\tau\right)\left[\left(\epsilon^{(n)}_{2,(k+1)\tau}\right)^2-1\right], \qquad A_{\mathsf{JP}}=\phi\left(\tau\right)+b_1\left(\tau\right)\left[\left(\epsilon^{(n)}_{2,(k+1)\tau}\right)^2-1\right].\end{equation*}
\begin{proposition}
\label{prop:garch_moment_existence}
For $\mathsf{M}\in\{\mathsf{VP},\mathsf{JP}\}$ and every integer $j\geq0$, the degree-$j$ diagonal block of the transition matrix of $h^{(n)}$ is the scalar $q^{(n)}_{jj}=\E^{\Pm^{(n)}}\left[A_{\mathsf M}^{\,j}\right]$. Under the corresponding restrictions of Propositions~\ref{prop:variance_polynomial_positivity} and~\ref{prop:joint_polynomial_positivity}, $A_{\mathsf M}\geq0$, and, for every integer $m\geq1$, Condition~\eqref{eq:spectral_stationarity_condition} holds through degree $m$ if and only if
\begin{equation}\E^{\Pm^{(n)}}\left[A_{\mathsf M}^{\,m}\right]<1.\label{eq:garch_moment_condition}\end{equation}
In that case, the unconditional moments of $h^{(n)}$ through order $m$ converge to finite limits independent of $h_0$.
\end{proposition}

\begin{proof}
See Supplementary Material~\ref{sm:proof_garch_moment_existence}.
\end{proof}

Under Condition~\eqref{eq:garch_moment_condition}, the unconditional moments of $h^{(n)}$ through order $m$ are uniformly bounded over time, so $\pi^{(n)}$ has finite moments of those orders. By invariance and Proposition~\ref{sm:prop:matryoshka_stationary_moments}, these coincide with the limiting moments. For $m\geq2$, the left-hand side of Condition~\eqref{eq:garch_moment_condition} depends on $|\eta\left(\tau\right)|$ under VPGARCH-SV, but is independent of $\eta\left(\tau\right)$ under JPGARCH-SV, where this loading enters only lower-degree coefficients. The parameters $a_0\left(\tau\right)$, $a_1\left(\tau\right)$, and $b_0\left(\tau\right)$ likewise do not enter the diagonal coefficients, although they can affect moment levels and the admissible parameter region.\footnote{For comparison, standard GARCH$\left(p,q\right)$ models admit a random-affine lag-vector representation $\bm Z_t=\bm b_t+\bm M_t\bm Z_{t-1}$ in squared return innovations and conditional variances, with an i.i.d. coefficient matrix $\bm M_t$. Under the usual positivity and innovation-moment assumptions, finiteness of the stationary $2m$th return-innovation moment is characterized by $\rho\left(\E\left[\bm M_t^{\otimes m}\right]\right)<1$, where $\otimes m$ denotes the $m$-fold Kronecker product; see \citet{LingMcAleer02a} and \citet[Section~2.4.1]{FrancqZakoian19}.} In particular, for $m=2$, Condition~\eqref{eq:garch_moment_condition} becomes $\phi\left(\tau\right)^2+d_2\left(\tau\right)<1$, where $d_2\left(\tau\right)=\eta\left(\tau\right)^2+\left(m_4-1\right)b_1\left(\tau\right)^2$ for VPGARCH-SV and $d_2\left(\tau\right)=\left(m_4-1\right)b_1\left(\tau\right)^2$ for JPGARCH-SV. Under this condition,
\begin{equation*}\E^{\Pm^{(n)}}\left[h^{(n)}_{k\tau}\right]\xrightarrow[k\to\infty]{}\sigma^2\left(\tau\right), \quad \E^{\Pm^{(n)}}\left[\left(h^{(n)}_{k\tau}\right)^2\right]\xrightarrow[k\to\infty]{}\frac{\omega\left(\tau\right)^2+d_0\left(\tau\right)+\left[2\omega\left(\tau\right)\phi\left(\tau\right)+d_1\left(\tau\right)\right]\sigma^2\left(\tau\right)}{1-\phi\left(\tau\right)^2-d_2\left(\tau\right)}.\end{equation*}
The same calculation through degree four gives the variance-moment forecasts needed for conditional skewness and kurtosis of future variance whenever the conditional variance is positive; Equation~\eqref{eq:garch_moment_condition} at $m=4$ ensures finite stationary fourth moments of the variance.

In the NGARCH-SV case, $A_{\mathsf{VP}}=\beta\left(\tau\right)+\alpha\left(\tau\right)\left[\epsilon^{(n)}_{2,(k+1)\tau}-\gamma\left(\tau\right)\right]^2$, so Condition~\eqref{eq:garch_moment_condition} recovers the familiar random-coefficient moment restriction; see \citet{HeTerasvirta99} and \citet{LingMcAleer02}. In the HNGARCH-SV case, $b_1\left(\tau\right)=0$ gives $A_{\mathsf{JP}}=\phi\left(\tau\right)$, so $\phi\left(\tau\right)<1$ ensures finite stationary variance moments of every order under the maintained innovation-moment assumptions.

\subsubsection{Cumulative quantities and return-based moments}\label{subsec:temporal_aggregates}

Variance polynomiality alone does not establish moment closure for return-based quantities. We therefore identify the appropriate augmented states for cumulative variance, squared conditionally demeaned returns, and cumulative returns.

For a fixed date $k\tau$, define cumulative variance and the conditionally demeaned return by
\begin{equation*}\mathcal{V}^{(n)}_{k,j}:=\tau\sum_{\ell=0}^{j-1}h^{(n)}_{(k+\ell)\tau}, \qquad \mathcal{V}^{(n)}_{k,0}:=0,\end{equation*}
\begin{equation*}\widetilde y^{(n)}_{(k+1)\tau}:=y^{(n)}_{(k+1)\tau}-\E^{\Pm^{(n)}}\left[y^{(n)}_{(k+1)\tau}\mid\F^{(n)}_{k\tau}\right]=\sqrt{\tau h^{(n)}_{k\tau}}\,\epsilon^{(n)}_{1,(k+1)\tau}.\end{equation*}
Let $\bm U^{(n,k)}_{j\tau}:=\left(\mathcal{V}^{(n)}_{k,j},h^{(n)}_{(k+j)\tau}\right)^\top$ for $j=0,\ldots,nL-k$, and let $\bm V^{(n,k)}_{j\tau}:=\left(\left(\widetilde y^{(n)}_{(k+j)\tau}\right)^2,h^{(n)}_{(k+j)\tau}\right)^\top$ for $j=1,\ldots,nL-k$, with $\bm V^{(n,k)}_0:=\left(0,h^{(n)}_{k\tau}\right)^\top$. Furthermore, define $\mathcal{R}^{(n)}_{k,j}:=Y^{(n)}_{(k+j)\tau}-Y^{(n)}_{k\tau}$, with $\mathcal{R}^{(n)}_{k,0}:=0$, and set $\bm Z^{(n,k)}_{j\tau}:=\left(\mathcal{R}^{(n)}_{k,j},h^{(n)}_{(k+j)\tau}\right)^\top$.

\begin{corollary}
\label{cor:polynomial_temporal_aggregates}
For any $k\in\{0,\ldots,nL\}$, the following statements hold.
\begin{enumerate}[label=(\roman*)]
\item Under either VPGARCH-SV or JPGARCH-SV, $\bm U^{(n,k)}$ and $\bm V^{(n,k)}$ are polynomial-preserving Markov chains on $\Rplus^2$.
\item Under JPGARCH-SV, and under VPGARCH-SV with independent innovations, $\bm Z^{(n,k)}$ is a polynomial-preserving Markov chain on $\mathbb{R}\times\Rplus$.
\end{enumerate}
\end{corollary}

\begin{proof}
See Supplementary Material~\ref{sm:proof_polynomial_temporal_aggregates}.
\end{proof}

The return-based transition coefficients generally involve mixed moments of $\left(\epsilon_1,\epsilon_2\right)$, which are not determined by the marginal laws and $\varrho$ alone. Part~\textnormal{(i)}, together with Proposition~\ref{prop:finite_dimensional_representation} and Corollary~\ref{cor:polynomial_cross_moments}, gives moments and serial cross-moments of cumulative variance and squared conditionally demeaned returns under both models. Under Equation~\eqref{eq:garch_moment_condition} with $m=2$, it also gives the stationary autocorrelation function of squared conditionally demeaned returns whenever their stationary variance is positive. Part~\textnormal{(ii)}, combined with Corollary~\ref{cor:polynomial_cross_moments}, gives cumulative-return moments jointly with variance, mixed moments across return dates, and moments of polynomial return aggregates, including realized variance.

Among the GARCH special cases at $\varrho=1$ in Section~\ref{subsec:polynomial_garch_models_p}, NGARCH inherits part~\textnormal{(i)}, while HNGARCH and QGARCH$\left(1,1\right)$ also inherit part~\textnormal{(ii)}. Exact second to fourth conditional moments of cumulative returns under QGARCH are derived by \citet{wongso2003}. Moment-based calculations for cumulative returns under NGARCH, GJR-GARCH, and EGARCH are developed in \citet{DuanGauthierSimonato99,Duanetal06,AlexanderLazarStanescu21}. The GJR-GARCH calculations in \citet{Duanetal06,AlexanderLazarStanescu21} use Taylor approximations for fractional variance moments entering return cross-moments. In contrast, under the maintained assumptions, Corollaries~\ref{cor:polynomial_cross_moments} and~\ref{cor:polynomial_temporal_aggregates} provide exact conditional moments of cumulative returns, their mixed moments with the variance state, and cross-date moments under JPGARCH-SV, and under VPGARCH-SV with independent innovations, at every finite order and horizon. These recursions require no fractional-moment approximations.

Appendix~\ref{sec:risk_neutral_dynamics} derives risk-neutral dynamics under a variance-dependent pricing kernel with equity and variance risk-preference parameters. We also characterize polynomial preservation under jointly Gaussian innovations. Retaining this property generally requires a particular state dependence of variance risk preferences and additional restrictions on the return drift and innovation loadings. Thus, polynomial preservation under $\Pm^{(n)}$ need not survive the change to $\Qm^{(n)}$.

\section{Polynomial weak limits}\label{sec:polynomial_diffusion_limits}

This section relates the discrete-time dynamics to continuous-time polynomial processes under the physical measure. We establish the diffusion moment connection and derive the physical diffusion limits, then consider an alternative scaling that yields polynomial jump-diffusions. State-space conditions distinguish local coefficient limits from globally admissible processes.{\footnote{Supplementary Material~\ref{subsec:risk_neutral_diffusion_limits} derives the corresponding risk-neutral diffusion limits under additional exponential-moment and risk-preference conditions and examines their polynomial structure. In the jointly Gaussian case, imposing the discrete-time restrictions in Appendix~\ref{subsec:q_polynomial_preservation} preserves the corresponding polynomial operators in the limit.}}

\subsection{Diffusion limit framework}
\label{subsec:general_diffusion_framework}

We first connect the discrete-time polynomial-preserving setting of Section~\ref{subsec:general_framework} to the continuous-time theory of polynomial diffusions. Consider the state space $E$, polynomial spaces $\Pol_m(E)$, and time step $\tau=1/n$ from that subsection. For the Markov chain $\bm{X}^{(n)}=\{\bm{X}^{(n)}_{k\tau}\}_{k=0}^{nL}$ defined therein, define its c\`adl\`ag extension on $[0,L]$ by
\begin{equation}\label{eq:cadlag_extension}
\bm{X}^{(n)}_t=\bm{X}^{(n)}_{k\tau}, \qquad k\tau\leq t<(k+1)\tau, \qquad k=0,\ldots,nL-1,
\end{equation}
with $\bm{X}^{(n)}_L=\bm{X}^{(n)}_{nL\tau}$.
We consider a filtered probability space $(\Omega,\F,\Ft,\Pm)$, where $\Ft=\{\F_t\}_{t\in[0,L]}$ is a filtration satisfying the usual conditions and $\Pm$ is the physical measure. Suppose that $\bm{X}^{(n)}_0$ converges in distribution under $\Pm^{(n)}$ to $\bm{X}_0$ under $\Pm$ and that the local drift, second-moment, and no-large-jumps conditions, together with the required compact-containment, well-posedness, and nonexplosion conditions, hold as in the diffusion-approximation framework of \citet{Stroock} and \citet[Theorem~2.2]{nelson1990arch}. Then the c\`adl\`ag extensions in Equation~\eqref{eq:cadlag_extension} converge weakly under $\Pm^{(n)}$, as random elements of $D([0,L],\mathbb{R}^d)$ equipped with the Skorokhod $\mathrm{J}_1$ topology, to an $E$-valued time-homogeneous diffusion $\bm{X}=\{\bm{X}_t\}_{t\in[0,L]}$ under $\Pm$, adapted to $\Ft$.

We write the limiting dynamics as
\begin{equation*}
d\bm{X}_t=\bm{A}(\bm{X}_t)\,dt+\bm{\Sigma}(\bm{X}_t)\,d\bm{W}_t,
\end{equation*}
where $\bm{A}:E\to\mathbb{R}^d$, $\bm{\Sigma}:E\to\mathbb{R}^{d\times d}$, and $\bm{W}=\{\bm{W}_t\}_{t\in[0,L]}$ is a $d$-dimensional standard Brownian motion under $\Pm$ adapted to $\Ft$. The covariance matrix is
\begin{equation*}
\bm{B}(\bm{x})=\bm{\Sigma}(\bm{x})\bm{\Sigma}(\bm{x})^\top.
\end{equation*}

For polynomial test functions $p\in\Pol(E)$, the differential operator associated with the limiting diffusion is
\begin{equation*}
\mathcal{A}p(\bm{x})=\sum_{i=1}^{d}A_i(\bm{x})\partial_i p(\bm{x})+\frac{1}{2}\sum_{i,j=1}^{d}B_{ij}(\bm{x})\partial_{ij}p(\bm{x}).
\end{equation*}
When $E$ has empty interior, derivatives are evaluated through polynomial extensions, and well-definedness means that the result is independent of the chosen extension. For the discrete-time chain, let $\trans^{(n)}$ be the one-step transition operator defined in Equation~\eqref{eq:transition_operator}, and define its scaled operator by
\begin{equation*}
\mathcal{A}^{(n)}p(\bm{x})=\frac{\trans^{(n)}p(\bm{x})-p(\bm{x})}{\tau}.
\end{equation*}

For compactly supported smooth test functions, the corresponding truncated local drift and second-moment conditions, together with the no-large-jumps condition, are equivalent to locally uniform convergence of the associated operators; see \citet[Lemma~11.2.1]{Stroock}. Since the relevant test functions here are polynomials and may be unbounded, Supplementary Material~\ref{sm:nelson_local_conditions} shows that a stronger local moment condition extends this operator convergence to $\Pol_m(E)$. Specifically, if the moment condition holds at some order $q=2+\delta\geq m$, then, for every $p\in\Pol_m(E)$ and every compact set $\mathcal{K}\subseteq E$, we have
\begin{equation}\label{eq:nelson_generator_convergence}
\lim_{\tau\to0}\sup_{\bm{x}\in\mathcal{K}}\left|\mathcal{A}^{(n)}p(\bm{x})-\mathcal{A}p(\bm{x})\right|=0.
\end{equation}

The following proposition transfers polynomial preservation through convergence of the scaled operators; weak convergence alone is not the condition used in the argument.\footnote{\citet{KallsenRichert25} establish a complementary connection between discrete- and continuous-time polynomial models: discrete observations of a continuous-time polynomial process form a polynomial state-space model. Proposition~\ref{prop:weak_limit_polynomial_preservation} instead considers the reverse approximation question, namely whether polynomial preservation of a sequence of discrete-time Markov chains is inherited by its weak diffusion limit.}

\begin{proposition}
\label{prop:weak_limit_polynomial_preservation}
Let $m\geq0$ and suppose that $\mathcal{A}$ is well defined on $\Pol_m(E)$, that Equation~\eqref{eq:nelson_generator_convergence} holds for every $p\in\Pol_m(E)$, and that $\bm{X}^{(n)}$ is $m$-polynomial preserving in the sense of Definition~\ref{def:pol_preserving_chain} for every $n$. Then
\begin{equation*}
\mathcal{A}\Pol_m(E)\subseteq\Pol_m(E).
\end{equation*}
If, in addition, $\mathcal{A}$ is well defined on $\Pol(E)$ and these hypotheses hold both for $m=1$ and for $m=2$, then
\begin{equation*}
A_i\in\Pol_1(E), \qquad B_{ij}\in\Pol_2(E), \qquad i,j=1,\ldots,d,
\end{equation*}
and hence $\mathcal{A}\Pol_k(E)\subseteq\Pol_k(E)$ for every integer $k\geq0$; that is, $\bm{X}$ is a polynomial diffusion.
\end{proposition}

\begin{proof}
See Appendix~\ref{app:proof_weak_limit_polynomial_preservation}.
\end{proof}

Proposition~\ref{prop:weak_limit_polynomial_preservation} reveals an asymmetry between discrete-time chains and diffusions. For diffusions, preservation of $\Pol_1(E)$ and $\Pol_2(E)$ forces preservation at every degree, because the generator acts through first and second derivatives alone. In discrete time no such implication holds.\footnote{For example, let $h_x^+=1+x+\sqrt{x}\left(\epsilon^2-1\right)$, where $\epsilon\sim N(0,1)$. This update remains in $\Rplus$ because $h_x^+\geq1+x-\sqrt{x}>0$ for every $x\geq0$. Moreover, $\E\left[h_x^+\right]=1+x$ and $\E\left[\left(h_x^+\right)^2\right]=(1+x)^2+2x$, so the corresponding one-step operator preserves $\Pol_1(\Rplus)$ and $\Pol_2(\Rplus)$. However, $\E\left[\left(h_x^+\right)^3\right]=(1+x)^3+6x(1+x)+8x^{3/2}$ is not polynomial. Thus, preservation at degrees one and two does not imply preservation at degree three in discrete time.}  The coefficient characterization applies to the covariance matrix $\bm{B}=\bm{\Sigma}\bm{\Sigma}^{\top}$ rather than to the diffusion loading $\bm{\Sigma}$ itself, which may involve non-polynomial square-root factorizations. The polynomial coefficient restrictions alone do not guarantee existence, uniqueness, nonexplosion, or stochastic invariance of $E$; these remain separate requirements for the limiting diffusion. See \citet{filipovic2016} for general conditions for polynomial diffusions on semialgebraic state spaces.

Operator convergence also connects the finite-dimensional moment calculations in discrete and continuous time. Using the notation of Section~\ref{subsec:general_framework}, let $\bm Q_m^{(n)}$ denote the coordinate matrix in Equation~\eqref{eq:basis_vector_transition}, representing the one-step transition operator in the basis $\bm{\HT}_m$ fixed across $n$, with $N_m=\dim\Pol_m(E)$. For $p\in\Pol_m(E)$, write $\bm p$ for its coordinate vector as in Equation~\eqref{eq:coordinate_representation}. For a polynomial diffusion, define the generator matrix $\bm G_m$ in the same basis by $\mathcal A\bm{\HT}_m=\bm G_m^\top\bm{\HT}_m$.

\begin{corollary}\label{cor:moment_matrix_convergence}
Suppose that $\bm X^{(n)}$ is polynomial preserving for every $n$ and that $\bm X$ is a globally defined polynomial diffusion on the same state space $E$. Fix $m\geq0$, and suppose that Equation~\eqref{eq:nelson_generator_convergence} holds for every $p\in\Pol_m(E)$.  Then
\begin{equation*}\lim_{\tau\to0}\frac{\bm Q_m^{(n)}-\bm I_{N_m}}{\tau}=\bm G_m, \qquad \lim_{\tau\to0}\sup_{0\leq t\leq L}\left\|\left(\bm Q_m^{(n)}\right)^{\lfloor t/\tau\rfloor}-e^{t\bm G_m}\right\|=0.\end{equation*}
Consequently, for $p\in\Pol_m(E)$ and a common deterministic initial state $\bm x\in E$,
\begin{equation*}\lim_{\tau\to0}\E_{\bm x}^{\Pm^{(n)}}\left[p\left(\bm X_t^{(n)}\right)\right]=\E_{\bm x}^{\Pm}\left[p\left(\bm X_t\right)\right]=\bm{\HT}_m(\bm x)^\top e^{t\bm G_m}\bm p,\end{equation*}
where the convergence is uniform for $t\in[0,L]$ and $\bm x\in\mathcal K$, for every compact set $\mathcal K\subseteq E$.
\end{corollary}

\begin{proof}
See Supplementary Material~\ref{sm:operator_convergence_polynomials}.
\end{proof}

Under the physical measure, the relevant polynomial states are the variance under both specifications and cumulative returns jointly with variance under JPGARCH-SV, for every $\varrho\in\left[0,1\right]$. Joint preservation also holds for VPGARCH-SV with independent innovations, as in Corollary~\ref{cor:polynomial_temporal_aggregates}.

The one-step forecast-error covariance $V^{(n)}_{p,q,1}$ defined in Section~\ref{subsec:general_framework} also has a local diffusion limit. For $p,q\in\Pol_m(E)$, suppose that $\mathcal A$ is well defined on $\Pol_{2m}(E)$ and that Equation~\eqref{eq:nelson_generator_convergence} holds on this space. Then
\begin{equation*}\lim_{\tau\to0}\frac{V^{(n)}_{p,q,1}(\bm x)}{\tau}=\nabla p(\bm x)^\top\bm B(\bm x)\nabla q(\bm x),\end{equation*}
locally uniformly on $E$, where $\bm B$ is the instantaneous covariance matrix defined above. The proof is given in Supplementary Material~\ref{sm:operator_convergence_polynomials}.

\subsection{Diffusion limits under the physical measure}
\label{subsec:physical_diffusion_limits}

We derive local diffusion limits of the VPGARCH-SV and JPGARCH-SV models under the physical measure. Local weak convergence is understood in $D([0,L],\mathbb{R}^2)$ with the Skorokhod $\mathrm{J}_1$ topology after stopping on increasing bounded domains whose closures lie in $\mathbb{R}\times(0,\infty)$ and whose union covers it, with almost surely continuous exit-time stopping maps for the limiting diffusion. Each discrete-time process is stopped at its first exit from the corresponding localizing domain.

In the general framework of Section~\ref{subsec:general_diffusion_framework}, we specialize the state process to
\begin{equation*}\bm{X}^{(n)}_{k\tau}=\left(Y^{(n)}_{k\tau},h^{(n)}_{k\tau}\right)^\top, \qquad k=0,1,\ldots,nL,\end{equation*}
and use its c\`adl\`ag extension defined in Equation~\eqref{eq:cadlag_extension}. The limiting characteristics depend on how the parameters behave as $\tau$ approaches zero.

\begin{assumption}\label{ass:p_scaling}
As $\tau\to0$, the parameters common to VPGARCH-SV and JPGARCH-SV satisfy
\begin{equation*}\kappa=\lim_{\tau\to0}\frac{1-\phi(\tau)}{\tau}, \quad \sigma^2=\lim_{\tau\to0}\sigma^2(\tau), \quad \lambda=\lim_{\tau\to0}\frac{\lambda(\tau)}{\tau}, \quad \eta=\lim_{\tau\to0}\frac{\eta(\tau)}{\sqrt{\tau}}, \quad b_i=\lim_{\tau\to0}\frac{b_i(\tau)}{\sqrt{\tau}}, \quad i=0,1.\end{equation*}
For VPGARCH-SV, in addition,
\begin{equation*}a_i=\lim_{\tau\to0}a_i(\tau), \qquad i=0,1.\end{equation*}
All limits are assumed to be finite.
\end{assumption}

The convergence rates in Assumption~\ref{ass:p_scaling} are of the standard type used to derive weak diffusion limits of GARCH models; see, for example, \citet{nelson1990arch}, \citet{Duan97}, and \citet{badescu2024discrete}. Such restrictions are generally not unique, and alternative scaling rates may produce different, possibly degenerate, diffusion limits, as demonstrated for the standard GARCH(1,1) model by \citet{corradi2000reconsidering}. The rates adopted here place the linear and centred quadratic innovation terms at the diffusion scale whenever their scaled loadings are nonzero.\footnote{In Section~\ref{subsec:polynomial_levy_limits}, we modify both the parameter scaling and the innovation law, using standardized L\'evy increments to obtain polynomial jump-diffusion limits.}

We assume that $\bm{X}^{(n)}_0=\bm{X}_0=(Y_0,h_0)^\top$ for every $n$, where $Y_0\in\mathbb{R}$ and $h_0>0$ are deterministic. To express the two limits jointly, let $\mathsf{M}\in\{\mathsf{VP},\mathsf{JP}\}$ identify VPGARCH-SV and JPGARCH-SV, respectively, and define
\begin{equation}\label{eq:p_limit_g_functions}g_{\mathsf{VP}}(x)=a_0+a_1x+x^2, \qquad g_{\mathsf{JP}}(x)=x.\end{equation}
The diffusion limits of the VPGARCH-SV and JPGARCH-SV models are given in the following proposition.

\begin{proposition}\label{prop:p_weak_limit}
Suppose the returns follow the VPGARCH-SV or JPGARCH-SV dynamics introduced in  Section~\ref{subsec:polynomial_garch_models_p} under $\Pm^{(n)}$, and the asymptotic constraints from  Assumption~\ref{ass:p_scaling} hold. Then, as $\tau\to0$, the c\`adl\`ag extensions of $\bm X^{(n)}$ converge weakly, locally on $\mathbb{R}\times(0,\infty)$, to the diffusion $\bm{X}=(Y,h)^\top$ satisfying
\begin{equation}\label{eq:p_limit_system}\begin{aligned}[b]dY_t&=(r+\lambda h_t)\,dt+\sqrt{h_t}\,dW_{1,t},\\ dh_t&=\kappa\left(\sigma^2-h_t\right)\,dt+\sqrt{v_{\mathsf{M}}(h_t)}\,dW_{2,t}.\end{aligned}\end{equation}
Here $W_1$ and $W_2$ are standard Brownian motions under $\Pm$, adapted to $\Ft$, and are stochastically correlated according to $d\left\langle W_1,W_2\right\rangle_t=\rho_{\mathsf{M}}(h_t)\,dt$, where $\rho_{\mathsf{M}}(x)=\varrho\eta\sqrt{g_{\mathsf{M}}(x)}/\sqrt{v_{\mathsf{M}}(x)}$ and $v_{\mathsf{M}}(x)=\eta^2g_{\mathsf{M}}(x)+(m_4-1)\left(b_0+b_1x\right)^2$, with $g_{\mathsf{M}}$ given in Equation~\eqref{eq:p_limit_g_functions}; when $v_{\mathsf{M}}(x)=0$, we set $\rho_{\mathsf{M}}(x)=0$.
\end{proposition}

\begin{proof}
See Appendix~\ref{app:p_weak_limit_proof}.
\end{proof}

Proposition~\ref{prop:p_weak_limit} gives the joint diffusion limit of log price and variance locally on $\mathbb{R}\times(0,\infty)$. Imposing the respective positivity restrictions in Propositions~\ref{prop:variance_polynomial_positivity} and~\ref{prop:joint_polynomial_positivity} for all sufficiently small $\tau$ forces $b_0=0$ in both models and $\eta^2a_0=0$ for VPGARCH-SV. If, in addition, $2\kappa\sigma^2\geq\eta^2a_1$ for VPGARCH-SV or $2\kappa\sigma^2\geq\eta^2$ for JPGARCH-SV, the limiting variance remains strictly positive and the diffusion is nonexplosive. Weak convergence then holds on the full horizon in $D([0,L],\mathbb{R}^2)$ with the Skorokhod $\mathrm{J}_1$ topology. The boundary conditions and localization argument are established in Supplementary Material~\ref{sm:variance_limit_boundary_conditions} and Appendix~\ref{app:p_weak_limit_proof}, respectively. Convergence through an attainable zero boundary requires a separate approximation argument.

In both models, the limiting variance equation has Pearson form \citep[see][]{FormanSorensen2008}, with affine drift and a squared diffusion coefficient $v_{\mathsf M}$ of degree at most two; its generator is therefore polynomial. The joint innovation law enters the limiting covariance matrix only through $\varrho$ and $m_4=m_{4,2}$. By joint symmetry, the centred quadratic variance innovation is uncorrelated with the return innovation, so only the linear variance innovation contributes to the instantaneous return--variance covariance. For JPGARCH-SV, this covariance is $\varrho\eta x$, giving a polynomial joint operator. For VPGARCH-SV with $\varrho\eta\neq0$, it is $\varrho\eta\sqrt{xg_{\mathsf{VP}}(x)}$, which is non-polynomial because its square has degree three. When $\varrho\eta=0$, the instantaneous return--variance covariance vanishes in both models, so the VPGARCH-SV joint operator is polynomial as well. Thus, zero innovation correlation suffices for joint polynomiality in the diffusion limit even when the discrete-time innovations are dependent. These operator properties yield polynomial diffusions and their moment formulas on an invariant state space where the solutions are globally defined. Stopping at an arbitrary interior boundary may destroy polynomial moment closure.

The determinant of the local limiting covariance matrix $\bm B_{\mathsf M}$ is
\begin{equation*}
\det\bm B_{\mathsf M}(x)=x\left[(1-\varrho^2)\eta^2g_{\mathsf M}(x)+(m_4-1)(b_0+b_1x)^2\right].
\end{equation*}
Since $m_4>1$, the matrix has full rank at a positive state $x$ if and only if $b_0+b_1x\neq0$ or both $\varrho<1$ and $\eta^2g_{\mathsf M}(x)>0$.

For $\varrho\eta\neq0$ and $(b_0,b_1)\neq(0,0)$, the JPGARCH-SV correlation is state dependent. Under the same restrictions, the VPGARCH-SV correlation is constant on states where $v_{\mathsf{VP}}(x)>0$ precisely when $b_1\neq0$ and $g_{\mathsf{VP}}(x)=(x+b_0/b_1)^2$, as in NGARCH-SV.  If $b_0=b_1=0$, both correlations equal $\varrho\operatorname{sgn}(\eta)$ wherever the variance diffusion coefficient is nonzero. HNGARCH-SV falls into this last case under the finite-step positivity restrictions, since $b_1=0$ by construction and those restrictions force $b_0=0$.

At $\varrho=1$ with Gaussian innovations, the NGARCH parameter mapping in Remark~\ref{rmk:ngarch_special_case} recovers the usual NGARCH diffusion coefficients, while the HNGARCH mapping in Remark~\ref{rmk:hngarch_special_case} recovers the local coefficient form discussed by \citet{badescu2024discrete}.

The JPGARCH-SV covariance structure can be compared with that of the Jacobi stochastic volatility model of \citet{Ackerer18}. To express the Jacobi covariance structure using the same Brownian factorization and physical return drift as in Equation~\eqref{eq:p_limit_system}, let $0\leq h_{\min}<\sigma^2<h_{\max}$ and define
\begin{equation}\label{eq:jacobi_q_function}Q(x)=\frac{(x-h_{\min})(h_{\max}-x)}{\left(\sqrt{h_{\max}}-\sqrt{h_{\min}}\right)^2}, \qquad x\in[h_{\min},h_{\max}].\end{equation}
For $\zeta_J>0$ and $\varrho_J\in[-1,1]$, an equivalent representation with the same covariance structure as the Jacobi model is
\begin{equation*}\begin{aligned}dY_t&=(r+\lambda h_t)\,dt+\sqrt{h_t}\,dW^{J}_{1,t},\\ dh_t&=\kappa\left(\sigma^2-h_t\right)\,dt+\zeta_J\sqrt{Q(h_t)}\,dW^{J}_{2,t},\end{aligned}\end{equation*}
where $W^{J}_1,W^{J}_2$ are $\Pm$-Brownian motions with instantaneous correlation $\varrho_J\sqrt{Q(h_t)/h_t}$, defined by its continuous extension $\varrho_J$ at $h_t=0$ when $h_{\min}=0$.

Both covariance structures are polynomial, but they are distinct for a genuine bounded Jacobi model. JPGARCH-SV combines the linear return--variance covariance $\varrho\eta h$ with the squared diffusion coefficient $v_{\mathsf{JP}}(h)$, whereas in the Jacobi model these functions are $\varrho_J\zeta_JQ(h)$ and $\zeta_J^2Q(h)$, respectively. The nonnegative quadratic coefficient of $v_{\mathsf{JP}}$ precludes matching the concave Jacobi quadratic. When $b_0=b_1=0$ and $\eta\neq0$, the JPGARCH-SV limit has the Heston covariance structure, with variance diffusion parameter $|\eta|$. In this case, the return--variance correlation $\varrho\operatorname{sgn}(\eta)$ spans $[-1,1]$ in the hybrid model and reduces to $\operatorname{sgn}(\eta)$ in its GARCH special case $\varrho=1$. Supplementary Material~\ref{sm:jacobi_garch_approximation} constructs a non-polynomial GARCH-type approximation of the Jacobi model of \citet{Ackerer18}, including a boundary-preserving version.

\subsection{Polynomial jump-diffusion limits}
\label{subsec:polynomial_levy_limits}

Standardized L\'evy increments provide an alternative scaling that connects the proposed specifications to polynomial jump-diffusions \citep[see, e.g.,][]{badescu2024discrete,FilipovicLarsson2020}. Within this subsection, we retain the VPGARCH-SV and JPGARCH-SV dynamics from  Section~\ref{subsec:polynomial_garch_models_p}, but replace their fixed-across-$n$ innovation law by the construction below and the parameter scaling of Assumption~\ref{ass:p_scaling} by Assumption~\ref{ass:p_levy_scaling}.

On the physical probability space of Section~\ref{subsec:general_diffusion_framework}, let $\bm J=\{\bm J_t=(J_{1,t},J_{2,t})^\top\}_{0\leq t\leq L}$ be a centred, jointly symmetric L\'evy process with respect to $\Ft$, starting at zero and having finite moments of all orders. Denote by $\bm C_J$ the covariance matrix per unit time of its Brownian component and by $\Pi_J$ its L\'evy measure, and assume
\begin{equation*}
\bm C_J+\int_{\mathbb R^2}\bm z\bm z^\top\Pi_J(d\bm z)=\begin{pmatrix}1&\varrho\\\varrho&1\end{pmatrix},\qquad \varrho\in[0,1].
\end{equation*}
For $i=1,2$ and $k=0,\ldots,nL-1$, define
\begin{equation}\label{eq:levy_increments}
\epsilon^{(n)}_{i,(k+1)\tau}=\frac{\Delta J_{i,(k+1)\tau}}{\sqrt{\tau}},
\end{equation}
where $\Delta J_{i,(k+1)\tau}=J_{i,(k+1)\tau}-J_{i,k\tau}$. With the L\'evy triplet fixed, the innovation pairs are i.i.d. over $k$, standardized and jointly symmetric with correlation $\varrho$ for each $n$, although their distribution generally varies with $\tau$. The return innovation is $\sqrt{h^{(n)}_{k\tau}}\Delta J_{1,(k+1)\tau}$, with conditional variance $\tau h^{(n)}_{k\tau}$.

\begin{assumption}\label{ass:p_levy_scaling}
For $\kappa,\sigma^2>0$, $r,\lambda,\eta\in\mathbb R$, and $b_0,b_1\geq0$, we assume that
\begin{equation}\label{eq:levy_parameter_scaling}
\begin{aligned}
\phi(\tau)&=1-\kappa\tau,\qquad \sigma^2(\tau)=\sigma^2,\qquad \lambda(\tau)=\lambda\tau,\\
b_i(\tau)&=b_i\tau,\quad i=0,1,\qquad \eta(\tau)=\eta\sqrt{\tau},
\end{aligned}
\end{equation}
for sufficiently small $\tau$. For VPGARCH-SV, we also set $a_1(\tau)=a_1\geq0$ and $a_0(\tau)=a_0\tau$, where $a_0\geq0$.\footnote{The exact parameter scalings simplify the discrete-time positivity conditions. Limiting assumptions, as in Assumption~\ref{ass:p_scaling}, can also be used, with finite-step admissibility imposed separately and corresponding adjustments to the convergence bounds.}
\end{assumption}

In this subsection, the limiting functions in Equation~\eqref{eq:p_limit_g_functions} are therefore $g_{\mathsf{VP}}(x)=a_1x+x^2$ and $g_{\mathsf{JP}}(x)=x$.\footnote{For $\eta\neq0$, nonnegativity at $h=0$ for every real innovation requires $\eta^2a_0(\tau)\leq4b_0(\kappa\sigma^2-b_0)\tau$. Hence a fixed positive value of $a_0(\tau)$ is incompatible with the all-real-innovation positivity requirement under this scaling, although $a_0(\tau)$ may remain positive at every finite time step.} The quadratic loading is now of order $\tau$. Indeed, if $\mu_{0,4}:=\int_{\mathbb R^2}z_2^4\Pi_J(d\bm z)$, the standardized fourth moment is $m_4^{(n)}=3+\mu_{0,4}/\tau$. This scaling therefore allows the quadratic innovation term to survive when the variance driver has jumps.

\begin{proposition}\label{prop:p_levy_limit}
Consider the VPGARCH-SV or JPGARCH-SV dynamics of Section~\ref{subsec:polynomial_garch_models_p} under $\Pm^{(n)}$, with innovations given by Equation~\eqref{eq:levy_increments} and parameters satisfying Assumption~\ref{ass:p_levy_scaling}. Let $(Y_0,h_0)^\top\in\mathbb R\times(0,\infty)$ be the common deterministic initial state and assume
\begin{equation}\label{eq:levy_global_conditions}
b_0\leq\kappa\sigma^2,\qquad \eta^2g_{\mathsf M}(x)\leq2x(b_0+b_1x),\quad x>0,
\end{equation}
and, for VPGARCH-SV, $\eta^2a_0\leq4b_0(\kappa\sigma^2-b_0)$.
For sufficiently small $\tau$, the conditional variances remain strictly positive, and, as $\tau\to0$, the c\`adl\`ag extensions of $(Y^{(n)},h^{(n)})$ converge weakly in $D([0,L],\mathbb R^2)$ with the Skorokhod $\mathrm{J}_1$ topology to the unique global strong solution
\begin{equation}\label{eq:p_levy_limit_system}
\begin{aligned}
dY_t&=(r+\lambda h_{t-})\,dt+\sqrt{h_{t-}}\,dJ_{1,t},\\
dh_t&=\kappa(\sigma^2-h_{t-})\,dt+\eta\sqrt{g_{\mathsf M}(h_{t-})}\,dJ_{2,t}+(b_0+b_1h_{t-})\left(d[J_2]_t-dt\right).
\end{aligned}
\end{equation}
Here, $[J_2]$ is the quadratic variation of $J_2$. The limiting variance process $h$ is strictly positive and polynomial on $(0,\infty)$ in both models. The process $(Y,h)$ is jointly polynomial on $\mathbb R\times(0,\infty)$ for JPGARCH-SV and for VPGARCH-SV when $J_1$ and $J_2$ are independent.
\end{proposition}

\begin{proof}
See Appendix~\ref{sm:proof_p_levy_limit}.{\footnote{Related approximation results include convergence in law of GARCH processes to COGARCH \citep{Kallsen2009} and first-jump approximations that converge in probability in the Skorokhod topology \citep{MallerMullerSzimayer2008}. Under the common L\'evy-driver construction in Equation~\eqref{eq:levy_increments}, the proof of Proposition~\ref{prop:p_levy_limit} also establishes convergence in probability in the Skorokhod $\mathrm{J}_1$ topology.}}
\end{proof}

Whereas Proposition~\ref{prop:p_weak_limit} states a local diffusion limit, the restrictions in Proposition~\ref{prop:p_levy_limit} ensure strict positivity of the limiting variance and allow localization to be removed, yielding joint weak convergence of log price and variance on the full horizon $[0,L]$.  In addition to $b_0\leq\kappa\sigma^2$, Condition~\eqref{eq:levy_global_conditions} requires $\eta^2\leq2b_0$ for JPGARCH-SV and $\eta^2a_1\leq2b_0$, $\eta^2\leq2b_1$ for VPGARCH-SV. For VPGARCH-SV, the additional bound on $a_0$ enters the proof of finite-step positivity. Propositions~\ref{prop:variance_polynomial_positivity} and~\ref{prop:joint_polynomial_positivity} give the sharp finite-step criteria for nonnegativity at every state and every real innovation under their respective parameter restrictions.\footnote{Under Assumption~\ref{ass:p_levy_scaling}, the discrete-time positivity conditions imply the limiting bound $\eta^2g_{\mathsf M}(x)\leq4x(b_0+b_1x)$. The stronger factor-$2$ bound in Condition~\eqref{eq:levy_global_conditions} keeps the post-jump variance at least half its pre-jump level and makes the paired logarithmic jump integral in the proof nonnegative.}

Under the standardization above, $[J_2]_t-t$ is the compensated squared-jump martingale:
\begin{equation*}
[J_2]_t-t=\sum_{0<s\leq t}(J_{2,s}-J_{2,s-})^2-t\int_{\mathbb R^2}z_2^2\Pi_J(d\bm z).
\end{equation*}
At pre-jump variance $h_{t-}=v$, a jump $\bm z=(z_1,z_2)^\top$ of the driver produces changes $\sqrt v\,z_1$ in log price and $\eta\sqrt{g_{\mathsf M}(v)}z_2+(b_0+b_1v)z_2^2$ in variance. The linear loading governs the signed response, while a positive quadratic loading allows sufficiently large jumps of $J_2$ of either sign to raise variance.\footnote{When $\eta=b_0=0$, $\varrho=1$, and the driver is pure jump, the variance equation becomes $dh_t=[\kappa\sigma^2-(\kappa+b_1)h_{t-}]dt+b_1h_{t-}d[J_2]_t$, a COGARCH variance specification (see, e.g., \citealp{KluppelbergLindnerMaller2004,Kallsen2009}). Under the present symmetry and moment assumptions, this special case is jointly polynomial as well. Related higher-moment formulas for COGARCH are developed by \citet{BibbonaNegri2015}.}

The limiting dynamics allow for pure-jump processes as well as Brownian components in log price and variance. At $h_{t-}=v$, the Brownian contributions to their instantaneous variance rates are $(\bm C_J)_{11}v$ and $(\bm C_J)_{22}\eta^2g_{\mathsf M}(v)$, respectively. When the driver is purely Brownian, $[J_2]_t=t$ and the compensated quadratic-variation term disappears. In this case, Assumption~\ref{ass:p_levy_scaling} satisfies the scaling requirements of Assumption~\ref{ass:p_scaling}, with $b_i(\tau)/\sqrt{\tau}\to0$ and, for VPGARCH-SV, $a_0(\tau)\to0$. Condition~\eqref{eq:levy_global_conditions} then implies the boundary-nonattainment conditions for global convergence to the corresponding diffusion in Proposition~\ref{prop:p_weak_limit}. Under the diffusion scaling of Section~\ref{subsec:physical_diffusion_limits}, the quadratic innovation term can instead generate an additional Brownian component.

To specify the generator entering the moment formulas, write $\bm{x}=(u,v)^\top$, where $u$ is the log-price state and $v>0$ is the current value of the variance process $h$, and set $b(v)=b_0+b_1v$. The drift vector and jump size of the limiting process in Equation~\eqref{eq:p_levy_limit_system} are
\begin{equation*}
\bm{A}(\bm{x})=\begin{pmatrix}r+\lambda v\\\kappa(\sigma^2-v)\end{pmatrix},\qquad \bm{D}_{\mathsf{M}}(v,\bm{z})=\begin{pmatrix}\sqrt{v}\,z_1\\\eta\sqrt{g_{\mathsf{M}}(v)}z_2+b(v)z_2^2\end{pmatrix}.
\end{equation*}
Its continuous covariance matrix is
\begin{equation*}
\bm{C}_{\mathsf{M}}(v)=\begin{pmatrix}(\bm{C}_J)_{11}v&(\bm{C}_J)_{12}\eta\sqrt{vg_{\mathsf{M}}(v)}\\(\bm{C}_J)_{12}\eta\sqrt{vg_{\mathsf{M}}(v)}&(\bm{C}_J)_{22}\eta^2g_{\mathsf{M}}(v)\end{pmatrix}.
\end{equation*}
For polynomial test functions $f$, the extended generator is
\begin{equation}\label{eq:levy_generator}
\begin{aligned}
\mathcal{A}f(\bm{x})={}&\bm{A}(\bm{x})^\top\nabla f(\bm{x})+\frac12\operatorname{Tr}\left[\bm{C}_{\mathsf{M}}(v)\nabla^2f(\bm{x})\right]\\
&+\int_{\mathbb{R}^2}\left[f\left(\bm{x}+\bm{D}_{\mathsf{M}}(v,\bm{z})\right)-f(\bm{x})-\bm{D}_{\mathsf{M}}(v,\bm{z})^\top\nabla f(\bm{x})\right]\Pi_J(d\bm{z}).
\end{aligned}
\end{equation}
For the variance state alone, the corresponding generator is obtained by applying this expression to functions independent of $u$.

The proof verifies the polynomial generator conditions of \citet[Lemma~1]{FilipovicLarsson2020} at every degree for the states identified in Proposition~\ref{prop:p_levy_limit}. For each such state, retain the basis $\bm{\HT}_m$ and one-step transition matrix $\bm{Q}_m^{(n)}$, and denote by $\bm{G}_m$ the coordinate matrix of the corresponding generator specified above, so that $\mathcal{A}\bm{\HT}_m=\bm{G}_m^\top\bm{\HT}_m$. For each fixed degree $m$, the same proof gives
\begin{equation}\label{eq:levy_moment_matrix_limit}
\bm{Q}_m^{(n)}=\bm{I}_{N_m}+\tau\bm{G}_m+\BigO{\tau^2},\qquad \sup_{0\leq t\leq L}\left\|\left(\bm{Q}_m^{(n)}\right)^{\lfloor t/\tau\rfloor}-e^{t\bm{G}_m}\right\|=\BigO{\tau}.
\end{equation}
Consequently, for every polynomial in the corresponding polynomial state, the moments of the c\`adl\`ag extensions converge uniformly over $t\in[0,L]$ and initial states in any compact subset of that state space. This extends the moment connection in Corollary~\ref{cor:moment_matrix_convergence} to jump limits, for which polynomial preservation at every degree also requires degree bounds on jump moments of order three and above.

\section{Empirical Illustration}
\label{sec:empirical_analysis}

This section investigates the empirical potential of the polynomial-preserving framework using S\&P~500 returns and options. We compare VPGARCH and JPGARCH with their respective nested benchmarks, NGARCH and HNGARCH, under Gaussian and symmetric normal inverse Gaussian (NIG) innovations. To simplify implementation, we use the GARCH special case $\varrho=1$ of the models in Section~\ref{subsec:polynomial_garch_models_p}, which permits direct evaluation of the return likelihood without latent-state filtering. We assume this specification for option calibration to keep the exercise focused on the additional flexibility in the variance dynamics. We assess in-sample fit and use exact conditional moments under $\Pm$ to interpret differences in the implied variance dynamics.

\subsection{Data}
\label{subsec:data}

Our empirical analysis uses daily S\&P~500 index returns from January 1996 to August 2023 obtained from the Center for Research in Security Prices (CRSP) through Wharton Research Data Services (WRDS). We compute daily excess log returns using the three-month Treasury bill rate from the Federal Reserve's H.15 release. The return sample contains 6,965 observations.

We use out-of-the-money call and put S\&P~500 index options from OptionMetrics through WRDS. To limit the computational cost of calibration, we use weekly cross-sections sampled on Wednesdays, with the preceding trading day used when Wednesday is not a trading day. We apply standard filters to limit the influence of illiquid contracts and unreliable quotes.\footnote{We exclude options with maturities less than 7 or greater than 365 calendar days, prices below \$0.375, non-positive trading volume or open interest, or prices that violate static no-arbitrage bounds.} The resulting sample contains 664,592 option observations, including 446,572 puts, on 1,443 weekly dates spanning the same period as the returns from January 1996 to August 2023. Table~\ref{tab:option_data} in the Supplementary Material summarizes the sample by maturity and moneyness, measured as the strike-to-forward ratio $K/F$. Approximately two-thirds of the observations have maturities below 90 calendar days, and average implied volatilities are highest for deep out-of-the-money puts.

\subsection{Return-based estimation}
\label{subsec:return_estimation}
\label{subsec:implied_variance_dynamics}

We estimate the four models by maximum likelihood using daily excess log returns $y_t-r_t$, under two distributional assumptions: Gaussian and standardized symmetric NIG innovations. We impose the parameter restrictions of Propositions~\ref{prop:variance_polynomial_positivity} and~\ref{prop:joint_polynomial_positivity}, estimating NGARCH and HNGARCH as special cases of VPGARCH and JPGARCH, respectively. One period corresponds to one trading day, so we set $\tau=1$ and suppress the superscript $(n)$ and the argument $\tau$.\footnote{For each estimation, $h_0$ is set equal to $\sigma^2$, $\eta$ is unrestricted in sign, and we set $a_0=0$ for VPGARCH based on the preliminary estimates. The parametrization of the symmetric NIG distribution is given in Appendix~\ref{app:nig}.}

\begin{table}
	\caption{\textbf{Maximum Likelihood Estimates from Daily Returns}}\label{tab:return_mle}
	\scriptsize
	\renewcommand{\arraystretch}{1.1}
	\setlength{\tabcolsep}{10pt}
	\begin{tabular*}{\textwidth}{@{\extracolsep{\fill}} l *{4}{c}}
		\toprule
		\multicolumn{5}{l}{\textbf{Panel A: Gaussian}} \\
		\midrule
		& HNGARCH & JPGARCH & NGARCH & VPGARCH \\
		\cmidrule(l){2-2} \cmidrule(l){3-3} \cmidrule(l){4-4} \cmidrule(l){5-5}
		$\lambda$ & 2.3566 & 2.9274 & 1.7382 & 1.8724 \\
		& (0.9950) & (0.9607) & (0.8165) & (0.7944) \\
		$\sigma^2\times10^{4}$ & 1.2248 & 1.1754 & 1.9100 & 1.5456 \\
		& (0.0567) & (0.0675) & (0.3835) & (0.2341) \\
		$\phi$ & 0.9516 & 0.9591 & 0.9848 & 0.9772 \\
		& (0.0030) & (0.0030) & (0.0034) & (0.0047) \\
		$\eta$ & -2.05E-03 & -1.85E-03 & -0.2023 & -0.2033 \\
		& (6.30E-05) & (6.83E-05) & (0.0082) & (0.0142) \\
		$a_1\times10^{6}$ & $-$ & $-$ & $-$ & 6.1416 \\
		& & & & (9.1657) \\
		$b_0\times10^{6}$ & 5.9317 & 2.7666 & $-$ & 2.0929 \\
		& (0.2778) & (0.2934) & & (0.3765) \\
		$b_1$ & $-$ & 0.0617 & 0.0943 & 0.0609 \\
		& & (0.0070) & (0.0045) & (0.0074) \\
		\addlinespace[1.5ex]
		Log-likelihood & 22,432.56 & 22,523.95 & 22,541.60 & 22,552.74 \\
		AIC & -44,855.11 & -45,035.89 & -45,073.21 & -45,091.48 \\
		BIC & -44,820.87 & -44,994.80 & -45,038.96 & -45,043.54 \\
		HQIC & -44,843.31 & -45,021.73 & -45,061.40 & -45,074.95 \\
		LR statistic & $-$ & 182.78 & $-$ & 22.27 \\
		& & [1.20E-41] & & [1.46E-05] \\
		\toprule
		\multicolumn{5}{l}{\textbf{Panel B: NIG}} \\
		\midrule
		& HNGARCH & JPGARCH & NGARCH & VPGARCH \\
		\cmidrule(l){2-2} \cmidrule(l){3-3} \cmidrule(l){4-4} \cmidrule(l){5-5}
		$\lambda$ & 4.9016 & 4.7644 & 2.4778 & 2.5825 \\
		& (0.9653) & (0.9117) & (0.6922) & (0.7194) \\
		$\sigma^2\times10^{4}$ & 1.0963 & 1.1077 & 2.0327 & 1.5021 \\
		& (0.0632) & (0.0914) & (0.5758) & (0.2714) \\
		$\phi$ & 0.9541 & 0.9611 & 0.9871 & 0.9782 \\
		& (0.0036) & (0.0039) & (0.0040) & (0.0053) \\
		$\eta$ & -2.00E-03 & -1.87E-03 & -0.2137 & -0.2052 \\
		& (9.71E-05) & (1.03E-04) & (0.0118) & (0.0181) \\
		$a_1\times10^{6}$ & $-$ & $-$ & $-$ & 11.5518 \\
		& & & & (12.6257) \\
		$b_0\times10^{6}$ & 5.0281 & 2.4441 & $-$ & 2.1051 \\
		& (0.4172) & (0.4831) & & (0.5401) \\
		$b_1$ & $-$ & 0.0603 & 0.0896 & 0.0559 \\
		& & (0.0100) & (0.0072) & (0.0099) \\
		$\nu$ & 0.4915 & 0.4150 & 0.3898 & 0.3893 \\
		& (0.0504) & (0.0447) & (0.0415) & (0.0417) \\
		\addlinespace[1.5ex]
		Log-likelihood & 22,568.06 & 22,634.39 & 22,651.17 & 22,659.86 \\
		AIC & -45,124.12 & -45,254.77 & -45,290.34 & -45,303.72 \\
		BIC & -45,083.03 & -45,206.83 & -45,249.25 & -45,248.93 \\
		HQIC & -45,109.96 & -45,238.25 & -45,276.18 & -45,284.83 \\
		LR statistic & $-$ & 132.65 & $-$ & 17.38 \\
		& & [1.08E-30] & & [1.68E-04] \\
		\bottomrule
	\end{tabular*}
	\footnotesize
	\emph{Notes:} This table reports maximum likelihood estimates obtained from 6,965 daily S\&P~500 excess returns from January 1996 to August 2023, with Gaussian innovations in Panel~A and NIG innovations in Panel~B. Standard errors, computed from the outer product of the gradients of the individual log-likelihood contributions, are in parentheses. Panel~B also reports the inverse shape parameter $\nu=1/\alpha$, which corresponds to an innovation kurtosis of $3+3\nu$.  For VPGARCH, $a_0$ is fixed at zero.  AIC, BIC and HQIC are the Akaike, Bayesian and Hannan--Quinn information criteria, $-2\ln L+2k$, $-2\ln L+k\ln T$ and $-2\ln L+2k\ln\ln T$, respectively, where $L$ is the maximized likelihood and $T=6{,}965$. The LR statistic tests each polynomial model against its nested benchmark, JPGARCH against HNGARCH with one restriction and VPGARCH against NGARCH with two restrictions; $p$-values are in square brackets.
\end{table}

Panel~A of Table~\ref{tab:return_mle} shows that both polynomial specifications improve the Gaussian return likelihood relative to their nested benchmarks, with the larger gain arising for JPGARCH. VPGARCH attains the highest likelihood and is preferred by all three information criteria.\footnote{The reported LR $p$-values use conventional $\chi^2$ reference distributions with one and two degrees of freedom for JPGARCH and VPGARCH, respectively. Since the nesting restrictions place parameters on the boundary, these need not be the correct asymptotic null distributions.} The additional flexibility is in the variance response across volatility regimes, since the affine loading $\psi_3(h)=b_0+b_1h$ accommodates both a constant and a proportional response to squared innovations, whereas HNGARCH and NGARCH retain only one of these components. All four models yield negative and statistically significant estimates of $\eta$, indicating leverage effects.

Panel~B shows that NIG innovations raise the log-likelihood of every model by 107--136 points at the cost of one additional parameter. The inverse shape parameter $\nu=1/\alpha$ controls tail thickness, with $\nu=0$ denoting the Gaussian limit. Since the innovations are standardized, their kurtosis equals the fourth moment $m_4=\E[\epsilon_t^4]=3+3\nu$, and their excess kurtosis is $3\nu$. The estimates of $\nu$ range from 0.39 to 0.49, implying kurtosis between 4.17 and 4.47. HNGARCH benefits most, narrowing its likelihood gap to JPGARCH, but the gains from the polynomial extensions persist. AIC and HQIC continue to favour the VPGARCH specification, whereas BIC is virtually identical for VPGARCH and NGARCH. Thus, the improvement over HNGARCH is more pronounced than that over NGARCH under either innovation distribution.

The polynomial property also permits an exact comparison of the fitted variance dynamics. Writing $\E_t$ and $\Var_t$ for conditional moments under $\Pm$ given $\F_t$, the representation in Section~\ref{subsec:moment_implications} gives
\begin{equation*}\E_t[h_{t+s}]=\sigma^2+\phi^s\left(h_t-\sigma^2\right),\qquad s\geq1.\end{equation*}
All four models therefore exhibit the standard geometric mean reversion, with different estimated speeds. Under Gaussian innovations, JPGARCH increases the mean-reversion half-life from 14 trading days under HNGARCH to 17, while VPGARCH reduces it from 45 under NGARCH to 30. The NIG estimates imply slightly greater persistence and preserve this ordering. Figure~\ref{fig:variance_term_structure} in the Supplementary Material reports the corresponding expected-variance profiles, normalized by each model's long-run level.

To examine how return news changes both the expected level and the uncertainty of future variance, we fix $h_t=\sigma^2$ and $\epsilon_{t+1}=\epsilon$. The expected-variance news impact curve defined below extends the one-period curve of \citet{engle1993measuring} across horizons:
\begin{equation}\label{eq:eovnic}\mathrm{EoVNIC}_s(\epsilon)=\E_{t+1}[h_{t+s}]=\sigma^2+\phi^{s-1}\left[\psi_2(\sigma^2)\epsilon+\psi_3(\sigma^2)\left(\epsilon^2-1\right)\right],\qquad s\geq1.\end{equation}
Its asymmetry is governed by $\psi_2$, its curvature by $\psi_3$, and its decay across horizons by the persistence parameter $\phi$; Figure~\ref{fig:eovnic} in the Supplementary Material illustrates these responses. For the variance-of-variance news impact curve, we denote $d(x)=\psi_2(x)^2+(m_4-1)\psi_3(x)^2=d_0+d_1x+d_2x^2$, with $d_0,d_1,d_2$ as defined in Section~\ref{subsec:moment_implications}. Here, $m_4=3$ under Gaussian innovations and $m_4=3+3\nu$ under NIG innovations. Since $d$ is quadratic, the law of total variance closes on the first two conditional moments and yields
\begin{equation}\label{eq:vovnic_closed_form}\mathrm{VoVNIC}_s(\epsilon)= \Var_{t+1}\left[h_{t+s}\right] =\sum_{j=0}^{s-2}\left(\phi^2+d_2\right)^{s-2-j}d\!\left(\mathrm{EoVNIC}_{j+1}(\epsilon)\right),\qquad s\geq2.\end{equation}
Both expressions are exact and require no simulation. The persistence of variance uncertainty depends on $\phi^2+d_2$, so it need not follow the same horizon pattern as expected variance.\footnote{The restriction $\phi^2+d_2<1$, which ensures a finite limiting second moment of the variance, is not imposed in estimation. At the NGARCH and VPGARCH estimates under both innovation distributions, $\phi^2+d_2>1$; finite-horizon conditional moments remain finite.}

\begin{figure}[!htbp]
	\centering
	\includegraphics{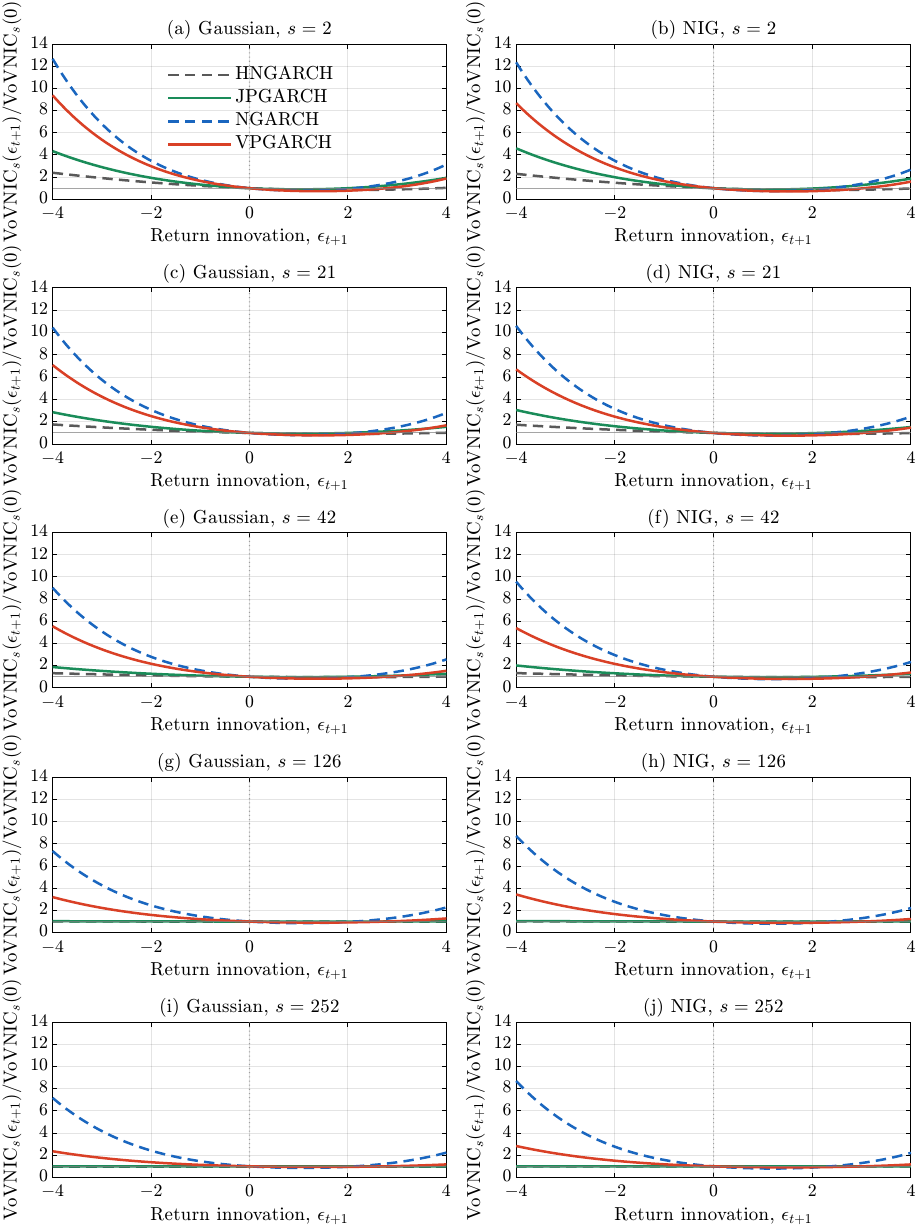}
	\caption{\textbf{Variance-of-Variance News Impact Curves}\newline
		\footnotesize
		\emph{Notes:} Each panel shows $\mathrm{VoVNIC}_s(\epsilon_{t+1})/\mathrm{VoVNIC}_s(0)$ for the horizon $s$ indicated, where $\mathrm{VoVNIC}_s(\epsilon)= \Var_{t+1}\left[h_{t+s}\right]$ is evaluated at $h_t=\sigma^2$ and $\epsilon_{t+1}=\epsilon$ under $\Pm$. The left panels use the Gaussian estimates and the right panels the NIG estimates in Table~\ref{tab:return_mle}. The horizontal line marks a ratio of one. All curves are computed exactly from Equation~\eqref{eq:vovnic_closed_form}.}
	\label{fig:vovnic}
\end{figure}

Figure~\ref{fig:vovnic} reports $\mathrm{VoVNIC}_s(\epsilon)/\mathrm{VoVNIC}_s(0)$, measuring the response relative to a zero return innovation. At the Gaussian estimates, $d(\sigma^2)/\sigma^4$ is similar for VPGARCH and NGARCH, at 0.054 and 0.059, respectively, yet a shock of $-4$ leaves their one-year variance-of-variance ratios at approximately 2.4 and 7.2. The constant and linear terms in VPGARCH accommodate substantial one-step variance uncertainty with a smaller quadratic coefficient; together with its lower $\phi$, this weakens the propagation of that uncertainty. Table~\ref{tab:vov_coefficients} in the Supplementary Material reports the corresponding coefficients. We notice that JPGARCH responds more strongly to negative innovations than HNGARCH, although both relative effects largely dissipate within six months. The NIG estimates yield the same qualitative patterns, with larger one-year relative effects under the NGARCH and VPGARCH specifications.

\subsection{Option-based estimation}
\label{subsec:option_pricing}

We next assess the ability of the four models to price options. On each of the 1,443 weekly dates described in Section~\ref{subsec:data}, we calibrate each model to the observed option cross-section by minimizing the implied volatility root mean squared error
\begin{equation}\label{eq:ivrmse}\mathrm{IVRMSE}_d=\sqrt{\frac{1}{N_d}\sum_{i=1}^{N_d}\left(\mathrm{IV}^{\mathrm{model}}_{i,d}-\mathrm{IV}^{\mathrm{mkt}}_{i,d}\right)^2},\end{equation}
where $N_d$ is the number of options on date $d$, and $\mathrm{IV}^{\mathrm{model}}_{i,d}$ and $\mathrm{IV}^{\mathrm{mkt}}_{i,d}$ are the model and market Black--Scholes implied volatilities.

For all models, we compute option prices by Monte Carlo simulation under the physical measure $\Pm$.\footnote{Orthogonal polynomial option-price expansions, such as those of \citet{AckererFilipovic2020}, use moments under the pricing measure. Our pricing kernel does not generally preserve polynomiality under $\Qm$, and the Gaussian polynomial-preservation restrictions in Appendix~\ref{subsec:q_polynomial_preservation} are not imposed in the calibration. The available physical moment recursions therefore do not directly supply the pricing-kernel-weighted moments required for these expansions.} Specifically, we simulate Equations~\eqref{eq:general_return_dynamics_p}--\eqref{eq:general_variance_dynamics_p} and average payoffs weighted by the product of the one-period pricing kernels in Equation~\eqref{eq:pricing_kernel}. For the affine Gaussian HNGARCH model, we validate the numerical implementation against its semi-closed-form European option prices. We use 50,000 simulated paths and, on each calibration date, hold the underlying random numbers fixed across parameter evaluations and models. Using common random numbers stabilizes the simulated calibration objective during optimization and facilitates consistent comparisons across models. The NIG innovations are generated using the variance-mixture representation in Equation~\eqref{eq:nig_mixture}.

We set the variance risk-preference parameter $\xi_{2,t}=0$, giving a pricing kernel that is monotonic in returns, as discussed in Appendix~\ref{subsec:pricing_kernel_q_dynamics}. This specification is convenient for NIG innovations because the normalization and martingale restrictions involve only the univariate cgf of the return innovation $\epsilon_{1,t}$. A nonzero variance risk-preference parameter would instead involve the joint cgf of $(\epsilon_{1,t},\epsilon_{2,t},\epsilon_{2,t}^2)$ and, for NIG innovations,  generally require numerical evaluation of this cgf within its domain of finiteness and numerical solution of the martingale restriction. Under the chosen specification, we determine the equity risk-preference parameter $\xi_{1,t}$ from Equation~\eqref{eq:pricing_kernel_no_arbitrage}.\footnote{For NIG innovations, we use the admissible negative branch of the closed-form expression in \citet[Corollary~A.3]{badescu2011comparison}. In our calibrations, $\xi_{1,t}$ satisfies the cgf-domain and martingale conditions at every simulated state. Appendix~\ref{app:nig} discusses the limitation of simulated-path feasibility for global admissibility.} Moreover, this common kernel specification keeps the comparison focused on variance dynamics and innovation distributions. With $\xi_{2,t}$ fixed at zero and $\xi_{1,t}$ determined by the martingale restriction, all model parameters, including the initial variance $h_0$ and, for VPGARCH, $a_0$, are re-estimated on each date under the positivity restrictions imposed in Section~\ref{subsec:return_estimation}. For each calibration, we retain the lowest objective value obtained across the available optimization runs and starting values.

Table~\ref{tab:weekly_ivrmse} reports the main calibration results. Both polynomial specifications reduce mean weekly IVRMSE and every reported error quantile relative to their nested benchmarks under both innovation distributions. Under Gaussian innovations, JPGARCH reduces the mean weekly IVRMSE of HNGARCH by 41.4\%, while VPGARCH improves on NGARCH by 11.2\%. Under NIG innovations, the corresponding gains are 7.5\% and 13.7\%. VPGARCH attains the lowest mean weekly IVRMSE in both cases, at 0.74\% under Gaussian innovations and 0.68\% under NIG innovations. The additional flexibility of the polynomial specifications therefore improves in-sample fit even when the innovation distribution accommodates fat tails.

\begin{table}[!htbp]
\caption{\textbf{Summary Statistics of Weekly Implied Volatility Root Mean Squared Errors}}
\label{tab:weekly_ivrmse}
\scriptsize
\renewcommand{\arraystretch}{1.25}
\newcommand{\mc}[1]{\multicolumn{1}{c}{#1}}
\centering
\begin{tabularx}{\linewidth}{X*{4}{d{5.5}}}
\toprule
\multicolumn{5}{l}{\textbf{Panel A: Gaussian}} \\

\midrule
\multicolumn{5}{l}{\textit{Summary Statistics of Weekly IVRMSE}} \\
\addlinespace[1ex]
& \multicolumn{1}{c}{HNGARCH} & \multicolumn{1}{c}{JPGARCH} & \multicolumn{1}{c}{NGARCH} & \multicolumn{1}{c}{VPGARCH} \\
\cmidrule(l){2-2}\cmidrule(l){3-3}\cmidrule(l){4-4}\cmidrule(l){5-5}

Mean ($\%$)
& 1.78 & 1.04 & 0.84 & 0.74 \\

Standard deviation ($\%$)
& 1.09 & 0.81 & 0.36 & 0.34 \\
\addlinespace[1.5ex]
Quantiles ($\%$) &&&& \\
$\quad$ 5th & 0.67 & 0.36 & 0.41 & 0.36 \\
$\quad$ 25th & 1.08 & 0.53 & 0.59 & 0.52 \\
$\quad$ 50th & 1.44 & 0.75 & 0.79 & 0.69 \\
$\quad$ 75th & 2.09 & 1.30 & 0.98 & 0.88 \\
$\quad$ 95th & 4.13 & 2.65 & 1.46 & 1.35 \\
$\quad$ 99th & 5.32 & 4.03 & 1.98 & 1.84 \\

\midrule
\multicolumn{5}{l}{\textit{Mean IVRMSE Improvement over the Row Model}} \\
\addlinespace[1ex]
& \multicolumn{1}{c}{HNGARCH} & \multicolumn{1}{c}{JPGARCH} & \multicolumn{1}{c}{NGARCH} & \multicolumn{1}{c}{VPGARCH} \\
\cmidrule(l){2-2}\cmidrule(l){3-3}\cmidrule(l){4-4}\cmidrule(l){5-5}

HNGARCH
& & \mc{(41.4\%)} & \mc{(53.0\%)} & \mc{(58.2\%)} \\

JPGARCH
& & & \mc{(19.7\%)} & \mc{(28.7\%)} \\

NGARCH
& & & & \mc{(11.2\%)} \\

\toprule
\multicolumn{5}{l}{\textbf{Panel B: NIG}} \\

\midrule
\multicolumn{5}{l}{\textit{Summary Statistics of Weekly IVRMSE}} \\
\addlinespace[1ex]
& \multicolumn{1}{c}{HNGARCH} & \multicolumn{1}{c}{JPGARCH} & \multicolumn{1}{c}{NGARCH} & \multicolumn{1}{c}{VPGARCH} \\
\cmidrule(l){2-2}\cmidrule(l){3-3}\cmidrule(l){4-4}\cmidrule(l){5-5}

Mean ($\%$)
& 0.78 & 0.72 & 0.79 & 0.68 \\

Standard deviation ($\%$)
& 0.53 & 0.51 & 0.36 & 0.34 \\
\addlinespace[1.5ex]
Quantiles ($\%$) &&&& \\
$\quad$ 5th & 0.31 & 0.29 & 0.36 & 0.32 \\
$\quad$ 25th & 0.44 & 0.40 & 0.55 & 0.46 \\
$\quad$ 50th & 0.62 & 0.57 & 0.74 & 0.61 \\
$\quad$ 75th & 0.95 & 0.88 & 0.94 & 0.81 \\
$\quad$ 95th & 1.74 & 1.64 & 1.40 & 1.29 \\
$\quad$ 99th & 2.53 & 2.37 & 1.89 & 1.76 \\

\midrule
\multicolumn{5}{l}{\textit{Mean IVRMSE Improvement over the Row Model}} \\
\addlinespace[1ex]
& \multicolumn{1}{c}{HNGARCH} & \multicolumn{1}{c}{JPGARCH} & \multicolumn{1}{c}{NGARCH} & \multicolumn{1}{c}{VPGARCH} \\
\cmidrule(l){2-2}\cmidrule(l){3-3}\cmidrule(l){4-4}\cmidrule(l){5-5}

HNGARCH
& & \mc{(7.5\%)} & \mc{($-$0.9\%)} & \mc{(13.0\%)} \\

JPGARCH
& & & \mc{($-$9.0\%)} & \mc{(5.9\%)} \\

NGARCH
& & & & \mc{(13.7\%)} \\
\addlinespace[1ex]
Gaussian
& \mc{(56.2\%)} & \mc{(30.8\%)} & \mc{(6.1\%)} & \mc{(8.7\%)} \\

\bottomrule
\end{tabularx}

\begin{minipage}{\linewidth}
\footnotesize
\emph{Notes:} This table reports the mean, the standard deviation and selected quantiles of the weekly implied volatility root mean squared errors (IVRMSEs), in percent, obtained by calibrating each model separately on each of the 1,443 weekly dates from January 1996 to August 2023 with $\xi_{2,t}=0$. The sample contains 664,592 out-of-the-money S\&P~500 index options from OptionMetrics. Panel~A reports the results for Gaussian innovations and Panel~B those for NIG innovations. The rows labelled with a model name report, in parentheses, the improvement of the column model over that model, computed as the percentage reduction in the mean weekly IVRMSE. The row labelled Gaussian in Panel~B reports the percentage reduction in the mean weekly IVRMSE relative to the same model with Gaussian innovations.
\end{minipage}

\end{table}

The NIG innovations lower the mean weekly IVRMSE by 56.2\% for HNGARCH and 30.8\% for JPGARCH, compared with 6.1\% for NGARCH and 8.7\% for VPGARCH. Variance dynamics can themselves generate fat tails in cumulative returns even under Gaussian innovations: shocks that raise variance increase the scale of subsequent return innovations. The smaller gains for NGARCH and VPGARCH suggest that heavier innovation tails provide less additional benefit to option fit in these specifications.\footnote{The median innovation kurtoses in the option calibrations are about 23 and 20 for HNGARCH and JPGARCH, compared with 4.4 and 4.1 for NGARCH and VPGARCH (see Tables~\ref{tab:option_parameters_gaussian} and~\ref{tab:option_parameters_nig} in the Supplementary Material). JPGARCH's median $b_1$ also falls from 0.147 under Gaussian innovations to 0.0043 under NIG innovations.} Under NIG innovations, JPGARCH has lower IVRMSE quantiles than NGARCH at every reported level through the 75th percentile and the lowest median among all four models, at 0.57\%. Its 95th and 99th percentiles, however, exceed those of NGARCH and VPGARCH. At the 99th percentile, the IVRMSE is 2.37\% for JPGARCH, compared with 1.89\% for NGARCH and 1.76\% for VPGARCH. Thus, JPGARCH's favourable central quantiles coexist with a heavier upper tail of weekly pricing errors. VPGARCH combines the lowest mean error with the lowest upper quantiles under both innovation distributions.

\begin{figure}[!htbp]
	\centering
	\includegraphics{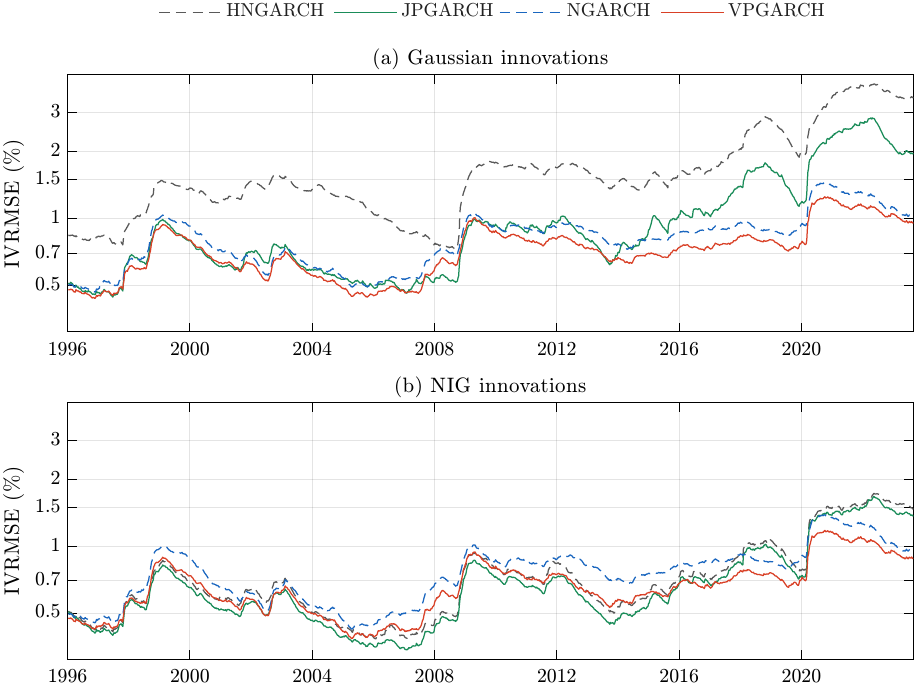}
	\caption{\textbf{Smoothed Weekly Implied Volatility Root Mean Squared Errors by Innovation Distribution}\newline
		\footnotesize
		\emph{Notes:} Panels~(a) and~(b) show exponentially weighted moving averages (EWMA) of the weekly IVRMSEs, in percent and on a logarithmic scale, of the four models under Gaussian and NIG innovations, respectively. The averages are computed using a smoothing parameter of $0.975$. Each model is calibrated separately on each of the 1,443 weekly dates from January 1996 to August 2023 with $\xi_{2,t}=0$. The raw weekly IVRMSEs are shown in Figure~\ref{fig:sm_weekly_ivrmse_distribution} of the Supplementary Material.}
	\label{fig:weekly_ivrmse_distribution}
\end{figure}

Figure~\ref{fig:weekly_ivrmse_distribution} illustrates these patterns using smoothed weekly IVRMSEs. We first notice that errors rise during periods of market stress, including the autumn of 1998, the financial crisis and the onset of the COVID-19 pandemic, and remain elevated after 2020. Under Gaussian innovations, VPGARCH generally provides the lowest errors. Under NIG innovations, JPGARCH performs particularly well over much of the pre-2020 period, whereas VPGARCH provides the lowest smoothed errors after 2020, when all models perform less well. Figure~\ref{fig:weekly_ivrmse_model} in the Supplementary Material compares Gaussian and NIG innovations model by model and shows the much larger gains from fat tails for HNGARCH and JPGARCH.

Tables~\ref{tab:ivrmse_moneyness_maturity_gaussian} and~\ref{tab:ivrmse_moneyness_maturity_nig} in Appendix~\ref{app:additional_results} provide a complementary comparison using pricing errors pooled across all options. Both polynomial models again improve on their benchmarks under both innovation distributions, and VPGARCH has the lowest overall IVRMSE. Under NIG innovations, however, NGARCH's overall IVRMSE of 1.09\% is lower than JPGARCH's 1.36\%, reversing their ordering by mean weekly IVRMSE; VPGARCH remains best at 0.97\%. The IVRMSE computed over all options gives greater weight to weeks with more observations and is more sensitive to exceptionally large weekly errors than the mean weekly IVRMSE. JPGARCH’s particularly large errors in its worst-fitting weeks are therefore consistent with its less favourable ranking by this measure.

We next examine how these improvements in pricing performance vary across moneyness and maturity. Under Gaussian innovations, JPGARCH delivers substantial gains over HNGARCH for out-of-the-money puts, reducing IVRMSE by about 39--50\% across the put moneyness groups. These gains persist under NIG innovations but narrow to about 6--13\%, as HNGARCH benefits more from the change in innovation distribution. For deep out-of-the-money puts, JPGARCH reduces the IVRMSE of HNGARCH from 4.34\% to 2.59\% under Gaussian innovations and from 1.59\% to 1.49\% under NIG innovations. VPGARCH improves on NGARCH on both sides of the smile, with lower IVRMSEs in every moneyness group and every maturity group when pooled over the other dimension, under both innovation distributions.  Its largest gains arise for slightly out-of-the-money calls, reaching about 20\%, and at intermediate maturities of 90--180 days, reaching about 15\%. The additional flexibility of VPGARCH therefore improves the fit beyond deep out-of-the-money contracts.

\section{Conclusions}
\label{sec:conclusions}

We develop a general discrete-time polynomial-preserving framework that combines flexible dynamics with exact conditional polynomial moments and cross-moments of future and cumulative quantities. The transition-operator formulation accommodates multivariate states and non-Gaussian innovations under the stated polynomial-preservation conditions. As its principal application, we introduce VPGARCH-SV and JPGARCH-SV, two hybrid GARCH--SV classes that nest standard option-pricing GARCH models. They allow greater flexibility in leverage and variance of variance while preserving polynomiality of the variance or joint return--variance state.  We establish their admissibility and moment conditions and identify when polynomiality survives risk neutralization under jointly Gaussian innovations. Under the stated conditions, the two scaling regimes provide a new discrete-time construction of continuous-time stochastic volatility models with polynomial variance dynamics. In both regimes, the discrete-time moment recursions converge to the matrix-exponential formulas of the limiting processes.

The empirical analysis combines daily S\&P~500 returns with 664,592 option observations on 1,443 weekly dates spanning more than 27 years. Both polynomial specifications improve return fit and option calibration relative to their nested benchmarks under Gaussian and NIG innovations. VPGARCH achieves the lowest mean weekly pricing error under both distributions. The much larger gains in option fit from NIG innovations for HNGARCH and JPGARCH show that the contribution of fat tails depends materially on the variance dynamics. Exact moment calculations also reveal differences in the persistence of expected variance and variance uncertainty.

Further work could develop polynomial specifications incorporating realized variance and fractional dynamics, and extend the jump-limit analysis to the pricing measure. Existing discrete-time models already incorporate several of these features \citep[see, e.g.,][]{chris14,ornth14,augustyniak2025fractional}. Our framework offers a route to characterizing their polynomial moment structure and developing more flexible alternatives beyond affine specifications.\footnote{Related continuous-time developments include polynomial jump-diffusions, polynomial Volterra processes, Gaussian polynomial volatility and signature-based models \citep[see, e.g.,][]{FilipovicLarsson2020,AbiJaberEtAl2024,AbiJaberIllandLi2025,CuchieroEtAl2025}. These constructions suggest further ways to incorporate nonlinear and path-dependent dynamics in discrete time through suitable state representations.}  Extending quadratic hedging methods to polynomial models would allow us to assess whether the improved option fit translates into better hedging performance \citep[see, e.g.,][]{augustyniak2022discrete}. Finally, the conditional moment formulas provide a basis for moment-based estimation of physical dynamics from returns and realized measures and, under polynomial-preserving pricing specifications, of risk-neutral dynamics from option-implied moments \citep[see, e.g.,][]{bolle02,FeunouOkou2018}.

\section*{Declaration of generative AI and AI-assisted technologies in the manuscript preparation process}

 During the preparation of the manuscript and its Supplementary Material, the authors used ChatGPT (OpenAI)  to assist with error checking, spelling, and language editing. The authors reviewed and edited the content as appropriate and take full responsibility for the manuscript.

	\setlength{\bibsep}{3pt}
	\titlespacing*{\section}{0pt}{6pt}{12pt}
	\singlespacing
	\begin{spacing}{0.0}
		\newcommand{\SortNoop}[1]{}

	\end{spacing}

	\appendix
	\numberwithin{equation}{section}

	\onehalfspacing

	\titlespacing*{\section}{0pt}{12pt}{6pt}
	\titlespacing*{\subsection}{0pt}{12pt}{6pt}
	\titlespacing*{\subsubsection}{0pt}{12pt}{6pt}

\section{Proofs}\label{ProofsofPropositions}

This appendix collects the proofs of the main characterization, polynomial diffusion connection, and physical diffusion and jump limits. The discrete-time risk-neutral results are given in Appendix~\ref{sec:risk_neutral_dynamics}, with proofs in Supplementary Material~\ref{app:proof_general_q_dynamics}--\ref{app:proof_q_polynomial_preservation}. Other proofs in the Supplementary Material verify the positivity, moment, and temporal-aggregation results and establish the risk-neutral diffusion limit.

\subsection{Proof of Proposition~\ref{prop:garch_polynomial_characterization}}
\label{app:proof_garch_polynomial_characterization}

\begin{proof}
For $x\in\Rplus$, let
\begin{equation*}y_x^+:=\psi_0\left(x;\tau\right)+\sqrt{\tau x}\,\epsilon_1, \qquad h_x^+:=\psi_1\left(x;\tau\right)+\psi_2\left(x;\tau\right)\epsilon_2+\psi_3\left(x;\tau\right)\left(\epsilon_2^2-1\right),\end{equation*}
where $\left(\epsilon_1,\epsilon_2\right)$ has the joint innovation law specified in Section~\ref{subsec:polynomial_garch_models_p}.

We first prove part~\textnormal{(i)}. Suppose that $\psi_1\in\Pol_1\left(\Rplus\right)$ and $\psi_2^2\in\Pol_2\left(\Rplus\right)$. For any integer $q\geq0$, each term in the multinomial expansion of $\left(h_x^+\right)^q$ is, up to a constant independent of $x$, of the form
\begin{equation*}\psi_1\left(x;\tau\right)^i\psi_2\left(x;\tau\right)^j\psi_3\left(x;\tau\right)^\ell\epsilon_2^j\left(\epsilon_2^2-1\right)^\ell, \qquad i+j+\ell=q.\end{equation*}
By symmetry, the terms with odd $j$ have zero expectation. For even $j$, the state-dependent coefficient is a polynomial of degree at most $i+j+\ell=q$. Hence, $\E^{\Pm^{(n)}}\left[\left(h_x^+\right)^q\right]\in\Pol_q\left(\Rplus\right)$ for every $q\geq0$, and $h^{(n)}$ is polynomial preserving.

Conversely, suppose that $h^{(n)}$ is polynomial preserving. Then
\begin{equation*}\E^{\Pm^{(n)}}\left[h_x^+\right]=\psi_1\left(x;\tau\right)\in\Pol_1\left(\Rplus\right),\end{equation*}
and symmetry gives
\begin{equation*}\E^{\Pm^{(n)}}\left[\left(h_x^+\right)^2\right]=\psi_1\left(x;\tau\right)^2+\psi_2\left(x;\tau\right)^2+\left(m_4-1\right)\psi_3\left(x;\tau\right)^2\in\Pol_2\left(\Rplus\right).\end{equation*}
Since $\psi_3\in\Pol_1\left(\Rplus\right)$ by assumption, it follows that $\psi_2^2\in\Pol_2\left(\Rplus\right)$. This proves part~\textnormal{(i)}.

For part~\textnormal{(ii)}, first suppose that $\varrho>0$ and $\left(y^{(n)},h^{(n)}\right)$ is polynomial preserving. Restricting the transition operator to polynomials in the variance coordinate and applying part~\textnormal{(i)} gives $\psi_1\in\Pol_1\left(\Rplus\right)$ and $\psi_2^2\in\Pol_2\left(\Rplus\right)$. Moreover,
\begin{align*}\E^{\Pm^{(n)}}\left[y_x^+\right]&=\psi_0\left(x;\tau\right)\in\Pol_1\left(\Rplus\right),\\ \E^{\Pm^{(n)}}\left[y_x^+h_x^+\right]&=\psi_0\left(x;\tau\right)\psi_1\left(x;\tau\right)+\varrho\sqrt{\tau x}\,\psi_2\left(x;\tau\right)\in\Pol_2\left(\Rplus\right).\end{align*}
Since $\varrho>0$, $q\left(x\right):=\sqrt{x}\,\psi_2\left(x;\tau\right)$ belongs to $\Pol_2\left(\Rplus\right)$. Since $q^2=x\psi_2^2\in\Pol_3\left(\Rplus\right)$, any nonzero $q$ satisfies $2\deg\left(q\right)\leq3$, and therefore $\deg\left(q\right)\leq1$. Since $q\left(0\right)=0$, there exists $c\in\mathbb{R}$ such that $q\left(x\right)=cx$. Hence, $\psi_2\left(x;\tau\right)=c\sqrt{x}$ for $x>0$. The polynomial identity $\psi_2^2\left(x;\tau\right)=c^2x$ then also holds at $x=0$, so $\psi_2\left(0;\tau\right)=0$ and
\begin{equation*}\psi_2\in\operatorname{span}\left\{x\mapsto\sqrt{x}\right\}.\end{equation*}

Conversely, suppose that $\psi_0,\psi_1\in\Pol_1\left(\Rplus\right)$ and $\psi_2\left(x;\tau\right)=c\sqrt{x}$ for some $c\in\mathbb{R}$. For integers $a,b\geq0$, each term in the expansion of $\left(y_x^+\right)^a\left(h_x^+\right)^b$ is, up to a constant independent of $x$, of the form
\begin{equation*}\psi_0\left(x;\tau\right)^{a-j}\psi_1\left(x;\tau\right)^i\psi_3\left(x;\tau\right)^\ell x^{\left(j+s\right)/2}\epsilon_1^j\epsilon_2^s\left(\epsilon_2^2-1\right)^\ell,\end{equation*}
where $0\leq j\leq a$ and $i+s+\ell=b$. Joint symmetry makes the terms with odd $j+s$ have zero expectation. When $j+s$ is even, the state-dependent coefficient is a polynomial of degree at most
\begin{equation*}\left(a-j\right)+i+\ell+\frac{j+s}{2}=a+b-\frac{j+s}{2}\leq a+b.\end{equation*}
Therefore, $\E^{\Pm^{(n)}}\left[\left(y_x^+\right)^a\left(h_x^+\right)^b\right]$ is a polynomial in the current variance state of degree at most $a+b$, and hence belongs to $\Pol_{a+b}\left(\mathbb{R}\times\Rplus\right)$. By linearity, $\left(y^{(n)},h^{(n)}\right)$ is polynomial preserving, proving sufficiency in part~\textnormal{(ii)} for every $\varrho\in\left[0,1\right]$.

For independent innovations, $\E^{\Pm^{(n)}}\left[\left(y_x^+\right)^a\left(h_x^+\right)^b\right]=\E^{\Pm^{(n)}}\left[\left(y_x^+\right)^a\right]\E^{\Pm^{(n)}}\left[\left(h_x^+\right)^b\right]$. Under $\psi_0\in\Pol_1\left(\Rplus\right)$, symmetry of $\epsilon_1$ makes the first factor a polynomial of degree at most $a$, while part~\textnormal{(i)} makes the second a polynomial of degree at most $b$. This proves sufficiency in the independent case; necessity follows by applying the joint transition operator to polynomials in each coordinate separately.

When $D_2\left(0,1\right)=N\left(0,1\right)$, Lemma~\ref{lem:hermite_loading_polynomiality} applied to the variance update shows that variance polynomial preservation itself implies $\psi_3\in\Pol_1\left(\Rplus\right)$.
\end{proof}

\subsection{Proof of Proposition~\ref{prop:weak_limit_polynomial_preservation}}
\label{app:proof_weak_limit_polynomial_preservation}

\begin{proof}
Fix $p\in\Pol_m(E)$. Since $\trans^{(n)}\Pol_m(E)\subseteq\Pol_m(E)$, we have $\mathcal{A}^{(n)}p=(\trans^{(n)}p-p)/\tau\in\Pol_m(E)$ for every $n$. By Equation~\eqref{eq:nelson_generator_convergence}, $\mathcal{A}^{(n)}p$ converges locally uniformly on $E$ to $\mathcal{A}p$. Since $\Pol_m(E)$ is finite dimensional and therefore closed under locally uniform convergence, it follows that $\mathcal{A}p\in\Pol_m(E)$. As $p$ was arbitrary, $\mathcal{A}\Pol_m(E)\subseteq\Pol_m(E)$.

Applying this with $m=1$ to the coordinate functions gives
\begin{equation*}
A_i=\mathcal{A}x_i\in\Pol_1(E),
\end{equation*}
while
\begin{equation*}
\mathcal{A}(x_ix_j)=x_jA_i+x_iA_j+B_{ij}\in\Pol_2(E),
\end{equation*}
where the second display uses $m=2$. Since $x_jA_i,x_iA_j\in\Pol_2(E)$, we obtain $B_{ij}\in\Pol_2(E)$ for all $i,j=1,\ldots,d$. For any $p\in\Pol_k(E)$, choose a polynomial extension of degree at most $k$. Each nonzero first-derivative term has degree at most $1+(k-1)=k$, and each nonzero second-derivative term has degree at most $2+(k-2)=k$. The additional well-definedness assumption on $\Pol(E)$ makes the result independent of the extension. Hence $\mathcal{A}p\in\Pol_k(E)$ for every $k\geq0$, and thus $\bm{X}$ is a polynomial diffusion with affine drift and quadratic covariance matrix.
\end{proof}

\subsection{Proof of Proposition~\ref{prop:p_weak_limit}}
\label{app:p_weak_limit_proof}

\begin{proof}
We verify the local diffusion conditions using the localization of Section~\ref{subsec:physical_diffusion_limits}. The increment law depends only on the current variance $h^{(n)}_{k\tau}=x$, so all limits below are uniform for $x$ in arbitrary compact sets $\mathcal{K}_h\subset(0,\infty)$. For $\mathsf{M}\in\{\mathsf{VP},\mathsf{JP}\}$, define
\begin{equation*}g_{\mathsf{VP}}(\tau,x)=a_0(\tau)+a_1(\tau)x+x^2, \qquad g_{\mathsf{JP}}(\tau,x)=x.\end{equation*}
By Assumption~\ref{ass:p_scaling}, $g_{\mathsf{M}}(\tau,x)$ converges locally uniformly to $g_{\mathsf{M}}(x)$ on $(0,\infty)$ as $\tau\to0$.

Conditional on $h^{(n)}_{k\tau}=x$, the one-step increments are
\begin{align*}\Delta Y^{(n)}_{(k+1)\tau}&=r\tau+\lambda(\tau)x+\sqrt{\tau x}\,\epsilon^{(n)}_{1,(k+1)\tau},\\ \Delta h^{(n)}_{(k+1)\tau}&=[1-\phi(\tau)]\left[\sigma^2(\tau)-x\right]+\eta(\tau)\sqrt{g_{\mathsf{M}}(\tau,x)}\,\epsilon^{(n)}_{2,(k+1)\tau}+\left[b_0(\tau)+b_1(\tau)x\right]\left[\left(\epsilon^{(n)}_{2,(k+1)\tau}\right)^2-1\right].\end{align*}
Hence, Assumption~\ref{ass:p_scaling} gives, locally uniformly on $(0,\infty)$,
\begin{align*}\lim_{\tau\to0}\frac{1}{\tau}\E^{\Pm^{(n)}}\left[\Delta Y^{(n)}_{(k+1)\tau}\mid h^{(n)}_{k\tau}=x\right]&=r+\lambda x,\\ \lim_{\tau\to0}\frac{1}{\tau}\E^{\Pm^{(n)}}\left[\Delta h^{(n)}_{(k+1)\tau}\mid h^{(n)}_{k\tau}=x\right]&=\kappa\left(\sigma^2-x\right).\end{align*}
Joint symmetry and standardization give $\E\left[\epsilon_1\epsilon_2\right]=\varrho$, $\E\left[\epsilon_1\epsilon_2^2\right]=\E\left[\epsilon_2^3\right]=0$, and $\Var\left[\epsilon_2^2-1\right]=m_4-1$. Hence,
\begin{align*}\lim_{\tau\to0}\frac{1}{\tau}\Var^{\Pm^{(n)}}\left[\Delta Y^{(n)}_{(k+1)\tau}\mid h^{(n)}_{k\tau}=x\right]&=x,\\ \lim_{\tau\to0}\frac{1}{\tau}\Cov^{\Pm^{(n)}}\left[\Delta Y^{(n)}_{(k+1)\tau},\Delta h^{(n)}_{(k+1)\tau}\mid h^{(n)}_{k\tau}=x\right]&=\varrho\eta\sqrt{xg_{\mathsf{M}}(x)},\\ \lim_{\tau\to0}\frac{1}{\tau}\Var^{\Pm^{(n)}}\left[\Delta h^{(n)}_{(k+1)\tau}\mid h^{(n)}_{k\tau}=x\right]&=\eta^2g_{\mathsf{M}}(x)+(m_4-1)\left(b_0+b_1x\right)^2=v_{\mathsf{M}}(x).\end{align*}
The conditional means are $O(\tau)$ locally uniformly, so the scaled conditional second moments have the same limits as the corresponding covariances. The limiting covariance matrix is therefore
\begin{equation*}\bm{B}_{\mathsf{M}}(x)=\begin{pmatrix}x&\varrho\eta\sqrt{xg_{\mathsf{M}}(x)}\\ \varrho\eta\sqrt{xg_{\mathsf{M}}(x)}&v_{\mathsf{M}}(x)\end{pmatrix}.\end{equation*}

The higher-moment condition follows directly from the scaling assumptions. Indeed, for every compact $\mathcal{K}_h\subset(0,\infty)$ and every $q>2$, the deterministic parts of both increments are $O(\tau)$ and their stochastic coefficients are $O(\sqrt{\tau})$ uniformly over $x\in\mathcal{K}_h$. Since both fixed marginal laws $D_1(0,1)$ and $D_2(0,1)$ have finite moments of all orders,
\begin{equation*}\lim_{\tau\to0}\sup_{x\in\mathcal{K}_h}\frac{1}{\tau}\E^{\Pm^{(n)}}\left[\left|\Delta Y^{(n)}_{(k+1)\tau}\right|^q+\left|\Delta h^{(n)}_{(k+1)\tau}\right|^q\mid h^{(n)}_{k\tau}=x\right]=0.\end{equation*}
Thus, the local moment condition in Supplementary Material~\ref{sm:nelson_local_conditions}, and hence the corresponding no-large-jumps condition, holds.

The limiting covariance matrix admits the factorization
\begin{equation*}\bm{\Sigma}_{\mathsf{M}}(x)=\begin{pmatrix}\sqrt{x}&0\\ \varrho\eta\sqrt{g_{\mathsf{M}}(x)}&\sqrt{v_{\mathsf{M}}(x)-\varrho^2\eta^2g_{\mathsf{M}}(x)}\end{pmatrix}, \qquad \bm{\Sigma}_{\mathsf{M}}(x)\bm{\Sigma}_{\mathsf{M}}(x)^\top=\bm{B}_{\mathsf{M}}(x).\end{equation*}
The coefficients are locally Lipschitz on every compact subset of $\mathbb{R}\times(0,\infty)$. For VPGARCH-SV, $g_{\mathsf{VP}}$ is nonnegative on the variance state space; any interior zero is a double root, so $\sqrt{g_{\mathsf{VP}}}$ remains locally Lipschitz. The same argument applies to the residual squared loading $v_{\mathsf{M}}-\varrho^2\eta^2g_{\mathsf{M}}$, which is a nonnegative polynomial of degree at most two, including the identically zero case. Thus, the localized limiting martingale problem is well posed.

The verified local moment conditions and well-posedness therefore yield Proposition~\ref{prop:p_weak_limit} by the localized diffusion-approximation theorem. Under the additional conditions stated after Proposition~\ref{prop:p_weak_limit}, strict positivity and nonexplosion of the limiting diffusion allow localization to be removed, yielding full-horizon weak convergence.
\end{proof}

\subsection{Proof of Proposition~\ref{prop:p_levy_limit}}
\label{sm:proof_p_levy_limit}

\bigskip

\begin{proof}
We establish discrete-time positivity and global existence of the limiting system, then prove weak convergence by coupling the discrete-time processes and their limit through the same L\'evy driver, adapting the stability approach of \citet[Chapter~V, Section~4]{Protter2005}. Localized mean-square estimates control the approximation error, while a Lyapunov bound controls the probability that the limiting variance leaves compact subsets of $(0,\infty)$ and allows localization to be removed. Sampling the interpolations at grid points then gives weak convergence in the Skorokhod $\mathrm{J}_1$ topology. Finally, we verify polynomial preservation and convergence of the moment matrices.

We construct the innovations for all $n$ from the same L\'evy process on a common physical probability space using Equation~\eqref{eq:levy_increments}. Throughout the proof, we set $q_\tau=1-(\kappa+b_1)\tau$, $b(v)=b_0+b_1v$, and
\begin{equation*}
g_{\mathsf{VP}}(\tau,v)=a_0\tau+a_1v+v^2,\qquad g_{\mathsf{JP}}(\tau,v)=v.
\end{equation*}
The limits are $g_{\mathsf{VP}}(v)=a_1v+v^2$ and $g_{\mathsf{JP}}(v)=v$; all expectations refer to the common physical measure.

\textit{Discrete-time positivity:} Conditional on the current variance $v>0$ and on $\Delta J_2=z$, the next variance is
\begin{equation*}
h^+=q_\tau v+(\kappa\sigma^2-b_0)\tau+\eta\sqrt{g_{\mathsf M}(\tau,v)}\,z+b(v)z^2.
\end{equation*}
For $0<\tau<1/[2(\kappa+b_1)]$ and $b(v)>0$, completing the square, using Condition~\eqref{eq:levy_global_conditions}, and, for VP, the bound on $a_0$, gives
\begin{equation*}
0<\left(q_\tau-\frac12\right)v\leq q_\tau v+(\kappa\sigma^2-b_0)\tau-\frac{\eta^2g_{\mathsf M}(\tau,v)}{4b(v)}\leq h^+.
\end{equation*}
If $b(v)=0$, then $b_0=b_1=\eta=0$ and the deterministic recursion is positive. Induction proves strict positivity.

\textit{Global existence:} Let $c_J=(\bm C_J)_{22}$ and let $\Pi_{J,2}$ be the marginal L\'evy measure of $J_2$, with mass at zero omitted. Standardization gives $\int z^2\Pi_{J,2}(dz)=1-c_J$. The process
\begin{equation*}
M_t:=\left[J_2\right]_t-t=\int_0^t\int_{\mathbb R^2}z_2^2\,\widetilde N_J(ds,d\bm z)
\end{equation*}
satisfies $\E\left[M_t^2\right]=t\mu_{0,4}$, where $\widetilde N_J$ is the compensated jump measure of $\bm J$. Hence $(J_1,J_2,M)$ is a square-integrable L\'evy martingale. The coefficients are locally Lipschitz on $\mathbb R\times(0,\infty)$ and bounded on compact variance intervals. Smooth cutoff extensions admit unique strong solutions by Picard iteration; pathwise uniqueness makes them agree until exit and defines the maximal local solution \citep[see, e.g.,][Chapter~V, Section~3]{Protter2005}. It remains to rule out explosion and approach to zero on $[0,L]$. By Condition~\eqref{eq:levy_global_conditions}, the post-jump variance at a jump of $J_2$ of size $z$ satisfies
\begin{equation*}
H(v,z)=v+\eta\sqrt{g_{\mathsf M}(v)}\,z+b(v)z^2\geq v/2,
\end{equation*}
so jumps cannot reach or cross zero.

Restricting Equation~\eqref{eq:levy_generator} to functions of the variance state and moving the quadratic jump compensator into the drift gives
\begin{equation*}
\begin{aligned}
\mathcal A_h f(v)={}&\left[\kappa\sigma^2-\kappa v-(1-c_J)b(v)\right]f'(v)+\frac{c_J\eta^2g_{\mathsf M}(v)}{2}f''(v)\\
&+\int_{\mathbb R}\left[f(H(v,z))-f(v)-\eta\sqrt{g_{\mathsf M}(v)}z f'(v)\right]\Pi_{J,2}(dz).
\end{aligned}
\end{equation*}

The marginal measure is symmetric, and
\begin{equation*}
\frac{H(v,z)H(v,-z)}{v^2}=1+\left[\frac{2b(v)}v-\frac{\eta^2g_{\mathsf M}(v)}{v^2}\right]z^2+\frac{b(v)^2}{v^2}z^4\geq1.
\end{equation*}
For $f=\log$, the compensated jump integrand is $\BigO{z^2}$ near zero and $\BigO{|z|}$ at infinity, hence $\Pi_{J,2}$-integrable, so it may be symmetrized: pairing $z$ and $-z$ makes the logarithmic jump integral nonnegative. Using $\eta^2g_{\mathsf M}(v)\leq2vb(v)$,
\begin{equation*}
-\kappa-b_1\leq\frac{\kappa\sigma^2-b_0}{v}-\kappa-b_1\leq\mathcal A_h\log v.
\end{equation*}
As $\mathcal A_h v=\kappa(\sigma^2-v)$, the function $V(v)=v-\log v\geq1$, which diverges as $v\downarrow0$ or $v\to\infty$, satisfies $\mathcal A_hV\leq\kappa\sigma^2+\kappa+b_1$. Stopping at the first exit $\zeta_N$ from $(1/N,N)$ and applying localized It\^o's formula gives
\begin{equation*}
\E[V(h_{t\wedge\zeta_N})]\leq V(h_0)+(\kappa\sigma^2+\kappa+b_1)t,\qquad t\leq L.
\end{equation*}
Outgoing jumps are integrable because $H(v,z)$ grows at most quadratically in $z$ and is bounded below by $v/2$. On $\{\zeta_N\leq L\}$, $V(h_{\zeta_N})\geq\min\{V(1/N),V(N)\}$. Consequently, for sufficiently large $N$,
\begin{equation*}
\Pm(\zeta_N\leq L)\leq\frac{V(h_0)+(\kappa\sigma^2+\kappa+b_1)L}{\min\{V(1/N),V(N)\}}\longrightarrow0.
\end{equation*}
Thus $h$ is global and bounded away from zero and infinity on $[0,L]$ almost surely, including through left limits. The first equation in~\eqref{eq:p_levy_limit_system} then defines the unique global process $Y$.

\textit{Convergence:} We first localize the GARCH--SV equations by modifying their coefficients outside a compact variance interval, so that standard mean-square estimates apply. Fix a compact interval $\mathcal K_h\subset(0,\infty)$ containing $h_0$ in its interior. Let $\chi:\mathbb R\to(0,\infty)$ be smooth and globally Lipschitz, equal to the identity on $\mathcal K_h$, with range in a larger compact subset of $(0,\infty)$. Replacing the variance argument in the drift and innovation loadings by $\chi$ gives bounded globally Lipschitz coefficients, unchanged on $\mathcal K_h$. The coefficients do not depend on the log-price coordinate. Denote the localized chain and limit by $\bm X^{(n),\chi}$ and $\bm X^\chi$, both starting from $(Y_0,h_0)^\top$.

We rewrite the localized GARCH--SV dynamics using Equation~\eqref{eq:levy_increments} and Assumption~\ref{ass:p_levy_scaling}, then interpolate each update using the same L\'evy driver as the limit. Let $t_i=i\tau$, $x_i=\chi(h^{(n),\chi}_{t_i})$, and $\Delta_iJ_{j,t}=J_{j,t}-J_{j,t_i}$, and define on $[t_i,t_{i+1}]$
\begin{align*}
\widehat Y_t^{(n),\chi}&=Y^{(n),\chi}_{t_i}+(r+\lambda x_i)(t-t_i)+\sqrt{x_i}\,\Delta_iJ_{1,t},\\
\widehat h_t^{(n),\chi}&=h^{(n),\chi}_{t_i}+\kappa(\sigma^2-x_i)(t-t_i)+\eta\sqrt{g_{\mathsf M}(\tau,x_i)}\,\Delta_iJ_{2,t}+b(x_i)\left[(\Delta_iJ_{2,t})^2-(t-t_i)\right].
\end{align*}
This agrees with the localized equation at grid points and holds coefficients fixed within each interval. We use It\^o's formula to separate the quadratic-variation term from the approximation remainder:
\begin{equation*}
(\Delta_iJ_{2,t})^2=[J_2]_t-[J_2]_{t_i}+2\int_{t_i}^t(J_{2,s-}-J_{2,t_i})\,dJ_{2,s}.
\end{equation*}
Relative to the equation driven by $(J_2,M)$ with coefficients fixed at $x_i$, the interpolation has martingale remainder
\begin{equation*}
R_t^{(n),\chi}=2\sum_{i:t_i<t}\int_{t_i}^{t\wedge t_{i+1}}b(x_i)(J_{2,s-}-J_{2,t_i})\,dJ_{2,s}.
\end{equation*}
Since $b(x_i)$ is bounded, martingale isometry and Doob's inequality yield
\begin{equation*}
\E\left[\sup_{0\leq t\leq L}|R_t^{(n),\chi}|^2\right]\leq C\sum_{i:t_i<L}\int_{t_i}^{t_{i+1}}(s-t_i)\,ds\leq C_L\tau.
\end{equation*}
Constants may depend on $\chi$ and $L$, but not on $n$. To control variation within each time step, note that
\begin{equation*}
\E\left[J_{2,u}^4\right]=\mu_{0,4}u+3u^2,\qquad \E\left[(J_{2,u}^2-u)^2\right]=\mu_{0,4}u+2u^2.
\end{equation*}
Boundedness of the coefficients then gives
\begin{equation*}
\sup_{0\leq s\leq L}\E\left[\left\|\widehat{\bm X}^{(n),\chi}_s-\widehat{\bm X}^{(n),\chi}_{\pi_\tau(s)}\right\|^2\right]\leq C_L\tau,\qquad \pi_\tau(s)=\tau\lfloor s/\tau\rfloor.
\end{equation*}
We now compare the interpolation with the localized limit. Since $g_{\mathsf M}$ is bounded away from zero on the range of $\chi$, $\eta\sqrt{g_{\mathsf M}(\tau,v)}-\eta\sqrt{g_{\mathsf M}(v)}=\BigO{\tau}$ uniformly for $v\in\chi(\mathbb R)$. Lipschitz bounds, isometry and Doob's inequality applied to the integral equations yield
\begin{equation*}
\E\left[\sup_{0\leq u\leq t}\left\|\widehat{\bm X}^{(n),\chi}_u-\bm X_u^\chi\right\|^2\right]\leq C_L\tau+C_L\int_0^t\E\left[\sup_{0\leq u\leq s}\left\|\widehat{\bm X}^{(n),\chi}_u-\bm X_u^\chi\right\|^2\right]\,ds.
\end{equation*}
The first term bounds the remainder, within-step variation and coefficient perturbation. Gronwall's inequality gives
\begin{equation*}
\E\left[\sup_{0\leq t\leq L}\left\|\widehat{\bm X}^{(n),\chi}_t-\bm X_t^\chi\right\|^2\right]\leq C_L\tau.
\end{equation*}

To return to the original dynamics, we enlarge the variance interval and use the boundary estimate proved above. Take $\mathcal K_{h,N}=[1/N,N]$, with a corresponding cutoff $\chi_N$, and let $A_N=\{h_t\in[2/N,N/2]\text{ for all }t\in[0,L]\}$. Then $\Pm(A_N^c)\to0$, and on $A_N$ pathwise uniqueness gives $\bm X^{\chi_N}=\bm X$. On this event, a uniform error below $1/(2N)$ between the localized interpolation and its localized limit keeps the grid variances of the localized chain in $\mathcal K_{h,N}$, so the original and localized recursions and their interpolations agree by induction. For the original interpolation $\widehat{\bm X}^{(n)}$, Markov's inequality gives
\begin{equation*}
\Pm\left(\sup_{0\leq t\leq L}\left\|\widehat{\bm X}^{(n)}_t-\bm X_t\right\|>\varepsilon\right)\leq\Pm(A_N^c)+\frac{C_{N,L}\tau}{\min\{\varepsilon,1/(2N)\}^2}.
\end{equation*}
First letting $\tau\to0$ and then $N\to\infty$ proves uniform convergence in probability of the interpolations. No global mean-square rate is asserted.

Finally, sampling the interpolations at grid points recovers the original step processes. Set $(S_\tau\bm z)(t)=\bm z(\pi_\tau(t))$, retaining the terminal value. For a c\`adl\`ag path continuous at $L$, $S_\tau\bm z\to\bm z$ in vector $\mathrm{J}_1$: a common time change aligns the finitely many jumps exceeding a fixed tolerance, while c\`adl\`ag regularity controls the remaining oscillations. The limit has no jump at deterministic time $L$. Since $\bm X^{(n)}=S_\tau\widehat{\bm X}^{(n)}$ and
\begin{equation*}
\sup_{0\leq t\leq L}\left\|(S_\tau\widehat{\bm X}^{(n)})(t)-(S_\tau\bm X)(t)\right\|\leq\sup_{0\leq t\leq L}\left\|\widehat{\bm X}^{(n)}_t-\bm X_t\right\|,
\end{equation*}
the $\mathrm{J}_1$ triangle inequality yields convergence in probability, hence weak convergence and tightness in $D([0,L],\mathbb R^2)$.

\textit{Polynomial preservation and moments:} We verify the degree bounds for the generator in Equation~\eqref{eq:levy_generator}, writing $\bm{x}=(u,v)^\top$ as in Section~\ref{subsec:polynomial_levy_limits}.
The jump kernel is the pushforward of $\Pi_J$ by $\bm D_{\mathsf M}(v,\cdot)$, omitting mass at zero. Jump moments of order at least two are finite. Write $\mu_{p,q}=\int z_1^pz_2^q\Pi_J(d\bm z)$ for $p+q\geq2$; symmetry makes odd-order moments zero. For integers $p,q\geq0$ with $p+q\geq2$,
\begin{equation*}
\int_{\mathbb R^2}D_{\mathsf M,1}(v,\bm z)^pD_{\mathsf M,2}(v,\bm z)^q\Pi_J(d\bm z)=v^{p/2}\sum_{j=0}^q\binom qj\eta^j g_{\mathsf M}(v)^{j/2}b(v)^{q-j}\mu_{p,2q-j}.
\end{equation*}
Only terms with even $p+j$ survive. For JP, each surviving term is proportional to $v^{(p+j)/2}b(v)^{q-j}$ and has degree at most $p+q$. For VP variance moments, $p=0$, so $j$ is even and the degree of $g_{\mathsf{VP}}(v)^{j/2}b(v)^{q-j}$ is at most $q$. The continuous covariance has the corresponding degree bounds. Combining the Brownian and jump contributions, the total instantaneous covariance is
\begin{equation*}
\bm C_{\mathsf M}(v)+\int\bm D_{\mathsf M}(v,\bm z)\bm D_{\mathsf M}(v,\bm z)^\top\Pi_J(d\bm z)=\begin{pmatrix}v&\varrho\eta\sqrt{vg_{\mathsf M}(v)}\\\varrho\eta\sqrt{vg_{\mathsf M}(v)}&\eta^2g_{\mathsf M}(v)+\mu_{0,4}b(v)^2\end{pmatrix}.
\end{equation*}
These satisfy the polynomial generator conditions of \citet[Lemma~1]{FilipovicLarsson2020}. When $J_1$ and $J_2$ are independent, the continuous cross-covariance vanishes and $\Pi_J$ is supported on the coordinate axes, so all mixed jump moments vanish, giving joint preservation for the VP specification. Zero correlation alone need not eliminate higher mixed jump moments. Discrete-time preservation follows from Proposition~\ref{prop:garch_polynomial_characterization} and Corollary~\ref{cor:polynomial_temporal_aggregates} for each sufficiently large $n$.

By \citet[Theorem~1]{FilipovicLarsson2020}, the corresponding global polynomial process has finite conditional moments of every order and satisfies
\begin{equation*}
\E^{\Pm}\left[p(\bm X_t)\mid\F_s\right]=\bm{\HT}_m(\bm X_s)^\top e^{(t-s)\bm G_m}\bm p,\qquad 0\leq s\leq t\leq L,
\end{equation*}
with the variance state substituted for $\bm X$ when only marginal preservation is used.

Finally, we compare the discrete and continuous moment operators. The moment--cumulant formula expresses each mixed L\'evy increment moment as a finite polynomial in $\tau$, since increment cumulants are $\tau$ times the unit-time cumulants. The latter vanish at first order, equal $1$ for the marginal second-order cumulants and $\varrho$ for the mixed second-order cumulant, and equal the corresponding jump moments at higher orders. Expanding the transition of each basis polynomial gives, coefficientwise,
\begin{equation*}
\trans^{(n)}p=p+\tau\mathcal Ap+\BigO{\tau^2}.
\end{equation*}
The centring $-\tau b(v)$ cancels $b(v)\E\left[(\Delta J_2)^2\right]=\tau b(v)$. For VP variance moments, symmetry leaves only even powers of the linear variance loading. Under independent drivers, the same cancellation holds for joint VP moments. Thus only integer powers of $g_{\mathsf{VP}}(\tau,v)$ remain, and $a_0\tau$ affects the transition coefficients only at order $\tau^2$.  This gives the matrix expansion in Equation~\eqref{eq:levy_moment_matrix_limit}. As in Corollary~\ref{cor:moment_matrix_convergence}, this gives a matrix-power error of $\BigO{\tau}$ uniformly on $[0,L]$. The conditional-moment formulas then imply moment convergence uniformly in time and over compact sets of initial states. These formulas concern the global process.
\end{proof}

\section{Risk-Neutral Dynamics and Polynomial Preservation}
\label{sec:risk_neutral_dynamics}

This appendix studies risk neutralization of the polynomial-preserving hybrid GARCH--SV models introduced in Section~\ref{subsec:polynomial_garch_models_p}. We first derive their dynamics under a variance-dependent pricing kernel for the general joint innovation law. We then examine, under jointly Gaussian innovations, whether VPGARCH-SV remains polynomial preserving in the variance state and whether JPGARCH-SV remains jointly polynomial preserving in returns and variance.

\subsection{Pricing kernel and risk-neutral dynamics}
\label{subsec:pricing_kernel_q_dynamics}

Consider the exponential-linear variance-dependent pricing kernel
\begin{equation}N^{(n)}_{(k+1)\tau}=\exp\left(\xi_{0,k\tau}(\tau)+\xi_{1,k\tau}(\tau)y^{(n)}_{(k+1)\tau}+\xi_{2,k\tau}(\tau)h^{(n)}_{(k+1)\tau}\right).\label{eq:pricing_kernel}\end{equation}
The coefficients $\xi_{j,k\tau}(\tau)$, $j=0,1,2$, are finite, real-valued, and $\F^{(n)}_{k\tau}$-measurable, with $\xi_{1,k\tau}(\tau)$ and $\xi_{2,k\tau}(\tau)$ being the equity and variance risk-preference parameters, respectively. Such kernels are widely used in discrete-time option pricing; see \citet{christoffersen2013capturing,corsi13,majew15,bades17,babaouglu2018option}.

For this kernel to price the risk-free and risky assets, we impose
\begin{equation}\E^{\Pm^{(n)}}\left[N^{(n)}_{(k+1)\tau}\mid\F^{(n)}_{k\tau}\right]=\exp\left(-r\tau\right), \qquad \E^{\Pm^{(n)}}\left[N^{(n)}_{(k+1)\tau}\exp\left(y^{(n)}_{(k+1)\tau}\right)\mid\F^{(n)}_{k\tau}\right]=1.\label{eq:pricing_kernel_no_arbitrage}\end{equation}
Let $\mathcal{C}^{\Pm^{(n)}}_{y^{(n)}_{(k+1)\tau},h^{(n)}_{(k+1)\tau}}$ denote the corresponding $\F^{(n)}_{k\tau}$-conditional joint cumulant generating function (cgf). The first restriction determines
\begin{equation*}\xi_{0,k\tau}(\tau)=-r\tau-\mathcal{C}^{\Pm^{(n)}}_{y^{(n)}_{(k+1)\tau},h^{(n)}_{(k+1)\tau}}\left(\xi_{1,k\tau}(\tau),\xi_{2,k\tau}(\tau)\right).\end{equation*}
Let $\mathcal{C}^{\Pm^{(n)}}$ denote the joint cgf of $\epsilon^{(n)}_{1,(k+1)\tau}$, $\epsilon^{(n)}_{2,(k+1)\tau}$, and $\left(\epsilon^{(n)}_{2,(k+1)\tau}\right)^2$. Since the innovation pairs are i.i.d., the same cgf applies conditionally on $\F^{(n)}_{k\tau}$.

\begin{samepage}
Using Equations~\eqref{eq:general_return_dynamics_p}--\eqref{eq:general_variance_dynamics_p}, the normalized kernel can be written in innovation form as
\begin{equation*}\log N^{(n)}_{(k+1)\tau}=-r\tau+\sum_{i=1}^2A_{i,k\tau}\left(\tau\right)\epsilon^{(n)}_{i,(k+1)\tau}+B_{k\tau}\left(\tau\right)\left(\epsilon^{(n)}_{2,(k+1)\tau}\right)^2-\mathcal{C}^{\Pm^{(n)}}\left(A_{1,k\tau}\left(\tau\right),A_{2,k\tau}\left(\tau\right),B_{k\tau}\left(\tau\right)\right),\end{equation*}
where
\begin{align*}A_{1,k\tau}\left(\tau\right)=\xi_{1,k\tau}\left(\tau\right)\sqrt{\tau h^{(n)}_{k\tau}},\quad A_{2,k\tau}\left(\tau\right)=\xi_{2,k\tau}\left(\tau\right)\psi_2\left(h^{(n)}_{k\tau};\tau\right),\quad B_{k\tau}\left(\tau\right)&=\xi_{2,k\tau}\left(\tau\right)\psi_3\left(h^{(n)}_{k\tau};\tau\right).\end{align*}
\end{samepage}
The exponential-moment condition in Proposition~\ref{prop:general_q_dynamics} is imposed throughout and is additional to the polynomial-moment assumptions of Section~\ref{subsec:polynomial_garch_models_p}. Under Equation~\eqref{eq:common_mean_parametrizations}, the risky-asset pricing restriction becomes
\begin{equation*}\mathcal{C}^{\Pm^{(n)}}\left(A_{1,k\tau}\left(\tau\right)+\sqrt{\tau h^{(n)}_{k\tau}},A_{2,k\tau}\left(\tau\right),B_{k\tau}\left(\tau\right)\right)-\mathcal{C}^{\Pm^{(n)}}\left(A_{1,k\tau}\left(\tau\right),A_{2,k\tau}\left(\tau\right),B_{k\tau}\left(\tau\right)\right)=-\lambda\left(\tau\right)h^{(n)}_{k\tau}.\end{equation*}
For $h^{(n)}_{k\tau}>0$, this restriction implicitly determines $\xi_{1,k\tau}(\tau)$ from a chosen $\xi_{2,k\tau}(\tau)$ whenever an admissible solution exists. Setting $\xi_{2,k\tau}(\tau)=0$ gives a pricing kernel that is monotonic in returns.

When $\varrho=1$, the log kernel at a positive variance state is quadratic in the return, with quadratic coefficient $B_{k\tau}\left(\tau\right)/\left(\tau h^{(n)}_{k\tau}\right)$, so an admissible $B_{k\tau}\left(\tau\right)>0$ gives a U-shaped kernel. In the hybrid setting, the kernel generally also depends on variance uncertainty not determined by the return. Under the positivity restrictions in Section~\ref{subsec:polynomial_garch_models_p}, $\psi_3\left(h^{(n)}_{k\tau};\tau\right)>0$ at positive states, so $B_{k\tau}\left(\tau\right)$ has the sign of $\xi_{2,k\tau}\left(\tau\right)$.

The pricing kernel induces the equivalent risk-neutral measure $\Qm^{(n)}$ through
\begin{equation*}\left.\frac{d\Qm^{(n)}}{d\Pm^{(n)}}\right|_{\F^{(n)}_{k\tau}}=\exp\left(rk\tau\right)\prod_{j=0}^{k-1}N^{(n)}_{(j+1)\tau}, \qquad k=1,\ldots,nL.\end{equation*}
For each $i=1,2$, let $P_{i,k\tau}\left(\tau\right)$ and $R_{i,k\tau}\left(\tau\right)$ denote the conditional mean and variance of $\epsilon^{(n)}_{i,(k+1)\tau}$ under $\Qm^{(n)}$. They are given by
\begin{equation}\begin{aligned}P_{i,k\tau}\left(\tau\right)&=\partial_i\mathcal{C}^{\Pm^{(n)}}\left(A_{1,k\tau}\left(\tau\right),A_{2,k\tau}\left(\tau\right),B_{k\tau}\left(\tau\right)\right),\\ R_{i,k\tau}\left(\tau\right)&=\partial_{ii}\mathcal{C}^{\Pm^{(n)}}\left(A_{1,k\tau}\left(\tau\right),A_{2,k\tau}\left(\tau\right),B_{k\tau}\left(\tau\right)\right).\end{aligned}\label{eq:general_variance_risk_ratio}\end{equation}
Here, $\partial_i$ and $\partial_{ii}$ denote the first and second derivatives with respect to the $i$th argument. Since both physical innovations have unit variance, $R_{i,k\tau}\left(\tau\right)$ also measures the change in their conditional variances. For each $i=1,2$ with $R_{i,k\tau}\left(\tau\right)>0$, set
\begin{equation*}\epsilon^{*\,(n)}_{i,(k+1)\tau}=\frac{\epsilon^{(n)}_{i,(k+1)\tau}-P_{i,k\tau}\left(\tau\right)}{\sqrt{R_{i,k\tau}\left(\tau\right)}}, \qquad \epsilon^{*\,(n)}_{i,(k+1)\tau}\mid\F^{(n)}_{k\tau}\overset{\Qm^{(n)}}{\sim}D_i^*\left(0,1\right),\end{equation*}
where $D_i^*\left(0,1\right)$ denotes the standardized conditional marginal law, with its date and history dependence suppressed. It need not coincide with the time-invariant physical law $D_i\left(0,1\right)$ or belong to the same parametric family; when parametrized, its shape parameters may depend on the current state and risk-preference coefficients and hence vary over time even after standardization. The joint cgf of the standardized innovations is given in Supplementary Material~\ref{app:proof_general_q_dynamics}.

\begin{proposition}
\label{prop:general_q_dynamics}
Suppose that the physical dynamics are given by Equations~\eqref{eq:general_return_dynamics_p}--\eqref{eq:general_variance_dynamics_p} and the pricing kernel in Equation~\eqref{eq:pricing_kernel} satisfies Equation~\eqref{eq:pricing_kernel_no_arbitrage}. For each $k=0,\ldots,nL-1$, assume that $\mathcal{C}^{\Pm^{(n)}}\left(z_1,z_2,B_{k\tau}\left(\tau\right)\right)$ is almost surely finite for $\left(z_1,z_2\right)$ in a neighbourhood of $\left(A_{1,k\tau}\left(\tau\right),A_{2,k\tau}\left(\tau\right)\right)$ and at $\left(A_{1,k\tau}\left(\tau\right)+\sqrt{\tau h^{(n)}_{k\tau}},A_{2,k\tau}\left(\tau\right)\right)$, with $R_{i,k\tau}\left(\tau\right)>0$ for $i=1,2$. Then, under $\Qm^{(n)}$,
\begin{align}y^{(n)}_{(k+1)\tau}&=\psi_0^*\left(h^{(n)}_{k\tau};\tau\right)+\sqrt{\tau h^{(n)}_{k\tau}R_{1,k\tau}\left(\tau\right)}\,\epsilon^{*\,(n)}_{1,(k+1)\tau},\label{eq:general_return_dynamics_q}\\ h^{(n)}_{(k+1)\tau}&=\psi_1^*\left(h^{(n)}_{k\tau};\tau\right)+\psi_2^*\left(h^{(n)}_{k\tau};\tau\right)\epsilon^{*\,(n)}_{2,(k+1)\tau}+\psi_3^*\left(h^{(n)}_{k\tau};\tau\right)\left[\left(\epsilon^{*\,(n)}_{2,(k+1)\tau}\right)^2-1\right],\label{eq:general_variance_dynamics_q}\end{align}
where date and history dependence of the $\F^{(n)}_{k\tau}$-measurable coefficients is suppressed:\footnote{Note that writing $\mathcal{C}^{\Qm^{(n)}}_{\epsilon^{*\,(n)}_{1,(k+1)\tau}}$ for the conditional cgf of the standardized return innovation, the martingale restriction also gives
\begin{equation*}\psi_0^*\left(h^{(n)}_{k\tau};\tau\right)=r\tau-\mathcal{C}^{\Qm^{(n)}}_{\epsilon^{*\,(n)}_{1,(k+1)\tau}}\left(\sqrt{\tau h^{(n)}_{k\tau}R_{1,k\tau}\left(\tau\right)}\right).\end{equation*}
}
\begin{align*}\psi_0^*\left(h^{(n)}_{k\tau};\tau\right)&=\psi_0\left(h^{(n)}_{k\tau};\tau\right)+\sqrt{\tau h^{(n)}_{k\tau}}\,P_{1,k\tau}\left(\tau\right),\\ \psi_1^*\left(h^{(n)}_{k\tau};\tau\right)&=\psi_1\left(h^{(n)}_{k\tau};\tau\right)+\psi_2\left(h^{(n)}_{k\tau};\tau\right)P_{2,k\tau}\left(\tau\right)+\psi_3\left(h^{(n)}_{k\tau};\tau\right)\left[P_{2,k\tau}\left(\tau\right)^2+R_{2,k\tau}\left(\tau\right)-1\right],\\ \psi_2^*\left(h^{(n)}_{k\tau};\tau\right)&=\sqrt{R_{2,k\tau}\left(\tau\right)}\left[\psi_2\left(h^{(n)}_{k\tau};\tau\right)+2\psi_3\left(h^{(n)}_{k\tau};\tau\right)P_{2,k\tau}\left(\tau\right)\right],\\ \psi_3^*\left(h^{(n)}_{k\tau};\tau\right)&=R_{2,k\tau}\left(\tau\right)\psi_3\left(h^{(n)}_{k\tau};\tau\right).\end{align*}
\end{proposition}

\begin{proof}
See Supplementary Material~\ref{app:proof_general_q_dynamics}.
\end{proof}

Since $\Qm^{(n)}$ is equivalent to $\Pm^{(n)}$ on the finite trading horizon, the variance state $h^{(n)}$ remains nonnegative under $\Qm^{(n)}$. The conditional variance rate of returns under $\Qm^{(n)}$, however, is
\begin{equation*}h^{*\,(n)}_{k\tau}=\frac{1}{\tau}\Var^{\Qm^{(n)}}\left[y^{(n)}_{(k+1)\tau}\mid\F^{(n)}_{k\tau}\right]=h^{(n)}_{k\tau}R_{1,k\tau}\left(\tau\right).\end{equation*}
Thus, $R_{1,k\tau}\left(\tau\right)$ is the variance risk ratio, while $R_{2,k\tau}\left(\tau\right)$ rescales the variance innovation. In general, these quantities may not coincide, and $R_{1,k\tau}\left(\tau\right)>1$ may not correspond to $B_{k\tau}\left(\tau\right)>0$.

\begin{samepage}
\begin{remark}[Jointly Gaussian innovations]
\label{rem:jointly_gaussian_q_innovations}
\normalfont
Suppose that the physical innovation pair is jointly Gaussian. For $B_{k\tau}\left(\tau\right)<1/2$, the tilted pair is conditionally Gaussian, with
\begin{align*}P_{1,k\tau}\left(\tau\right)&=\left(1-\varrho^2\right)A_{1,k\tau}\left(\tau\right)+\varrho P_{2,k\tau}\left(\tau\right), & P_{2,k\tau}\left(\tau\right)&=\frac{A_{2,k\tau}\left(\tau\right)+\varrho A_{1,k\tau}\left(\tau\right)}{1-2B_{k\tau}\left(\tau\right)},\\ R_{1,k\tau}\left(\tau\right)&=1-\varrho^2+\varrho^2R_{2,k\tau}\left(\tau\right), & R_{2,k\tau}\left(\tau\right)&=\frac{1}{1-2B_{k\tau}\left(\tau\right)}.\end{align*}
The standardized marginals satisfy $D_i^*\left(0,1\right)=D_i\left(0,1\right)=N\left(0,1\right)$, whereas their conditional correlation is
\begin{equation*}\varrho^*_{k\tau}\left(\tau\right):=\Corr^{\Qm^{(n)}}\left[\epsilon^{*\,(n)}_{1,(k+1)\tau},\epsilon^{*\,(n)}_{2,(k+1)\tau}\mid\F^{(n)}_{k\tau}\right]=\varrho\sqrt{\frac{R_{2,k\tau}\left(\tau\right)}{R_{1,k\tau}\left(\tau\right)}},\end{equation*}
which can depend on the date and history through $B_{k\tau}\left(\tau\right)$.
\end{remark}
\end{samepage}

General adapted risk-preference parameters need not preserve the time-homogeneous Markov structure. Moreover, outside the jointly Gaussian case, the standardized conditional innovation law need not be jointly symmetric, and its higher moments may depend on the state and history. Proposition~\ref{prop:garch_polynomial_characterization} therefore cannot generally be applied under $\Qm^{(n)}$. The next subsection imposes state-dependent risk-preference coefficients and jointly Gaussian innovations to characterize when polynomial preservation survives.

\subsection{Polynomial preservation under the risk-neutral measure}
\label{subsec:q_polynomial_preservation}

Throughout this subsection, suppose that the physical innovation pair is jointly Gaussian with correlation $\varrho\in\left[0,1\right]$, that $h_0>0$, and that the variance dynamics preserve $\Rplus$, as required in Section~\ref{subsec:polynomial_garch_models_p}. Assume $b_0\left(\tau\right)\geq0$ and $b_1\left(\tau\right)>0$ for VPGARCH-SV, and $b_0\left(\tau\right),b_1\left(\tau\right)\geq0$ with $b_0\left(\tau\right)+b_1\left(\tau\right)>0$ for JPGARCH-SV. The variance then remains strictly positive almost surely at every trading date. Polynomial preservation below is understood on $\left(0,\infty\right)$ for the variance state and on $\mathbb{R}\times\left(0,\infty\right)$ for the joint return--variance state.

To preserve time-homogeneous Markov dynamics, let $\xi_{2,k\tau}\left(\tau\right)=\xi_2\left(h^{(n)}_{k\tau};\tau\right)$ and write $P_i\left(x;\tau\right)$ and $R_i\left(x;\tau\right)$ for the corresponding state functions from Appendix~\ref{subsec:pricing_kernel_q_dynamics}. The loading $\psi_3\left(x;\tau\right)$ is strictly positive for every $x>0$ under the maintained restrictions. For a scalar $\xi_2\left(\tau\right)<1/2$, define
\begin{equation}\label{eq:variance_risk_ratio}R_2\left(\tau\right)=\frac{1}{1-2\xi_2\left(\tau\right)}, \qquad R_1\left(\tau\right)=1-\varrho^2+\varrho^2R_2\left(\tau\right).\end{equation}
The following proposition identifies the risk-preference parameters compatible with polynomial preservation when the innovations are correlated.

\begin{proposition}
\label{prop:q_polynomial_preservation}
Under the assumptions above, suppose that $\varrho>0$, that the pricing kernel in Equation~\eqref{eq:pricing_kernel} satisfies the no-arbitrage conditions in Equation~\eqref{eq:pricing_kernel_no_arbitrage}, that $2\xi_2\left(x;\tau\right)\psi_3\left(x;\tau\right)<1$ for every $x>0$, and that $R_2\left(x;\tau\right)$ admits a finite limit as $x\downarrow0$. In both model classes, the corresponding polynomial-preserving property is preserved under $\Qm^{(n)}$ if and only if there exists $\xi_2\left(\tau\right)<1/2$ such that
\begin{equation*}\xi_2\left(x;\tau\right)=\frac{\xi_2\left(\tau\right)}{\psi_3\left(x;\tau\right)}, \qquad x>0,\end{equation*}
together with the following model-specific conditions:
\begin{enumerate}[label=\textnormal{(\roman*)}]
\item For VPGARCH-SV,
\begin{equation*}\lambda\left(\tau\right)=-\frac{\tau}{2}R_1\left(\tau\right),\end{equation*}
and
\begin{equation*}\eta\left(\tau\right)\left(1-\varrho^2\right)\xi_2\left(\tau\right)\left[b_1\left(\tau\right)^2a_0\left(\tau\right)-b_0\left(\tau\right)b_1\left(\tau\right)a_1\left(\tau\right)+b_0\left(\tau\right)^2\right]=0.\end{equation*}
\item For JPGARCH-SV,
\begin{equation*}b_1\left(\tau\right)\left[\lambda\left(\tau\right)+\frac{\tau}{2}R_1\left(\tau\right)\right]=0, \qquad b_0\left(\tau\right)b_1\left(\tau\right)\left(1-\varrho^2\right)\eta\left(\tau\right)\xi_2\left(\tau\right)=0.\end{equation*}
\end{enumerate}
\end{proposition}

\begin{proof}
See Supplementary Material~\ref{app:proof_q_polynomial_preservation}.
\end{proof}

The state dependence of $\xi_2\left(x;\tau\right)$ must therefore offset that of the centred-quadratic loading, giving the constant innovation variances in Equation~\eqref{eq:variance_risk_ratio} and the constant risk-neutral correlation $\varrho^*\left(\tau\right)=\varrho\sqrt{R_2\left(\tau\right)/R_1\left(\tau\right)}$. When $\varrho=1$, the additional product restrictions vanish and the proposition recovers the Gaussian GARCH characterization. For $0<\varrho<1$, nonzero leverage and a nonzero variance risk-preference parameter impose further restrictions: in VPGARCH-SV, $b_0\left(\tau\right)+b_1\left(\tau\right)x$ must divide $a_0\left(\tau\right)+a_1\left(\tau\right)x+x^2$; in JPGARCH-SV, either $b_0\left(\tau\right)=0$ or $b_1\left(\tau\right)=0$. The HNGARCH-SV case $b_1\left(\tau\right)=0$ therefore retains joint polynomial preservation without an additional restriction on $\lambda\left(\tau\right)$. When $\varrho=0$, the innovations remain independent under the kernel, and $R_1\left(x;\tau\right)=1$. If $\eta\left(\tau\right)\neq0$, the same characterization holds with the restrictions involving $\lambda\left(\tau\right)$ omitted. If $\eta\left(\tau\right)=0$, the corresponding polynomial property holds if and only if $\psi_3\left(x;\tau\right)R_2\left(x;\tau\right)$ is affine; the variance innovation ratio need not be constant. These statements retain the admissibility and finite-limit assumptions of the proposition.

For VPGARCH-SV with $\varrho>0$, $\lambda\left(\tau\right)=-\tau R_1\left(\tau\right)/2$ implies $P_1\left(x;\tau\right)=0$: the change of measure leaves the conditional log-return mean unchanged. Under $\Pm^{(n)}$, the conditional expected gross return is then $\exp\left(r\tau+\tau h^{(n)}_{k\tau}\left[1-R_1\left(\tau\right)\right]/2\right)$. Its excess over the risk-free gross return is negative, zero, or positive according as $R_1\left(\tau\right)$ is greater than, equal to, or less than one. Thus, a positive variance risk-preference parameter is incompatible with a positive expected excess simple return under these simultaneous polynomial-preservation restrictions. In the GARCH case $\varrho=1$, this is also the incompatibility with a U-shaped kernel.%
The same conclusion applies to JPGARCH-SV when $\varrho>0$ and $b_1\left(\tau\right)>0$. These conclusions include the zero-leverage case $\eta\left(\tau\right)=0$.

Since $h^{*\,(n)}_{k\tau}=R_1\left(\tau\right)h^{(n)}_{k\tau}$ under Proposition~\ref{prop:q_polynomial_preservation}, the corresponding polynomial property also holds in the risk-neutral variance state. Supplementary Material~\ref{sm:q_preservation_hstar} extends the analysis to affine relations between the physical and risk-neutral variance states for $\varrho\in\left[0,1\right]$, allowing the innovation variance ratio to diverge near zero, and gives the additional restrictions and their proof.

These results concern  polynomial preservation under both physical and risk-neutral measures for the pricing kernel in Equation~\eqref{eq:pricing_kernel}.  Alternative martingale measures, or models specified directly under $\Qm^{(n)}$ without requiring physical-measure polynomiality, require a separate analysis.

\section{Parametrization of the NIG distribution}\label{app:nig}

The NIG innovations used in Section~\ref{sec:empirical_analysis} are normal variance mixtures,
\begin{equation}\label{eq:nig_mixture}
	\epsilon_t=\sqrt{V_t}\,Z_t,\qquad Z_t\sim N(0,1),\qquad V_t\sim\mathrm{IG}(1,\alpha),
\end{equation}
where $V_t$ is independent of $Z_t$ and follows an inverse Gaussian distribution with mean one and shape parameter $\alpha>0$, with density
\begin{equation*}
	f_V(v)=\sqrt{\frac{\alpha}{2\pi v^3}}\exp\left(-\frac{\alpha(v-1)^2}{2v}\right),\qquad v>0.
\end{equation*}
The density of $\epsilon_t$ is
\begin{equation*}
	f_\epsilon(x)=\frac{\alpha\,e^{\alpha}}{\pi}\,\frac{K_1\left(\sqrt{\alpha\left(\alpha+x^2\right)}\right)}{\sqrt{\alpha+x^2}},
\end{equation*}
where $K_1$ is the modified Bessel function of the third kind with index one. In the notation of \citet{badescu2011comparison}, this is the generalized hyperbolic distribution $\mathrm{GH}(-\frac{1}{2},\sqrt{\alpha},0,\sqrt{\alpha},0)$. Since $\E[V_t]=1$ and $\Var[V_t]=1/\alpha$, the first four moments are
\begin{equation*}
	\E[\epsilon_t]=0,\qquad \E[\epsilon_t^2]=1,\qquad \E[\epsilon_t^3]=0,\qquad \E[\epsilon_t^4]=m_4=3+\frac{3}{\alpha}.
\end{equation*}
The distribution is therefore symmetric, with kurtosis $3+3/\alpha$, and converges to the standard normal as $\alpha\to\infty$. All its moments are finite. Its cgf is
\begin{equation*}\mathcal{C}_\epsilon(s)=\log\E[e^{s\epsilon_t}]=\alpha-\sqrt{\alpha^2-\alpha s^2},\qquad |s|\leq\sqrt{\alpha},\end{equation*}
and the corresponding exponential moment is infinite outside this interval. With $\xi_{2,t}=0$, the pricing restrictions in Appendix~\ref{subsec:pricing_kernel_q_dynamics} require both $\xi_{1,t}\sqrt{\tau h_t}$ and $(\xi_{1,t}+1)\sqrt{\tau h_t}$ to belong to this domain. Hence $\tau h_t\leq4\alpha$ is necessary, with strict inequality under the local-finiteness assumption of Proposition~\ref{prop:general_q_dynamics}. For fixed finite $\alpha$, the variance equations can reach states violating this bound, so the kernel with $\xi_{2,t}=0$ is not globally admissible. In our empirical implementation, however, the bound is satisfied at every simulated state in the calibrations.

\clearpage
\begingroup
\newif\ifFirstAppendixTable
\FirstAppendixTabletrue
\let\OriginalSidewaystable\sidewaystable
\renewcommand{\sidewaystable}[1][p]{%
  \OriginalSidewaystable[#1]
  \ifFirstAppendixTable
    \global\FirstAppendixTablefalse
    \titlespacing*{\section}{0pt}{0pt}{3pt}%
    \section{Additional tables}\label{app:additional_results}
    This appendix collects additional tables referred to in Section~\ref{sec:empirical_analysis}.
    \par
  \fi
}
\begin{sidewaystable}
	\caption{\textbf{Implied Volatility Root Mean Squared Errors by Moneyness and Maturity under Gaussian Innovations}\label{tab:ivrmse_moneyness_maturity_gaussian}}
	\scriptsize
	\renewcommand{\arraystretch}{1.25}
	\newcommand{\mc}[1]{\multicolumn{1}{c}{#1}}
	\newcommand{\ov}[1]{\multicolumn{1}{c}{\textit{#1}}}
	\centering
	\begin{tabular*}{\linewidth}{@{\extracolsep{\fill}} cc*{9}{d{1.2}}}
		\toprule
		Maturity & Model & \multicolumn{9}{c}{Moneyness} \\
		\cmidrule(l){1-1}\cmidrule(l){2-2}\cmidrule(l){3-11}
		& & \mc{$[0,0.900)$} & \mc{$[0.900,0.950)$} & \mc{$[0.950,0.975)$} & \mc{$[0.975,1.000)$} & \mc{$[1.000,1.025)$} & \mc{$[1.025,1.050)$} & \mc{$[1.050,1.100)$} & \mc{$[1.100,\infty)$} & \mc{Overall} \\
		\cmidrule(l){3-3}\cmidrule(l){4-4}\cmidrule(l){5-5}\cmidrule(l){6-6}\cmidrule(l){7-7}\cmidrule(l){8-8}\cmidrule(l){9-9}\cmidrule(l){10-10}\cmidrule(l){11-11}
		\multirow{4}{*}{$[7,30)$}   & HNGARCH & 7.61 & 2.09 & 1.94 & 2.30 & 2.12 & 2.23 & 3.71 & 6.36 & \ov{4.17} \\
		& JPGARCH & 5.09 & 1.46 & 1.11 & 1.31 & 1.65 & 2.25 & 2.91 & 3.58 & \ov{2.85} \\
		& NGARCH & 1.95 & 0.87 & 0.67 & 0.77 & 0.95 & 0.88 & 1.20 & 2.31 & \ov{1.23} \\
		& VPGARCH & 1.90 & 0.86 & 0.67 & 0.67 & 0.76 & 0.78 & 1.05 & 2.01 & \ov{1.16} \\[1.5ex]
		\multirow{4}{*}{$[30,90)$}  & HNGARCH & 4.02 & 1.77 & 1.90 & 1.88 & 1.66 & 1.36 & 2.42 & 4.64 & \ov{2.98} \\
		& JPGARCH & 2.16 & 0.88 & 0.80 & 0.84 & 1.13 & 1.63 & 2.24 & 3.17 & \ov{1.80} \\
		& NGARCH & 1.24 & 0.72 & 0.67 & 0.63 & 0.72 & 0.78 & 0.74 & 1.29 & \ov{0.96} \\
		& VPGARCH & 1.16 & 0.68 & 0.64 & 0.60 & 0.55 & 0.56 & 0.58 & 1.06 & \ov{0.87} \\[1.5ex]
		\multirow{4}{*}{$[90,180)$} & HNGARCH & 3.47 & 1.54 & 1.64 & 1.70 & 1.72 & 1.55 & 1.48 & 3.08 & \ov{2.69} \\
		& JPGARCH & 1.91 & 1.02 & 0.84 & 0.84 & 1.08 & 1.50 & 2.00 & 3.10 & \ov{1.82} \\
		& NGARCH & 1.33 & 1.13 & 0.91 & 0.69 & 0.58 & 0.79 & 0.98 & 1.21 & \ov{1.13} \\
		& VPGARCH & 1.16 & 0.97 & 0.81 & 0.65 & 0.53 & 0.68 & 0.84 & 1.05 & \ov{0.99} \\[1.5ex]
		\multirow{4}{*}{$[180,270)$} & HNGARCH & 3.57 & 1.51 & 1.66 & 1.82 & 1.93 & 2.05 & 2.07 & 2.78 & \ov{2.92} \\
		& JPGARCH & 2.28 & 1.23 & 1.10 & 1.12 & 1.22 & 1.52 & 2.05 & 3.50 & \ov{2.29} \\
		& NGARCH & 1.55 & 1.35 & 1.01 & 0.82 & 0.64 & 0.79 & 1.20 & 1.88 & \ov{1.46} \\
		& VPGARCH & 1.36 & 1.14 & 0.86 & 0.72 & 0.62 & 0.79 & 1.15 & 1.76 & \ov{1.31} \\[1.5ex]
		\multirow{4}{*}{$[270,365]$} & HNGARCH & 3.46 & 1.72 & 1.83 & 2.13 & 2.25 & 2.48 & 2.66 & 3.38 & \ov{3.02} \\
		& JPGARCH & 2.38 & 1.42 & 1.26 & 1.47 & 1.57 & 1.88 & 2.38 & 4.01 & \ov{2.57} \\
		& NGARCH & 1.78 & 1.46 & 1.15 & 1.01 & 0.94 & 1.06 & 1.46 & 2.30 & \ov{1.71} \\
		& VPGARCH & 1.60 & 1.27 & 1.01 & 0.92 & 0.94 & 1.08 & 1.47 & 2.24 & \ov{1.60} \\[1.5ex]
		\multirow{4}{*}{Overall}    & HNGARCH & \ov{4.34} & \ov{1.80} & \ov{1.86} & \ov{1.98} & \ov{1.84} & \ov{1.70} & \ov{2.46} & \ov{3.90} & \ov{3.16} \\
		& JPGARCH & \ov{2.59} & \ov{1.10} & \ov{0.93} & \ov{1.02} & \ov{1.30} & \ov{1.78} & \ov{2.28} & \ov{3.36} & \ov{2.10} \\
		& NGARCH & \ov{1.42} & \ov{0.91} & \ov{0.75} & \ov{0.70} & \ov{0.77} & \ov{0.81} & \ov{0.94} & \ov{1.62} & \ov{1.14} \\
		& VPGARCH & \ov{1.31} & \ov{0.84} & \ov{0.71} & \ov{0.64} & \ov{0.63} & \ov{0.67} & \ov{0.82} & \ov{1.47} & \ov{1.04} \\
		\bottomrule
	\end{tabular*}

	\begin{minipage}{\linewidth}
		\footnotesize
		\emph{Notes:} This table reports the implied volatility root mean squared errors (IVRMSEs), in percent, by maturity and moneyness for the calibrations of Table~\ref{tab:weekly_ivrmse} under Gaussian innovations, with the bins of Table~\ref{tab:option_data}. Moneyness is the strike price divided by the forward price, and maturity is measured in calendar days. Each entry is the IVRMSE computed from the pricing errors of all options in the cell over the 1,443 weekly dates. The overall entries are illustrated in italics.
	\end{minipage}
\end{sidewaystable}

\begin{sidewaystable}
	\caption{\textbf{Implied Volatility Root Mean Squared Errors by Moneyness and Maturity under NIG Innovations}\label{tab:ivrmse_moneyness_maturity_nig}}
	\scriptsize
	\renewcommand{\arraystretch}{1.25}
	\newcommand{\mc}[1]{\multicolumn{1}{c}{#1}}
	\newcommand{\ov}[1]{\multicolumn{1}{c}{\textit{#1}}}
	\centering
	\begin{tabular*}{\linewidth}{@{\extracolsep{\fill}} cc*{9}{d{1.2}}}
		\toprule
		Maturity & Model & \multicolumn{9}{c}{Moneyness} \\
		\cmidrule(l){1-1}\cmidrule(l){2-2}\cmidrule(l){3-11}
		& & \mc{$[0,0.900)$} & \mc{$[0.900,0.950)$} & \mc{$[0.950,0.975)$} & \mc{$[0.975,1.000)$} & \mc{$[1.000,1.025)$} & \mc{$[1.025,1.050)$} & \mc{$[1.050,1.100)$} & \mc{$[1.100,\infty)$} & \mc{Overall} \\
		\cmidrule(l){3-3}\cmidrule(l){4-4}\cmidrule(l){5-5}\cmidrule(l){6-6}\cmidrule(l){7-7}\cmidrule(l){8-8}\cmidrule(l){9-9}\cmidrule(l){10-10}\cmidrule(l){11-11}
		\multirow{4}{*}{$[7,30)$}   & HNGARCH & 2.15 & 1.16 & 1.12 & 1.47 & 1.51 & 2.07 & 2.76 & 3.71 & \ov{1.81} \\
		& JPGARCH & 2.19 & 1.08 & 1.02 & 1.17 & 1.26 & 1.96 & 2.55 & 3.51 & \ov{1.70} \\
		& NGARCH & 1.85 & 0.81 & 0.65 & 0.78 & 0.97 & 0.90 & 1.16 & 2.12 & \ov{1.18} \\
		& VPGARCH & 1.75 & 0.80 & 0.63 & 0.69 & 0.79 & 0.76 & 1.04 & 1.96 & \ov{1.09} \\[1.5ex]
		\multirow{4}{*}{$[30,90)$}  & HNGARCH & 1.34 & 0.68 & 0.67 & 0.70 & 0.72 & 1.03 & 1.62 & 2.17 & \ov{1.20} \\
		& JPGARCH & 1.26 & 0.66 & 0.65 & 0.67 & 0.68 & 0.98 & 1.57 & 2.11 & \ov{1.14} \\
		& NGARCH & 1.16 & 0.70 & 0.66 & 0.62 & 0.70 & 0.76 & 0.70 & 1.19 & \ov{0.91} \\
		& VPGARCH & 1.08 & 0.65 & 0.62 & 0.59 & 0.53 & 0.54 & 0.54 & 0.99 & \ov{0.82} \\[1.5ex]
		\multirow{4}{*}{$[90,180)$} & HNGARCH & 1.41 & 0.64 & 0.61 & 0.63 & 0.61 & 0.76 & 1.18 & 1.84 & \ov{1.21} \\
		& JPGARCH & 1.26 & 0.58 & 0.58 & 0.61 & 0.61 & 0.74 & 1.15 & 1.95 & \ov{1.15} \\
		& NGARCH & 1.27 & 1.10 & 0.89 & 0.68 & 0.57 & 0.76 & 0.94 & 1.11 & \ov{1.08} \\
		& VPGARCH & 1.07 & 0.91 & 0.77 & 0.63 & 0.52 & 0.65 & 0.79 & 0.96 & \ov{0.92} \\[1.5ex]
		\multirow{4}{*}{$[180,270)$} & HNGARCH & 1.91 & 0.93 & 0.98 & 1.03 & 0.98 & 1.12 & 1.40 & 2.24 & \ov{1.73} \\
		& JPGARCH & 1.71 & 0.86 & 0.90 & 0.95 & 0.91 & 1.06 & 1.36 & 2.39 & \ov{1.65} \\
		& NGARCH & 1.46 & 1.30 & 0.98 & 0.80 & 0.62 & 0.77 & 1.16 & 1.78 & \ov{1.39} \\
		& VPGARCH & 1.24 & 1.05 & 0.80 & 0.68 & 0.61 & 0.77 & 1.11 & 1.67 & \ov{1.22} \\[1.5ex]
		\multirow{4}{*}{$[270,365]$} & HNGARCH & 2.22 & 1.27 & 1.22 & 1.48 & 1.51 & 1.67 & 1.92 & 2.88 & \ov{2.14} \\
		& JPGARCH & 1.98 & 1.16 & 1.12 & 1.38 & 1.41 & 1.59 & 1.86 & 3.00 & \ov{2.04} \\
		& NGARCH & 1.68 & 1.42 & 1.12 & 0.98 & 0.93 & 1.05 & 1.43 & 2.21 & \ov{1.64} \\
		& VPGARCH & 1.45 & 1.18 & 0.94 & 0.87 & 0.92 & 1.06 & 1.43 & 2.15 & \ov{1.50} \\[1.5ex]
		\multirow{4}{*}{Overall}    & HNGARCH & \ov{1.59} & \ov{0.85} & \ov{0.84} & \ov{0.99} & \ov{1.02} & \ov{1.33} & \ov{1.74} & \ov{2.31} & \ov{1.43} \\
		& JPGARCH & \ov{1.49} & \ov{0.80} & \ov{0.79} & \ov{0.86} & \ov{0.90} & \ov{1.27} & \ov{1.66} & \ov{2.35} & \ov{1.36} \\
		& NGARCH & \ov{1.35} & \ov{0.88} & \ov{0.73} & \ov{0.70} & \ov{0.77} & \ov{0.81} & \ov{0.90} & \ov{1.52} & \ov{1.09} \\
		& VPGARCH & \ov{1.21} & \ov{0.78} & \ov{0.67} & \ov{0.64} & \ov{0.63} & \ov{0.65} & \ov{0.78} & \ov{1.39} & \ov{0.97} \\
		\bottomrule
	\end{tabular*}

	\begin{minipage}{\linewidth}
		\footnotesize
		\emph{Notes:} This table reports the same IVRMSEs as Table~\ref{tab:ivrmse_moneyness_maturity_gaussian} for the calibrations under NIG innovations. See the notes to Table~\ref{tab:ivrmse_moneyness_maturity_gaussian}.
	\end{minipage}
\end{sidewaystable}
\endgroup

\clearpage
\bookmark[level=0,page=\arabic{page}]{Supplementary Material}
\ArxivBibScope{.SM}
\makeatletter
\let\maketitle\ArxivMakeTitle
\let\@maketitle\ArxivTitleBlock
\let\title\ArxivTitle
\let\author\ArxivAuthor
\let\date\ArxivDate
\let\thanks\ArxivThanks
\renewcommand{\ther@tfl@t}{SM.\arabic{r@tfl@t}}
\makeatother
\singlespacing
\counterwithout*{equation}{section}
\setcounter{page}{1}
\setcounter{section}{0}
\setcounter{subsection}{0}
\setcounter{subsubsection}{0}
\setcounter{equation}{0}
\setcounter{table}{0}
\setcounter{figure}{0}
\setcounter{footnote}{0}
\setcounter{theorem}{0}
\setcounter{assumption}{0}
\setcounter{proposition}{0}
\setcounter{lemma}{0}
\setcounter{corollary}{0}
\setcounter{remark}{0}
\setcounter{definition}{0}
\setcounter{example}{0}
\setcounter{condition}{0}
\setcounter{r@tfl@t}{0}
\renewcommand{\thesection}{SM.\Alph{section}}
\renewcommand{\thesubsection}{\thesection.\arabic{subsection}}
\renewcommand{\theequation}{SM.\arabic{equation}}
\renewcommand{\thetable}{SM.\arabic{table}}
\renewcommand{\thefigure}{SM.\arabic{figure}}
\renewcommand*{\thepage}{SM-\arabic{page}}

\renewcommand{\thetheorem}{SM.\arabic{theorem}}
\renewcommand{\theassumption}{SM.\arabic{assumption}}
\renewcommand{\theproposition}{SM.\arabic{proposition}}
\renewcommand{\thelemma}{SM.\arabic{lemma}}
\renewcommand{\thecorollary}{SM.\arabic{corollary}}
\renewcommand{\theremark}{SM.\arabic{remark}}
\renewcommand{\thedefinition}{SM.\arabic{definition}}

\title{\vspace{-0.5cm}\textbf{Supplementary Material}\\[2ex] \Title}
\date{\vspace{-0.5cm}}
\author{\Authors}
\makeatletter
\g@addto@macro{\normalsize}{%
    \setlength{\abovedisplayskip}{4pt}
    \setlength{\abovedisplayshortskip}{4pt}
    \setlength{\belowdisplayskip}{4pt}
    \setlength{\belowdisplayshortskip}{4pt}}
\makeatother

\maketitle

\titlespacing*{\section}{0pt}{0pt}{0pt}
\onehalfspacing

 \noindent Throughout the Supplementary Material, $\|\cdot\|$ denotes an arbitrary norm on the relevant finite-dimensional space, applied to vectors and to matrices alike. Since all norms on a finite-dimensional space are equivalent, none of the statements below depends on this choice.

\bigskip

\section{Recursive moment computation of polynomial-preserving Markov chains}
\label{sm:matryoshka_details}

\bigskip

This appendix records a degree-block implementation of the moment formulas in Section~\ref{subsec:general_framework} of the main paper. The block-triangular transition matrices express the degree restriction in the polynomial state-space models of \citet{KallsenRichert25}. Following the Matryoshkan matrix approach of \citet{DawPender2023}, we give finite-horizon recursions and long-run moment limits in the notation used for the hybrid GARCH--SV applications.

Fix $n$ and $m\geq0$, and use the nested bases from Section~\ref{subsec:general_framework}. Write
\begin{equation*}\bm{\HT}_m(\bm{x})=\left(\bm{\HT}^{(0)}(\bm{x})^\top,\bm{\HT}^{(1)}(\bm{x})^\top,\ldots,\bm{\HT}^{(m)}(\bm{x})^\top\right)^\top,\end{equation*}
where $\bm{\HT}^{(0)}\equiv1$ and the entries of $\bm{\HT}^{(r)}$ extend the basis of $\Pol_{r-1}(E)$ to that of $\Pol_r(E)$. The transition-operator convention is
\begin{equation}\label{sm:eq:basis_vector_transition}\trans^{(n)}\bm{\HT}_m(\bm{x})=\left(\bm{Q}^{(n)}_m\right)^\top\bm{\HT}_m(\bm{x}).\end{equation}
Let $\bm{Q}^{(n)}_{ij}$ denote the block containing the coefficients of $\bm{\HT}^{(i)}$ in the images of $\bm{\HT}^{(j)}$. Degree preservation implies $\bm{Q}^{(n)}_{ij}=\bm{0}$ for $i>j$, and $\bm{Q}^{(n)}_{00}=1$. The coordinate matrix on $\Pol_r(E)$ is the leading principal block $\bm{Q}^{(n)}_r$ of $\bm{Q}^{(n)}_m$, with
\begin{equation}\label{sm:eq:matryoshka_block}\bm{Q}^{(n)}_r=\begin{pmatrix}\bm{Q}^{(n)}_{r-1}&\bm{B}^{(n)}_r\\\bm{0}&\bm{Q}^{(n)}_{rr}\end{pmatrix},\qquad \bm{B}^{(n)}_r=\left(\left(\bm{Q}^{(n)}_{0r}\right)^\top,\left(\bm{Q}^{(n)}_{1r}\right)^\top,\ldots,\left(\bm{Q}^{(n)}_{r-1,r}\right)^\top\right)^\top,\qquad r\geq1.\end{equation}
This nesting is the Matryoshka structure used below.

\begin{proposition}\label{sm:prop:matryoshka_moment_recursion}
Suppose that $\bm{X}^{(n)}$ is polynomial preserving. Fix $k_0\in\{0,\ldots,nL\}$ and, for $k=0,\ldots,nL-k_0$, define
\begin{equation*}\bm{m}_k^{(r)}:=\E^{\Pm^{(n)}}\left[\bm{\HT}^{(r)}\left(\bm{X}^{(n)}_{(k_0+k)\tau}\right)\,\middle|\,\F^{(n)}_{k_0\tau}\right],\qquad r=0,\ldots,m,\end{equation*}
and collect the moments through degree $r$ in
\begin{equation*}\bar{\bm{m}}_k^{(r)}=\left(\left(\bm{m}_k^{(0)}\right)^\top,\left(\bm{m}_k^{(1)}\right)^\top,\ldots,\left(\bm{m}_k^{(r)}\right)^\top\right)^\top.\end{equation*}
Then $\bar{\bm{m}}_0^{(r)}=\bm{\HT}_r\left(\bm{X}^{(n)}_{k_0\tau}\right)$ and, for $k=0,\ldots,nL-k_0-1$,
\begin{equation}\label{sm:eq:cumulative_moment_recursion}\bar{\bm{m}}_{k+1}^{(r)}=\left(\bm{Q}^{(n)}_r\right)^\top\bar{\bm{m}}_k^{(r)},\qquad r=0,\ldots,m.\end{equation}
In particular, for $r=1,\ldots,m$,
\begin{equation}\label{sm:eq:matryoshka_degree_recursion}\bm{m}_{k+1}^{(r)}=\left(\bm{Q}^{(n)}_{rr}\right)^\top\bm{m}_k^{(r)}+\left(\bm{B}^{(n)}_r\right)^\top\bar{\bm{m}}_k^{(r-1)}.\end{equation}
Equivalently, for $k=0,\ldots,nL-k_0$,
\begin{equation}\label{sm:eq:matryoshka_closed_form}\bm{m}_{k}^{(r)}=\left[\left(\bm{Q}^{(n)}_{rr}\right)^\top\right]^{k}\bm{m}_0^{(r)}+\sum_{j=0}^{k-1}\left[\left(\bm{Q}^{(n)}_{rr}\right)^\top\right]^{k-1-j}\left(\bm{B}^{(n)}_r\right)^\top\bar{\bm{m}}_j^{(r-1)},\end{equation}
where the sum is zero when $k=0$.
\end{proposition}

\begin{proof}
Proposition~\ref{prop:finite_dimensional_representation} gives Equation~\eqref{sm:eq:cumulative_moment_recursion}. Taking the last block row in the transpose of Equation~\eqref{sm:eq:matryoshka_block} gives Equation~\eqref{sm:eq:matryoshka_degree_recursion}, whose iteration yields Equation~\eqref{sm:eq:matryoshka_closed_form}.
\end{proof}

Equation~\eqref{sm:eq:matryoshka_degree_recursion} computes consecutive-horizon moments using block matrix--vector products and reuses lower-degree moment sequences. Taking $k_0=0$ gives unconditional moments when the initial state is deterministic; otherwise, take expectations over its distribution when the required moments are integrable.

For the long-run analysis, extend the chain to all $k\geq0$ using the same transition kernel, and let $\rho(\cdot)$ denote the spectral radius. Since the matrix acting on nonconstant moments is block triangular, its spectral radius is the largest spectral radius of the diagonal blocks $\bm{Q}^{(n)}_{rr}$, $r=1,\ldots,m$. The following condition is therefore the degree-block form of the spectral-radius condition for the full nonconstant moment system.

\begin{proposition}\label{sm:prop:matryoshka_stationary_moments}
Suppose that $\bm{X}^{(n)}$ is polynomial preserving and fix $m\geq1$. Let $\bm{m}_k^{(r)}$ be defined as in Proposition~\ref{sm:prop:matryoshka_moment_recursion}, now for all $k\geq0$. If
\begin{equation}\label{sm:eq:matryoshka_spectral_condition}\rho\left(\bm{Q}^{(n)}_{rr}\right)<1,\qquad r=1,\ldots,m,\end{equation}
then, for each $r=0,\ldots,m$, $\bm{m}_k^{(r)}$ converges almost surely as $k\to\infty$ to a deterministic limit $\bm{m}_\infty^{(r)}$ independent of the origin date and state. These limits are obtained from $\bm{m}_\infty^{(0)}=1$ by
\begin{equation}\label{sm:eq:matryoshka_stationary_moments}\bm{m}_\infty^{(r)}=\left[\bm{I}-\left(\bm{Q}^{(n)}_{rr}\right)^\top\right]^{-1}\left(\bm{B}^{(n)}_r\right)^\top\bar{\bm{m}}_\infty^{(r-1)},\qquad r=1,\ldots,m.\end{equation}
If, in addition, the chain admits an invariant probability distribution $\pi^{(n)}$ with finite moments through degree $m$, then
\begin{equation*}\bm{m}_\infty^{(r)}=\int_E\bm{\HT}^{(r)}(\bm{x})\,\pi^{(n)}(d\bm{x}),\qquad r=0,\ldots,m,\end{equation*}
so the limiting moments coincide with the corresponding stationary moments.
\end{proposition}

\begin{proof}
Partition the full moment system as
\begin{equation*}\left(\bm{Q}^{(n)}_m\right)^\top=\begin{pmatrix}1&\bm{0}\\\bm{b}&\bm{C}\end{pmatrix},\qquad \bm{z}_k=\left(\left(\bm{m}_k^{(1)}\right)^\top,\ldots,\left(\bm{m}_k^{(m)}\right)^\top\right)^\top.\end{equation*}
Then $\bm{z}_{k+1}=\bm{C}\bm{z}_k+\bm{b}$ and $\rho\left(\bm{C}\right)=\max_{1\leq r\leq m}\rho\left(\bm{Q}^{(n)}_{rr}\right)<1$. Consequently,
\begin{equation*}\bm{z}_k=\bm{C}^k\bm{z}_0+\sum_{j=0}^{k-1}\bm{C}^j\bm{b},\qquad \lim_{k\to\infty}\bm{z}_k=\left(\bm{I}-\bm{C}\right)^{-1}\bm{b}\quad\text{almost surely}.\end{equation*}
The limit is deterministic and independent of the origin date and state. Solving its fixed-point equation successively by degree gives Equation~\eqref{sm:eq:matryoshka_stationary_moments}. If $\pi^{(n)}$ is invariant with finite moments through degree $m$, invariance applied to the absolute values of the basis polynomials and Fubini's theorem justify integrating Equation~\eqref{sm:eq:basis_vector_transition} against $\pi^{(n)}$. Its nonconstant moment vector therefore solves the same uniquely solvable fixed-point equation. Hence the limiting and stationary moments coincide.
\end{proof}

For a scalar chain with deterministic initial state and monomial basis $\bm{\HT}_m(x)=(1,x,\ldots,x^m)^\top$, reindex the matrix entries by polynomial degree and write
\begin{equation*}\trans^{(n)}x^j=\sum_{i=0}^{j}q^{(n)}_{ij}x^i,\qquad m_{k,j}:=\E^{\Pm^{(n)}}\left[\left(X^{(n)}_{k\tau}\right)^j\right].\end{equation*}
Equation~\eqref{sm:eq:matryoshka_degree_recursion} reduces to
\begin{equation*}m_{k+1,j}=q^{(n)}_{jj}m_{k,j}+\sum_{i=0}^{j-1}q^{(n)}_{ij}m_{k,i},\qquad j=1,\ldots,m,\end{equation*}
with $m_{k,0}=1$. The spectral condition becomes $|q^{(n)}_{jj}|<1$ for $j=1,\ldots,m$. Section~\ref{subsec:moment_implications} identifies these coefficients for the VPGARCH-SV and JPGARCH-SV variance recursions and expresses this condition in terms of the model parameters through Equation~\eqref{eq:garch_moment_condition}.

\bigskip

\section{Local diffusion conditions and operator convergence}
\label{sm:nelson_local_conditions}

\bigskip

This section records the local conditions underlying the diffusion approximation and establishes convergence of the scaled one-step operators on polynomial test functions. Additional moment control is required because polynomials may be unbounded when the state space is unbounded. Supplementary Material~\ref{subsec:risk_neutral_diffusion_limits} derives the risk-neutral diffusion limits and their polynomial implications. For the physical local conditions below, conditional expectations given $\bm{X}^{(n)}_{k\tau}=\bm{x}$ are understood with respect to the transition kernel $\Pi^{(n)}(\bm{x},d\bm{z})$ introduced in Section~\ref{subsec:general_framework} of the main paper.  Define the one-step increment by
\begin{equation*}
\Delta\bm{X}^{(n)}_{(k+1)\tau}=\bm{X}^{(n)}_{(k+1)\tau}-\bm{X}^{(n)}_{k\tau}.
\end{equation*}

\begin{assumption}
\label{sm:ass:nelson_polynomial_degree_m}
Fix $m \geq 0$. Suppose that there exist continuous functions $\bm{A}:E\to\mathbb{R}^d$ and $\bm{B}:E\to\mathbb{R}^{d\times d}$, where $\bm{B}(\bm{x})$ is symmetric and positive semidefinite for every $\bm{x}\in E$. Suppose further that there exists $\delta>0$ such that, with $q:=2+\delta$, one has $q\geq m$. For every compact set $\mathcal{K}\subseteq E$, assume
\begin{align}
\lim_{\tau\to0}\sup_{\bm{x}\in\mathcal{K}}\left\|\frac{1}{\tau}\E^{\Pm^{(n)}}\left[\Delta\bm{X}^{(n)}_{(k+1)\tau}\,\middle|\,\bm{X}^{(n)}_{k\tau}=\bm{x}\right]-\bm{A}(\bm{x})\right\|&=0, \label{sm:eq:nelson_drift_condition}\\
\lim_{\tau\to0}\sup_{\bm{x}\in\mathcal{K}}\left\|\frac{1}{\tau}\E^{\Pm^{(n)}}\left[\Delta\bm{X}^{(n)}_{(k+1)\tau}\Delta\bm{X}^{(n)\top}_{(k+1)\tau}\,\middle|\,\bm{X}^{(n)}_{k\tau}=\bm{x}\right]-\bm{B}(\bm{x})\right\|&=0. \label{sm:eq:nelson_covariance_condition}
\end{align}
For each component $i=1,\ldots,d$, also assume
\begin{equation}
\label{sm:eq:nelson_moment_condition}
\lim_{\tau\to0}\sup_{\bm{x}\in\mathcal{K}}\frac{1}{\tau}\E^{\Pm^{(n)}}\left[\left|\Delta X^{(n)}_{(k+1)\tau,i}\right|^q\,\middle|\,\bm{X}^{(n)}_{k\tau}=\bm{x}\right]=0.
\end{equation}
\end{assumption}

The first two conditions identify the drift vector and covariance matrix of the limiting diffusion through the scaled conditional first and second moments. The third is a stronger local moment condition tailored to polynomial test functions. By finite-dimensional norm equivalence and Markov's inequality, it implies the usual no-large-jumps condition, while $q\geq m$ provides the additional control required for polynomials of degree at most $m$. The conditions are imposed uniformly over arbitrary compact subsets of $E$. Since Proposition~\ref{prop:weak_limit_polynomial_preservation} requires the operator convergence only at $m=1$ and $m=2$, the restriction $q\geq m$ is not binding in the applications in Section~\ref{subsec:physical_diffusion_limits} of the main paper and Supplementary Material~\ref{subsec:risk_neutral_diffusion_limits} below, where any $\delta>0$ suffices.

Together with convergence of the initial distributions, the required compact-containment condition, and well-posedness and nonexplosion of the limiting diffusion, the local conditions above yield the standard diffusion approximation; see \citet[Theorem~2.2]{nelson1990arch}. The stronger moment condition in Equation~\eqref{sm:eq:nelson_moment_condition} is used below to extend the corresponding operator convergence from compactly supported smooth functions to polynomial test functions.

\subsection{Operator convergence on polynomial test functions}
\label{sm:operator_convergence_polynomials}

For a polynomial $p$ in the domain of $\trans^{(n)}$, define the scaled one-step operator by
\begin{equation*}
\mathcal{A}^{(n)}p(\bm{x})=\frac{\trans^{(n)}p(\bm{x})-p(\bm{x})}{\tau}.
\end{equation*}
Suppose that the limiting diffusion has drift vector $\bm{A}$ and covariance matrix $\bm{B}=\bm{\Sigma}\bm{\Sigma}^\top$. For a polynomial test function $p$, define
\begin{equation*}
\mathcal{A}p(\bm{x})=\sum_{i=1}^{d}A_i(\bm{x})\partial_i p(\bm{x})+\frac{1}{2}\sum_{i,j=1}^{d}B_{ij}(\bm{x})\partial_{ij}p(\bm{x}).
\end{equation*}
When $p\in\Pol_m(E)$ is the restriction of a polynomial on $\mathbb{R}^d$, the expression above is understood through a polynomial extension. We assume that $\mathcal{A}$ is well defined on $\Pol_m(E)$.

For compactly supported smooth test functions, the corresponding truncated local drift and second-moment conditions, together with the no-large-jumps condition, are equivalent to locally uniform convergence of the associated operators; see \citet[Lemma~11.2.1, pp.~268--270]{Stroock}. The following proposition extends this conclusion to polynomial test functions.

\begin{proposition}
\label{sm:prop:generator_stability}
Suppose that Assumption~\ref{sm:ass:nelson_polynomial_degree_m} holds and that $\mathcal{A}$ is well defined on $\Pol_m(E)$. Then, for every $p\in\Pol_m(E)$ and every compact set $\mathcal{K}\subseteq E$,
\begin{equation*}
\lim_{\tau\to0}\sup_{\bm{x}\in\mathcal{K}}\left|\mathcal{A}^{(n)}p(\bm{x})-\mathcal{A}p(\bm{x})\right|=0.
\end{equation*}
\end{proposition}

\begin{proof}
Fix $p\in\Pol_m(E)$ and a compact set $\mathcal{K}\subseteq E$, and choose a polynomial extension of $p$ to $\mathbb{R}^d$ of degree at most $m$. Its second-order Taylor expansion is
\begin{equation*} p(\bm{x}+\bm{z})-p(\bm{x})=\sum_{i=1}^{d}\partial_i p(\bm{x})z_i+\frac{1}{2}\sum_{i,j=1}^{d}\partial_{ij}p(\bm{x})z_i z_j+R_p(\bm{x},\bm{z}). \end{equation*}
If $m\leq2$, then $R_p\equiv0$. Otherwise, the exact polynomial remainder satisfies
\begin{equation*} |R_p(\bm{x},\bm{z})|\leq C_{p,\mathcal{K}}\sum_{j=3}^{m}\|\bm{z}\|^j, \qquad \bm{x}\in\mathcal{K},\quad \bm{z}\in\mathbb{R}^d, \end{equation*}
for some $C_{p,\mathcal{K}}<\infty$.

Equation~\eqref{sm:eq:nelson_covariance_condition} bounds the scaled conditional second moments uniformly on $\mathcal{K}$, while Equation~\eqref{sm:eq:nelson_moment_condition} and finite-dimensional norm equivalence imply that the scaled conditional $q$th moments of the increment norm tend uniformly to zero. For $2<j<q$, conditional H\"older's inequality gives
\begin{align*} \frac{1}{\tau}\E^{\Pm^{(n)}}\left[\left\|\Delta\bm{X}^{(n)}_{(k+1)\tau}\right\|^j\,\middle|\,\bm{X}^{(n)}_{k\tau}=\bm{x}\right]&\leq\left(\frac{1}{\tau}\E^{\Pm^{(n)}}\left[\left\|\Delta\bm{X}^{(n)}_{(k+1)\tau}\right\|^2\,\middle|\,\bm{X}^{(n)}_{k\tau}=\bm{x}\right]\right)^{\frac{q-j}{q-2}}\\ &\quad\times\left(\frac{1}{\tau}\E^{\Pm^{(n)}}\left[\left\|\Delta\bm{X}^{(n)}_{(k+1)\tau}\right\|^q\,\middle|\,\bm{X}^{(n)}_{k\tau}=\bm{x}\right]\right)^{\frac{j-2}{q-2}}. \end{align*}
Consequently, every scaled conditional moment of order $j=3,\ldots,m$ tends uniformly to zero; the case $j=q$ follows directly from the moment condition. The remainder bound therefore gives the same conclusion for the scaled conditional expectation of $|R_p|$.

Since $q\geq m$, the moment condition also ensures conditional integrability of the polynomial and all Taylor terms on $\mathcal{K}$ for sufficiently small $\tau$. Conditioning the Taylor expansion, dividing by $\tau$, and applying Equations~\eqref{sm:eq:nelson_drift_condition} and~\eqref{sm:eq:nelson_covariance_condition}, with the derivatives bounded on $\mathcal{K}$, proves the asserted uniform convergence.
\end{proof}

Proposition~\ref{sm:prop:generator_stability} supplies the operator convergence on polynomial test functions used in the main paper. The finite-dimensional argument showing that polynomial preservation passes to the limiting generator is given in the main appendix.

\begin{proof}[Proof of Corollary~\ref{cor:moment_matrix_convergence}]
Choose $N_m$ states in $E$ whose basis-evaluation matrix is invertible. Evaluating Equation~\eqref{eq:nelson_generator_convergence} at these states yields the first matrix limit. Set $\bm D_n=(\bm Q_m^{(n)}-\bm I_{N_m})/\tau$ and $\delta_n=\|\bm D_n-\bm G_m\|$. In an induced matrix norm, boundedness of $\bm D_n$ and the exponential expansion give
\begin{equation*}\left\|\bm Q_m^{(n)}-e^{\tau\bm G_m}\right\|\leq\tau\delta_n+C\tau^2, \qquad \left\|\left(\bm Q_m^{(n)}\right)^j\right\|,\ \left\|e^{j\tau\bm G_m}\right\|\leq e^{Mj\tau},\end{equation*}
for constants $C,M$ independent of sufficiently small $\tau$. Writing the difference of the $k$th powers as a sum therefore gives
\begin{equation*}\sup_{k\tau\leq L}\left\|\left(\bm Q_m^{(n)}\right)^k-e^{k\tau\bm G_m}\right\|\leq C_{L,m}(\delta_n+\tau).\end{equation*}
Taking $k=\lfloor t/\tau\rfloor$ and using $\|e^{t\bm G_m}-e^{k\tau\bm G_m}\|\leq C_{L,m}\tau$ proves the second limit. Norm equivalence gives the same conclusion for any matrix norm.

For the diffusion, affine drift and quadratic covariance give $\mathcal A(1+\sum_i x_i^2)^r\leq C_r(1+\sum_i x_i^2)^r$ for each integer $r\geq1$. Localization and Gronwall's inequality yield finite moments of every order on bounded time intervals from a deterministic initial state. These bounds justify taking expectations in It\^o's formula for the basis polynomials, giving
\begin{equation*}\E_{\bm x}^{\Pm}\left[\bm{\HT}_m(\bm X_t)\right]=e^{t\bm G_m^\top}\bm{\HT}_m(\bm x).\end{equation*}
Combine this identity with Proposition~\ref{prop:finite_dimensional_representation}. For every compact set $\mathcal K\subseteq E$, the conditional-moment formulas give, in Euclidean and induced operator norms,
\begin{align*}\sup_{\substack{0\leq t\leq L\\\bm x\in\mathcal K}}\left|\E_{\bm x}^{\Pm^{(n)}}\left[p(\bm X_t^{(n)})\right]-\E_{\bm x}^{\Pm}\left[p(\bm X_t)\right]\right|\leq{}&\sup_{\bm x\in\mathcal K}\|\bm{\HT}_m(\bm x)\|\,\|\bm p\| \times\sup_{0\leq t\leq L}\left\|\left(\bm Q_m^{(n)}\right)^{\lfloor t/\tau\rfloor}-e^{t\bm G_m}\right\|.\end{align*}
Boundedness of the polynomial basis on $\mathcal K$ and the matrix limit prove uniform convergence jointly in time and initial state. The same bound applies whenever the two processes have these finite-dimensional moment representations, including the jump limits in Section~\ref{subsec:polynomial_levy_limits} of the main paper.
\end{proof}

For the forecast-error covariance limit in Section~\ref{subsec:general_diffusion_framework}, the identity $\trans^{(n)}f=f+\tau\mathcal A^{(n)}f$ gives
\begin{equation*}\frac{V^{(n)}_{p,q,1}}{\tau}=\mathcal A^{(n)}(pq)-p\mathcal A^{(n)}q-q\mathcal A^{(n)}p-\tau\left(\mathcal A^{(n)}p\right)\left(\mathcal A^{(n)}q\right).\end{equation*}
Operator convergence on $\Pol_{2m}(E)$ gives locally uniform convergence of the first three terms and local boundedness of $\mathcal A^{(n)}p$ and $\mathcal A^{(n)}q$, so the last term vanishes. The diffusion product rule gives
\begin{equation*}\mathcal A(pq)-p\mathcal A q-q\mathcal A p=\nabla p^\top\bm B\nabla q,\end{equation*}
which proves the asserted limit.

\bigskip

\subsection{State-space conditions for the diffusion limits}
\label{sm:variance_limit_boundary_conditions}

\bigskip

We first examine the physical variance diffusion in Proposition~\ref{prop:p_weak_limit}. Its state-space conditions are independent of $\varrho$, since the marginal variance dynamics are unchanged by the hybrid extension. Its autonomous dynamics are
\begin{equation*}dh_t=\kappa(\sigma^2-h_t)\,dt+\sqrt{v_{\mathsf{M}}(h_t)}\,dW_t,\qquad \mathsf{M}\in\{\mathsf{VP},\mathsf{JP}\},\end{equation*}
where
\begin{equation*}v_{\mathsf{VP}}(x)=\eta^2(a_0+a_1x+x^2)+(m_4-1)(b_0+b_1x)^2,\qquad v_{\mathsf{JP}}(x)=\eta^2x+(m_4-1)(b_0+b_1x)^2.\end{equation*}
Assume $\kappa\geq0$, $\sigma^2\geq0$, and that the squared diffusion coefficient is nonnegative on $\Rplus$. For VPGARCH-SV with $\eta\neq0$, also require $a_0+a_1x+x^2\geq0$ on $\Rplus$. Stochastic invariance below means existence of a nonnegative solution from every nonnegative initial state, without adding reflection or a killing rule.

\begin{proposition}
\label{sm:prop:physical_limit_state_space}
The following statements hold.
\begin{enumerate}[label=\textnormal{(\roman*)}]
\item For the JPGARCH-SV limit, $\Rplus$ is stochastically invariant if and only if $b_0=0$.
\item For the VPGARCH-SV limit, $\Rplus$ is stochastically invariant if and only if $b_0=0$ and $\eta^2a_0=0$. In particular, when $\eta\neq0$, this requires $a_0=b_0=0$.
\item Under the respective invariance conditions and $h_0>0$, zero is unattainable if and only if $2\kappa\sigma^2\geq\eta^2$ for JPGARCH-SV and $2\kappa\sigma^2\geq\eta^2a_1$ for VPGARCH-SV.
\end{enumerate}
Under the invariance conditions, the physical variance diffusion is nonexplosive on $\Rplus$.
\end{proposition}

\begin{proof}
Invariance requires zero diffusion coefficient and inward-pointing drift at zero. The drift there is $\kappa\sigma^2\geq0$, while
\begin{equation*}v_{\mathsf{JP}}(0)=(m_4-1)b_0^2,\qquad v_{\mathsf{VP}}(0)=\eta^2a_0+(m_4-1)b_0^2.\end{equation*}
Since $m_4>1$ and the two terms in $v_{\mathsf{VP}}(0)$ are nonnegative, the stated restrictions are necessary. Under them the squared diffusion coefficients become
\begin{equation*}v_{\mathsf{JP}}(x)=x[\eta^2+(m_4-1)b_1^2x],\qquad v_{\mathsf{VP}}(x)=x\{\eta^2a_1+[\eta^2+(m_4-1)b_1^2]x\}.\end{equation*}
Thus $v(x)=x(c_1+c_2x)$ for nonnegative constants $c_1,c_2$. The square-root coefficient satisfies a local $1/2$-H\"older estimate, and the affine drift is Lipschitz. The standard one-dimensional existence, uniqueness, and comparison arguments, with the coefficients extended using $x^+=\max\{x,0\}$ in the diffusion term, give a nonnegative solution. This proves sufficiency in parts~\textnormal{(i)} and~\textnormal{(ii)}.

For $c_1>0$, put $\alpha=2\kappa\sigma^2/c_1$. The scale and speed densities behave near zero as $s'(x)\asymp x^{-\alpha}$ and $m(x)\asymp x^{\alpha-1}$. When $\alpha\geq1$, the scale integral diverges and zero is unattainable. When $\alpha<1$, the scale distance from zero is of order $x^{1-\alpha}$, and its product with the speed density is integrable at zero, so the boundary is attainable. If $c_1=0$, variation of constants using a stochastic exponential gives a strictly positive solution from $h_0>0$; this also covers $c_2=0$. Substituting $c_1=\eta^2$ or $c_1=\eta^2a_1$ proves part~\textnormal{(iii)}. Finally, the physical drift and diffusion coefficient have at most linear growth, excluding finite-time explosion.
\end{proof}

These continuous-time conditions constrain the sequences of globally nonnegative discrete-time models. Under Proposition~\ref{prop:joint_polynomial_positivity} and Assumption~\ref{ass:p_scaling},
\begin{equation*}0\leq b_0(\tau)\leq\omega(\tau)=[1-\phi(\tau)]\sigma^2(\tau)=\kappa\sigma^2\tau+o(\tau),\end{equation*}
so $b_0=0$. Under Proposition~\ref{prop:variance_polynomial_positivity}, $\delta_0(\tau)\leq0$ gives
\begin{equation*}0\leq\eta(\tau)^2a_0(\tau)\leq4b_0(\tau)[\omega(\tau)-b_0(\tau)].\end{equation*}
It follows that $0\leq b_0(\tau)\leq\omega(\tau)$ and, on dividing by $\tau$ and taking limits, that $b_0=0$ and $\eta^2a_0=0$.

The same finite-step restrictions also imply
\begin{equation*}\eta^2\leq4\kappa\sigma^2\quad\text{for JPGARCH-SV},\qquad \eta^2a_1\leq4\kappa\sigma^2\quad\text{for VPGARCH-SV}.\end{equation*}
For JPGARCH-SV, divide Equation~\eqref{eq:joint_positivity_leverage_restriction} by $\tau$ and use $b_0\left(\tau\right)\leq\omega\left(\tau\right)=\kappa\sigma^2\tau+o\left(\tau\right)$ and $b_1\left(\tau\right)=\BigO{\sqrt{\tau}}$. For VPGARCH-SV, $\delta_0\left(\tau\right)\leq0$ gives $\left|\delta_0\left(\tau\right)\right|=\BigO{\tau^2}$, while $\left|\delta_2\left(\tau\right)\right|=\BigO{\sqrt{\tau}}$. Equation~\eqref{eq:positivity_delta_restrictions} therefore bounds $\delta_1\left(\tau\right)$ above by $C\tau^{5/4}$ for some $C\geq0$, and
\begin{equation*}\frac{\delta_1\left(\tau\right)}{\tau}=\eta^2a_1-\frac{4b_0\left(\tau\right)}{\tau}+o\left(1\right)\end{equation*}
gives the second inequality. These are necessary consequences of the finite-step restrictions, not a complete characterization of admissible parameter sequences. They are distinct from the zero-boundary nonattainment conditions in Proposition~\ref{sm:prop:physical_limit_state_space}. In particular, imposing Proposition~\ref{prop:joint_polynomial_positivity} at every small time step excludes a nondegenerate HNGARCH-SV local limit with $b_1=0$ and $b_0>0$.

For the risk-neutral limit in Supplementary Material~\ref{subsec:risk_neutral_diffusion_limits}, write its variance drift as $\kappa(\sigma^2-x)+\delta_{\mathsf{M}}(x)$, where
\begin{equation*}\delta_{\mathsf{M}}(x)=-\varrho\eta\left(\lambda+\frac{1}{2}\right)\sqrt{xg_{\mathsf{M}}(x)}+\xi_2(x)v_{\mathsf{M}}^{\perp}(x).\end{equation*}
Under the tangency restrictions above, $\eta^2g_{\mathsf{M}}(x)=c_1x+\BigO{x^2}$ and $v_{\mathsf{M}}^{\perp}(x)=\left(1-\varrho^2\right)c_1x+\BigO{x^2}$ as $x\downarrow0$, with $c_1=\eta^2$ for JPGARCH-SV and $c_1=\eta^2a_1$ for VPGARCH-SV. Hence, the condition $\xi_2(x)v_{\mathsf{M}}^{\perp}(x)=\BigO{x}$ implies $\delta_{\mathsf{M}}(x)=\BigO{x}$. Bounded $\xi_2$ near zero is sufficient; the weaker bound $\xi_2(x)=\BigO{x^{-1}}$ remains sufficient when $\left(1-\varrho^2\right)c_1=0$.

Under this condition, for $c_1>0$, the scale densities of the physical and risk-neutral equations satisfy
\begin{equation*}\frac{s'_{\Qm}(x)}{s'_{\Pm}(x)}=\exp\left(-\int_{x_0}^{x}\frac{2\delta_{\mathsf{M}}(y)}{v_{\mathsf{M}}(y)}\,dy\right),\end{equation*}
up to a positive normalization constant. The integral has a finite limit at zero, so the scale and speed tests give the same classification. If $c_1=0$, the drift is $\kappa\sigma^2+\BigO{x}$ near zero and the diffusion coefficient is proportional to $x$; comparison with a linear equation again excludes attainment from a positive state. Thus the zero-boundary tests are unchanged, with boundary drift $\kappa\sigma^2$.

The bound on $\xi_2v_{\mathsf{M}}^{\perp}$ is sufficient, not necessary. For $c_1>0$, convergence of the scale-density correction integral is also sufficient; it is not a necessary condition for the two boundary classifications to agree. Under Proposition~\ref{prop:q_polynomial_preservation} and Assumption~\ref{ass:q_scaling}, the nonnegative finite-step loadings give, for $0<x\leq1$,
\begin{equation*}\left|\xi_2(x;\tau)\right|=\left|\xi_2(1;\tau)\right|\frac{b_0(\tau)+b_1(\tau)}{b_0(\tau)+b_1(\tau)x}\leq\frac{\left|\xi_2(1;\tau)\right|}{x}.\end{equation*}
Taking limits gives $\xi_2(x)=\BigO{x^{-1}}$, which alone need not preserve the boundary classification when $\varrho<1$ and $c_1>0$. At the diffusion level, for example, in JPGARCH-SV with $b_0=b_1=0$, $\eta\neq0$, and $\xi_2(x)=\ell/x$, the risk-neutral boundary drift is $\beta_{\Qm}=\kappa\sigma^2+\left(1-\varrho^2\right)\eta^2\ell$. Provided $\beta_{\Qm}\geq0$, a nonnegative variance diffusion exists and zero is unattainable from a positive state if and only if $2\beta_{\Qm}\geq\eta^2$.

These are local statements about zero. Assumption~\ref{ass:q_scaling} imposes no growth bound at infinity, so nonexplosion under $\Qm$ does not follow from the physical result. For example, an additional bound $\mu_{\Qm}(x)\leq C(1+x)$ for all sufficiently large $x$, where $\mu_{\Qm}$ is the full risk-neutral variance drift, is sufficient for nonexplosion of a nonnegative solution by a Lyapunov argument. A global risk-neutral diffusion requires this or a suitable test at infinity.

\bigskip

\subsection{A GARCH-type approximation with Jacobi diffusion limit}
\label{sm:jacobi_garch_approximation}

\bigskip

We construct a GARCH-type approximation of the Jacobi model with a polynomial diffusion limit but without discrete-time polynomial preservation.

Let $Q$ be the quadratic function defined in Equation~\eqref{eq:jacobi_q_function}, and suppose that $0\leq h_{\min}<\sigma^2<h_{\max}$, $\kappa>0$, $\zeta_J>0$, and $\varrho_J\in[-1,1]$. As shown by \citet{Ackerer18},
\begin{equation*}
0\leq Q(x)\leq x, \qquad x\in[h_{\min},h_{\max}].
\end{equation*}
For $x\in(h_{\min},h_{\max})$, define
\begin{equation*}
\psi_2^J(x;\tau)=\sqrt{\tau}\,\varrho_J\zeta_J\frac{Q(x)}{\sqrt{x}}, \qquad \psi_3^J(x;\tau)=\sqrt{\frac{\tau}{2}}\,\zeta_J\sqrt{Q(x)-\varrho_J^2\frac{Q(x)^2}{x}}.
\end{equation*}
The second square root is well defined because $0\leq Q(x)\leq x$ and $|\varrho_J|\leq1$. The loading functions extend continuously to $[h_{\min},h_{\max}]$ and vanish at both endpoints, including $x=0$ when $h_{\min}=0$.

Consider, while $h^{(n)}_{k\tau}\in(h_{\min},h_{\max})$, the GARCH-type recursion
\begin{equation}
\label{sm:eq:jacobi_garch_recursion}
\begin{aligned}[b]
y^{(n)}_{(k+1)\tau}&=r\tau+\lambda h^{(n)}_{k\tau} \tau+\sqrt{\tau h^{(n)}_{k\tau}}\,\epsilon^{(n)}_{(k+1)\tau},\qquad \epsilon^{(n)}_{(k+1)\tau}\mid\F^{(n)}_{k\tau}\overset{\Pm^{(n)}}{\sim}N(0,1), \\
h^{(n)}_{(k+1)\tau}&=h^{(n)}_{k\tau}+\kappa\left(\sigma^2-h^{(n)}_{k\tau}\right)\tau+\psi_2^J\left(h^{(n)}_{k\tau};\tau\right)\epsilon^{(n)}_{(k+1)\tau}+\psi_3^J\left(h^{(n)}_{k\tau};\tau\right)\left[\left(\epsilon^{(n)}_{(k+1)\tau}\right)^2-1\right].
\end{aligned}
\end{equation}
Since the Gaussian innovation has unbounded support, the recursion need not remain in $[h_{\min},h_{\max}]$ for fixed $\tau$. Accordingly, the raw recursion is used only while the variance remains in the interior; it is stopped on exit. Local weak convergence is understood through regular exhausting domains, as in Section~\ref{subsec:physical_diffusion_limits}, now in $\mathbb{R}\times(h_{\min},h_{\max})$. A globally boundary-preserving version is given below.

\begin{proposition}
\label{sm:prop:jacobi_garch_limit}
Suppose that $\bm{X}^{(n)}_0=\bm{X}_0=(Y_0,h_0)^\top$ for every $n$, where $h_0\in(h_{\min},h_{\max})$. Then, as $\tau\to0$, the c\`adl\`ag extensions generated by Equation~\eqref{sm:eq:jacobi_garch_recursion} converge weakly, locally on $\mathbb{R}\times(h_{\min},h_{\max})$, to
\begin{equation*}
\begin{aligned}[b]
dY_t&=(r+\lambda h_t)\,dt+\sqrt{h_t}\,dW^J_{1,t},\\
dh_t&=\kappa\left(\sigma^2-h_t\right)\,dt+\zeta_J\sqrt{Q(h_t)}\,dW^J_{2,t},
\end{aligned}
\end{equation*}
where $W^J_1$ and $W^J_2$ are standard Brownian motions under $\Pm$ satisfying
\begin{equation*}
d\left\langle W^J_1,W^J_2\right\rangle_t=\varrho_J\sqrt{\frac{Q(h_t)}{h_t}}\,dt.
\end{equation*}
If, in addition,
\begin{equation}
\label{sm:eq:jacobi_boundary_condition}
\frac{\zeta_J^2(h_{\max}-h_{\min})}{\left(\sqrt{h_{\max}}-\sqrt{h_{\min}}\right)^2}\leq2\kappa\min\left\{h_{\max}-\sigma^2,\sigma^2-h_{\min}\right\},
\end{equation}
then the limiting variance process remains in $(h_{\min},h_{\max})$ for every finite time.
\end{proposition}

\begin{proof}
Dividing the conditional mean increments by $\tau$ gives $r+\lambda x$ and $\kappa\left(\sigma^2-x\right)$. The identities
\begin{equation*}\frac{\sqrt{\tau x}}{\tau}\psi_2^J\left(x;\tau\right)=\varrho_J\zeta_JQ\left(x\right),\qquad \frac{\left(\psi_2^J\left(x;\tau\right)\right)^2+2\left(\psi_3^J\left(x;\tau\right)\right)^2}{\tau}=\zeta_J^2Q\left(x\right)\end{equation*}
give the scaled conditional covariance matrix
\begin{equation*}\bm B_J\left(x\right)=\begin{pmatrix}x&\varrho_J\zeta_JQ\left(x\right)\\ \varrho_J\zeta_JQ\left(x\right)&\zeta_J^2Q\left(x\right)\end{pmatrix}.\end{equation*}
The conditional means are $\BigO{\tau}$ locally uniformly, so the scaled second moments have the same limits. On interior compact sets, the deterministic increments are $\BigO{\tau}$ and their stochastic loadings are $\BigO{\sqrt{\tau}}$. Gaussian moments therefore give the higher-moment condition exactly as in Appendix~\ref{app:p_weak_limit_proof}. Local well-posedness of the Jacobi system and the localized diffusion-approximation theorem yield the stated convergence. Equation~\eqref{sm:eq:jacobi_boundary_condition} is the endpoint nonattainment condition of \citet[Theorem~2.1]{Ackerer18}, with their long-run variance and variance diffusion parameters replaced by $\sigma^2$ and $\zeta_J$.
\end{proof}

The raw recursion need not preserve polynomials. For example, when $\varrho_J=0$, its centred variance innovation $U_x^{(n)}=\psi_3^J\left(x;\tau\right)\left[\left(\epsilon^{(n)}_{(k+1)\tau}\right)^2-1\right]$ satisfies
\begin{equation*}\E^{\Pm^{(n)}}\left[\left(U_x^{(n)}\right)^3\mid h^{(n)}_{k\tau}=x\right]=2\sqrt{2}\,\tau^{3/2}\zeta_J^3Q\left(x\right)^{3/2},\end{equation*}
which is non-polynomial for a genuine Jacobi quadratic. The conditional mean and variance are polynomial, so the third raw conditional moment is non-polynomial as well.

If one additionally wishes the discrete-time variance itself to remain in the bounded Jacobi state space, define
\begin{equation*}
\mathcal{P}_J(z)=\min\left\{h_{\max},\max\left\{h_{\min},z\right\}\right\},
\end{equation*}
and replace the variance recursion in Equation~\eqref{sm:eq:jacobi_garch_recursion} by
\begin{equation*}
\begin{aligned}
\widehat h^{(n)}_{(k+1)\tau}&=h^{(n)}_{k\tau}+\kappa\left(\sigma^2-h^{(n)}_{k\tau}\right)\tau+\psi_2^J\left(h^{(n)}_{k\tau};\tau\right)\epsilon^{(n)}_{(k+1)\tau}+\psi_3^J\left(h^{(n)}_{k\tau};\tau\right)\left[\left(\epsilon^{(n)}_{(k+1)\tau}\right)^2-1\right],\\
h^{(n)}_{(k+1)\tau}&=\mathcal{P}_J\left(\widehat h^{(n)}_{(k+1)\tau}\right).
\end{aligned}
\end{equation*}
Together with the same return recursion, this defines a boundary-preserving approximation on $[h_{\min},h_{\max}]$. Its local limiting characteristics are unchanged. Indeed, for a compact variance set $\mathcal{K}_h\subset(h_{\min},h_{\max})$, let $d>0$ be its distance from the endpoints. If $\widehat h^+_x$ is the unprojected update, then for $j=1,2$ and any $q>2$, $\left|\mathcal{P}_J\left(\widehat h^+_x\right)-\widehat h^+_x\right|^j\leq d^{j-q}\left|\widehat h^+_x-x\right|^q$. The higher-moment bound therefore gives
\begin{equation*}\lim_{\tau\to0}\sup_{x\in\mathcal{K}_h}\frac{1}{\tau}\E\left[\left|\mathcal{P}_J\left(\widehat h^+_x\right)-\widehat h^+_x\right|^j\right]=0.\end{equation*}
The first bound controls the drift error, while the second and Cauchy--Schwarz control the covariance errors. Projection is a contraction relative to the current variance state, so the higher-moment bound is retained. Under Condition~\eqref{sm:eq:jacobi_boundary_condition}, the limiting variance never reaches the endpoints. Localization can then be removed on every finite horizon: its continuous paths stay inside an interior compact interval with probability tending to one as the interval expands, and the return martingale has uniformly bounded second moments because $h^{(n)}\leq h_{\max}$. Thus the projected scheme converges weakly on $D([0,L],\mathbb{R}^2)$ to the same Jacobi diffusion.

The projected chain is also non-polynomial. For $0<\kappa\tau<1$, the conditional mean at either endpoint equals $\mu_\tau(x):=x+\kappa(\sigma^2-x)\tau$, since both innovation loadings vanish there. Any affine conditional mean on $[h_{\min},h_{\max}]$ must therefore equal $\mu_\tau$ everywhere. But for an interior $x$ sufficiently close to $h_{\max}$, the unprojected quadratic update has a minimum above $h_{\min}$ and is unbounded above, because $\psi_3^J(x;\tau)>0$. Lower clipping is then inactive and upper clipping strictly reduces its expectation, giving $\E[\mathcal{P}_J(\widehat h^+_x)]<\mu_\tau(x)$. Hence even first-degree preservation fails for the projected chain.

\bigskip

\subsection{Diffusion limits under the risk-neutral measure}
\label{subsec:risk_neutral_diffusion_limits}
\label{sm:proof_q_weak_limit}

\bigskip

We now derive the continuous-time limits of VPGARCH-SV and JPGARCH-SV under the risk-neutral measures $\Qm^{(n)}$ characterized in Appendix~\ref{subsec:pricing_kernel_q_dynamics}. We retain the parameter scaling and deterministic initial state of Section~\ref{subsec:physical_diffusion_limits} of the main paper, and impose an additional limit for the variance risk-preference process. Convergence is local in the same sense as in that subsection, in $D([0,L],\mathbb{R}^2)$ equipped with the Skorokhod $\mathrm{J}_1$ topology.

\begin{assumption}\label{ass:q_scaling}
Suppose that Assumption~\ref{ass:p_scaling} holds. Assume also that the variance risk-preference parameter is a time-homogeneous state function, $\xi_{2,k\tau}(\tau)=\xi_2(h^{(n)}_{k\tau};\tau)$, and that there exists a finite locally Lipschitz function $\xi_2:(0,\infty)\to\mathbb{R}$ such that, for every compact set $\mathcal{K}\subset(0,\infty)$,
\begin{equation*}\lim_{\tau\to0}\sup_{x\in\mathcal{K}}\left|\xi_2(x;\tau)-\xi_2(x)\right|=0.\end{equation*}
Assume further that the joint innovation cgf introduced in Appendix~\ref{subsec:pricing_kernel_q_dynamics} is finite on an open neighbourhood of the origin. The limiting variance risk-preference process is defined by $\xi_{2,t}=\xi_2(h_t)$.
\end{assumption}

Assumption~\ref{ass:q_scaling} supplements the physical diffusion scaling with a finite limit for the variance risk-preference process and the exponential-moment regularity used for the risk-neutral diffusion approximation. In particular, it requires $\E\left[\exp\left(c\epsilon_2^2\right)\right]<\infty$ for some $c>0$.\footnote{Proposition~\ref{prop:q_weak_limit} does not cover the NIG variance innovation specification used in our empirical analysis, even when $\xi_{2,t}\equiv0$. This exponential-moment requirement is stronger than the cgf-domain condition in Appendix~\ref{subsec:pricing_kernel_q_dynamics} and does not rule out an admissible NIG pricing kernel.} Define the instantaneous variance contribution orthogonal to the return shock as
\begin{equation*}v_{\mathsf{M}}^{\perp}(x)=\left(1-\varrho^2\right)\eta^2g_{\mathsf{M}}(x)+(m_4-1)\left(b_0+b_1x\right)^2.\end{equation*}

\begin{proposition}\label{prop:q_weak_limit}
Suppose the returns follow the VPGARCH-SV or JPGARCH-SV dynamics  under the risk-neutral measure $\Qm^{(n)}$ introduced in  Appendix~\ref{subsec:pricing_kernel_q_dynamics}, and the limiting constraints from  Assumptions~\ref{ass:p_scaling}  and~\ref{ass:q_scaling} hold. Then, as $\tau\to0$, their c\`adl\`ag extensions converge weakly, locally on $\mathbb{R}\times(0,\infty)$, to the diffusion $\bm{X}=(Y,h)^\top$ satisfying
\begin{equation*}\begin{aligned}[b]dY_t&=\left(r-\frac{1}{2}h_t\right)dt+\sqrt{h_t}\,dW^*_{1,t},\\ dh_t&=\left[\kappa\left(\sigma^2-h_t\right)-\varrho\eta\left(\lambda+\frac{1}{2}\right)\sqrt{h_tg_{\mathsf{M}}(h_t)}+\xi_2(h_t)v_{\mathsf{M}}^{\perp}(h_t)\right]dt+\sqrt{v_{\mathsf{M}}(h_t)}\,dW^*_{2,t}.\end{aligned}\end{equation*}
Here $W^*_1$ and $W^*_2$ are standard Brownian motions under $\Qm$, adapted to $\Ft$, and are stochastically correlated according to $d\left\langle W^*_1,W^*_2\right\rangle_t=\rho_{\mathsf{M}}(h_t)\,dt$, where $g_{\mathsf{M}}$, $v_{\mathsf{M}}$, and $\rho_{\mathsf{M}}$ are the same functions as in Proposition~\ref{prop:p_weak_limit}.
\end{proposition}

\begin{proof}
Under Assumption~\ref{ass:q_scaling}, we verify the risk-neutral counterparts of the local conditions in Appendix~\ref{app:p_weak_limit_proof}. All conditional moments depend only on the current variance $h^{(n)}_{k\tau}=x$, and all limits and remainder estimates below are uniform on arbitrary compact sets $\mathcal{K}_h\subset(0,\infty)$.

For $\mathsf{M}\in\{\mathsf{VP},\mathsf{JP}\}$, define
\begin{equation*}g_{\mathsf{VP}}(\tau,x)=a_0(\tau)+a_1(\tau)x+x^2, \qquad g_{\mathsf{JP}}(\tau,x)=x.\end{equation*}
Assumption~\ref{ass:p_scaling} implies, locally uniformly on $(0,\infty)$,
\begin{equation*}\lim_{\tau\to0}g_{\mathsf{M}}(\tau,x)=g_{\mathsf{M}}(x), \qquad \lim_{\tau\to0}\frac{\eta(\tau)}{\sqrt{\tau}}=\eta, \qquad \lim_{\tau\to0}\frac{b_0(\tau)+b_1(\tau)x}{\sqrt{\tau}}=b_0+b_1x.\end{equation*}

Write $\mathcal{C}(z_1,z_2,z_3)$ for the joint cgf of $\left(\epsilon_1,\epsilon_2,\epsilon_2^2\right)$ from Appendix~\ref{subsec:pricing_kernel_q_dynamics}. Write $A_i(x;\tau)$, $B(x;\tau)$, $P_i(x;\tau)$, and $R_i(x;\tau)$ for the corresponding state functions, suppressing their arguments where convenient. Since
\begin{equation*}A_2(x;\tau)=\xi_2(x;\tau)\eta(\tau)\sqrt{g_{\mathsf{M}}(\tau,x)}, \qquad B(x;\tau)=\xi_2(x;\tau)\left[b_0(\tau)+b_1(\tau)x\right],\end{equation*}
both $A_2$ and $B$ are $\BigO{\sqrt{\tau}}$. Put $u_\tau(x)=\sqrt{\tau x}$ and
\begin{equation*}F_\tau(a,x)=\frac{\mathcal{C}\left(a+u_\tau(x),A_2,B\right)-\mathcal{C}\left(a,A_2,B\right)}{u_\tau(x)}.\end{equation*}
On a fixed neighborhood of the origin, $\partial_{11}\mathcal{C}$ is bounded above and bounded away from zero, since $\partial_{11}\mathcal{C}(0,0,0)=1$. Moreover, $F_\tau(0,x)=\BigO{\sqrt{\tau}}$, and the martingale equation is
\begin{equation*}F_\tau\left(A_1(x;\tau),x\right)=-\frac{\lambda(\tau)x}{u_\tau(x)}=\BigO{\sqrt{\tau}}.\end{equation*}
The lower derivative bound gives a solution $A_1(x;\tau)=\BigO{\sqrt{\tau}}$. Strict convexity in the first cgf argument makes it the unique admissible solution, so the tilted coefficients are time-homogeneous state functions.

Joint symmetry gives $\E\left[\epsilon_1\epsilon_2^2\right]=\E\left[\epsilon_2^3\right]=0$. Taylor expansion of the first cgf derivatives at the origin therefore yields
\begin{equation*}P_1=A_1+\varrho A_2+o\left(\sqrt{\tau}\right), \qquad P_2=\varrho A_1+A_2+o\left(\sqrt{\tau}\right).\end{equation*}
Expanding the second derivatives, and using $\partial_{223}\mathcal{C}(0,0,0)=m_4-1$, gives
\begin{equation*}\lim_{\tau\to0}R_i(x;\tau)=1,\quad i=1,2, \qquad R_2=1+(m_4-1)B+o\left(\sqrt{\tau}\right).\end{equation*}
In particular, both $R_i$ are bounded above and bounded away from zero on every compact subset for sufficiently small $\tau$.

Expanding the martingale restriction in the first cgf argument gives
\begin{equation*}\sqrt{\tau x}\,P_1+\frac{1}{2}\tau xR_1+o(\tau)=-\lambda(\tau)x.\end{equation*}
Combining this with the preceding expansions yields
\begin{align*}\lim_{\tau\to0}\frac{P_1(x;\tau)}{\sqrt{\tau}}&=-\left(\lambda+\frac{1}{2}\right)\sqrt{x},\\ \lim_{\tau\to0}\frac{P_2(x;\tau)}{\sqrt{\tau}}&=-\varrho\left(\lambda+\frac{1}{2}\right)\sqrt{x}+\left(1-\varrho^2\right)\xi_2(x)\eta\sqrt{g_{\mathsf{M}}(x)},\end{align*}
and
\begin{equation*}\lim_{\tau\to0}\frac{R_2(x;\tau)-1}{\sqrt{\tau}}=(m_4-1)\xi_2(x)\left(b_0+b_1x\right).\end{equation*}

We now compute the limiting conditional drift. From Proposition~\ref{prop:general_q_dynamics},
\begin{equation*}\E^{\Qm^{(n)}}\left[\Delta Y^{(n)}_{(k+1)\tau}\mid h^{(n)}_{k\tau}=x\right]=r\tau+\lambda(\tau)x+\sqrt{\tau x}\,P_1(x;\tau),\end{equation*}
so
\begin{equation*}\lim_{\tau\to0}\frac{1}{\tau}\E^{\Qm^{(n)}}\left[\Delta Y^{(n)}_{(k+1)\tau}\mid h^{(n)}_{k\tau}=x\right]=r-\frac{1}{2}x.\end{equation*}
For the variance process,
\begin{equation*}\begin{aligned}\E^{\Qm^{(n)}}\left[\Delta h^{(n)}_{(k+1)\tau}\mid h^{(n)}_{k\tau}=x\right]={}&\left[1-\phi(\tau)\right]\left[\sigma^2(\tau)-x\right]+\eta(\tau)\sqrt{g_{\mathsf{M}}(\tau,x)}P_2(x;\tau)\\ &+\left[b_0(\tau)+b_1(\tau)x\right]\left[P_2(x;\tau)^2+R_2(x;\tau)-1\right].\end{aligned}\end{equation*}
Using the limits above gives
\begin{equation*}\begin{aligned}\lim_{\tau\to0}\frac{1}{\tau}\E^{\Qm^{(n)}}\left[\Delta h^{(n)}_{(k+1)\tau}\mid h^{(n)}_{k\tau}=x\right]={}&\kappa\left(\sigma^2-x\right)-\varrho\eta\left(\lambda+\frac{1}{2}\right)\sqrt{xg_{\mathsf{M}}(x)}+\xi_2(x)v_{\mathsf{M}}^{\perp}(x).\end{aligned}\end{equation*}

The cgf condition in Assumption~\ref{ass:q_scaling} implies that $\E\left[\exp\left(2c\left[|\epsilon_1|+|\epsilon_2|+\epsilon_2^2\right]\right)\right]<\infty$ for some $c>0$. Since $A_1,A_2,B$ converge uniformly to zero on $\mathcal{K}_h$, the normalized tilted densities are bounded there by $C\exp\left(c\left[|\epsilon_1|+|\epsilon_2|+\epsilon_2^2\right]\right)$ for sufficiently small $\tau$. This envelope controls every fixed polynomial joint moment. Dominated convergence therefore gives locally uniform convergence of all joint innovation moments to their physical values and uniform bounds on their absolute moments. The same holds after standardization because $P_i$ converge to zero and $R_i$ converge to one.
Applying the original increment formulas in Appendix~\ref{app:p_weak_limit_proof}, with $\epsilon_1$ driving returns and $\epsilon_2$ driving variance, the scaled conditional covariance matrix converges locally uniformly to $\bm{B}_{\mathsf{M}}(x)$ in that appendix. The conditional means are $\BigO{\tau}$ locally uniformly, so the scaled conditional second moments have the same limits.

The deterministic parts of the original increments are $\BigO{\tau}$ and all stochastic loadings are $\BigO{\sqrt{\tau}}$ locally uniformly. The uniform moment bounds thus give, for every $q>2$,
\begin{equation*}\lim_{\tau\to0}\sup_{x\in\mathcal{K}_h}\frac{1}{\tau}\E^{\Qm^{(n)}}\left[\left|\Delta Y^{(n)}_{(k+1)\tau}\right|^q+\left|\Delta h^{(n)}_{(k+1)\tau}\right|^q\mid h^{(n)}_{k\tau}=x\right]=0.\end{equation*}
Thus, the higher-moment and no-large-jumps conditions hold.

The limiting covariance matrix admits the same factorization as in the proof of Proposition~\ref{prop:p_weak_limit}. The additional terms in the risk-neutral variance drift are locally Lipschitz on $(0,\infty)$ under Assumption~\ref{ass:q_scaling}, while the diffusion coefficients have the same local regularity as under $\Pm$. Consequently, the martingale problem is well posed up to the lifetime in $\mathbb{R}\times(0,\infty)$. The local moment bounds, including control of exit overshoots, give convergence along regular localizing domains to the diffusion in Proposition~\ref{prop:q_weak_limit}.

Finally,
\begin{equation*}h^{*\,(n)}_{k\tau}=h^{(n)}_{k\tau}R_1\left(h^{(n)}_{k\tau};\tau\right), \qquad \lim_{\tau\to0}\sup_{x\in\mathcal{K}_h}\left|xR_1(x;\tau)-x\right|=0.\end{equation*}
Hence, under the same localization, the transformation $x\mapsto xR_1(x;\tau)$ converges uniformly to the identity. It follows from the local weak convergence of $h^{(n)}$ that $h^{*\,(n)}$ converges weakly under $\Qm^{(n)}$ to the same limiting variance state $h$.
\end{proof}

Although the standardized risk-neutral innovation pair need not be jointly symmetric for fixed $\tau$, its joint moments converge locally uniformly to their physical values. In particular, the third and fourth moments of $\epsilon^{*\,(n)}_{2,(k+1)\tau}$ converge to $0$ and $m_4$, respectively, and the innovation correlation converges to $\varrho$. Hence, the finite-step asymmetry induced by the pricing kernel vanishes at the diffusion scale.

The proof also gives $\lim_{\tau\to0}R_1(x;\tau)=1$ locally uniformly, so $h^{*\,(n)}$ and $h^{(n)}$ have the same local limiting variance state. The covariance characteristics are consequently the same under the physical and risk-neutral limits. Under the tangency conditions in Supplementary Material~\ref{sm:variance_limit_boundary_conditions}, the bound $\xi_2(x)v_{\mathsf{M}}^{\perp}(x)=\BigO{x}$ as $x\downarrow0$ preserves the zero-boundary classification; bounded $\xi_2$ near zero is sufficient. The weaker bound $\xi_2(x)=\BigO{x^{-1}}$ need not suffice in the hybrid setting, because the variance-risk adjustment can change the boundary drift. None of these local conditions controls explosion at infinity, which requires an additional growth or boundary test.

The discounted price associated with the limiting return equation is a local martingale. Local weak convergence from the discrete-time risk-neutral models does not, by itself, establish a true martingale on the full horizon or global equivalence of the limiting measures. Such conclusions require the corresponding nonexplosion and martingale conditions. No convergence result for unbounded derivative payoffs is asserted here.

The following conditions characterize polynomiality of the limiting operators; the corresponding processes are polynomial when the state-space and existence requirements are satisfied.  For VPGARCH-SV, the marginal variance operator is polynomial if and only if its drift in Proposition~\ref{prop:q_weak_limit} is affine, since $v_{\mathsf{VP}}$ is quadratic. Thus, the combined drift adjustment
\begin{equation*}-\varrho\eta\left(\lambda+\frac{1}{2}\right)\sqrt{xg_{\mathsf{VP}}(x)}+\xi_2(x)v_{\mathsf{VP}}^{\perp}(x)\end{equation*}
must be affine in $x$. Joint polynomiality additionally requires $\varrho\eta=0$, as under the physical measure. For JPGARCH-SV, the leverage contribution is linear for every $\lambda$, because $g_{\mathsf{JP}}(x)=x$. Its joint operator is therefore polynomial if and only if $\xi_2(x)v_{\mathsf{JP}}^{\perp}(x)$ is affine.

In the jointly Gaussian case, imposing the discrete-time polynomial-preserving restrictions of Appendix~\ref{subsec:q_polynomial_preservation} at every time step makes each scaled conditional variance drift affine. Under Assumption~\ref{ass:q_scaling}, its locally uniform limit in Proposition~\ref{prop:q_weak_limit} is therefore affine as well. Together with the covariance conditions above, this verifies inheritance of the corresponding polynomial operator in both model classes, including zero scaled loadings and the independent-innovation case $\varrho=0$.

For VPGARCH-SV with $\varrho>0$, the restrictions of Proposition~\ref{prop:q_polynomial_preservation} imply $\lambda=-1/2$, since $\lim_{\tau\to0}R_1(x;\tau)=1$. The non-polynomial leverage term then disappears. This restriction has a direct economic interpretation: under $\Pm$, it implies $dS_t/S_t=r\,dt+\sqrt{h_t}\,dW_{1,t}$, so the equity risk premium vanishes and the return drift coincides with its risk-neutral counterpart. It does not, however, imply $\Pm=\Qm$, since variance risk may still be priced through $\xi_2(h_t)v_{\mathsf{VP}}^{\perp}(h_t)$. If this adjustment also vanishes, in particular when $\xi_2\equiv0$, the limiting state equations coincide and determine the same state-process law whenever the martingale problem is well posed.

The measure change thus preserves the local covariance structure but can destroy polynomiality through the drift, as in \citet{filipovic2016,Ackerer18}. The limiting requirements are weaker than their discrete-time counterparts: $\lambda$ does not restrict joint polynomiality of JPGARCH-SV, and a constant $\xi_2$ suffices for HNGARCH-SV because $b_1=0$ makes $v_{\mathsf{JP}}^{\perp}$ affine. VPGARCH-SV instead requires cancellation of a non-polynomial leverage term when $\varrho\eta\left(\lambda+1/2\right)\neq0$.

\bigskip

\section{Higher-order Hermite extensions}
\label{sm:sec:higher_order_hermite}

\bigskip

This section considers jointly Gaussian hybrid GARCH--SV dynamics with $\varrho\in\left[0,1\right]$ and provides a higher-order counterpart of Proposition~\ref{prop:garch_polynomial_characterization} for variance recursions driven by Hermite polynomials through order $p$, under the additional assumption that the loadings on the even-order Hermite polynomials are polynomial. For every integer $j\geq0$, define
\begin{equation*}H_j\left(z\right):=\left(-1\right)^j\exp\left(\frac{z^2}{2}\right)\frac{d^j}{dz^j}\exp\left(-\frac{z^2}{2}\right).\end{equation*}
Thus, $H_0\left(z\right)=1$, $H_1\left(z\right)=z$, $H_2\left(z\right)=z^2-1$, and $H_3\left(z\right)=z^3-3z$, with $H_j\left(-z\right)=\left(-1\right)^jH_j\left(z\right)$. If $\epsilon\overset{\Pm^{(n)}}{\sim}N\left(0,1\right)$, then
\begin{equation*}\E^{\Pm^{(n)}}\left[H_j\left(\epsilon\right)\right]=0\ \text{for every }j\geq1, \qquad \E^{\Pm^{(n)}}\left[H_i\left(\epsilon\right)H_j\left(\epsilon\right)\right]=j!\,\mathbf{1}_{\left\{i=j\right\}}.\end{equation*}

Let $p\geq2$ and define
\begin{equation*}\mathcal{O}_p:=\left\{j\in\left\{1,\ldots,p\right\}:j\text{ is odd}\right\}, \qquad \mathcal{E}_p:=\left\{j\in\left\{1,\ldots,p\right\}:j\text{ is even}\right\}.\end{equation*}
For every $j\in\mathcal{O}_p$, let $c_j\left(\tau\right)\in\mathbb{R}$, with the normalization $c_1\left(\tau\right)=1$. Consider the dynamics
\begin{align*}
y^{(n)}_{(k+1)\tau}&=\psi_0\left(h^{(n)}_{k\tau};\tau\right)+\sqrt{\tau h^{(n)}_{k\tau}}\,\epsilon^{(n)}_{1,(k+1)\tau}, \\
h^{(n)}_{(k+1)\tau}&=\psi_1\left(h^{(n)}_{k\tau};\tau\right)+\psi_2\left(h^{(n)}_{k\tau};\tau\right)\sum_{j\in\mathcal{O}_p}c_j\left(\tau\right)H_j\left(\epsilon^{(n)}_{2,(k+1)\tau}\right)+\sum_{j\in\mathcal{E}_p}\psi_{j+1}\left(h^{(n)}_{k\tau};\tau\right)H_j\left(\epsilon^{(n)}_{2,(k+1)\tau}\right),
\end{align*}
where
\begin{equation*}\left(\epsilon^{(n)}_{1,(k+1)\tau},\epsilon^{(n)}_{2,(k+1)\tau}\right)^\top\mid\F^{(n)}_{k\tau}\overset{\Pm^{(n)}}{\sim}N_2\left(\bm{0},\begin{pmatrix}1&\varrho\\\varrho&1\end{pmatrix}\right), \qquad \varrho\in\left[0,1\right].\end{equation*}
The odd-order terms share the loading function $\psi_2$, with $c_j\left(\tau\right)$ determining their relative contributions, whereas each even-order term has its own loading function. The normalization $c_1\left(\tau\right)=1$ identifies $\psi_2$ as the loading on $H_1$. When $p=2$, the preceding dynamics reduce to the jointly Gaussian specialization of Equations~\eqref{eq:general_return_dynamics_p}--\eqref{eq:general_variance_dynamics_p}. The characterization below concerns polynomial preservation; positivity remains a separate requirement. As in the main paper, for each fixed $\tau$ we suppress the second argument of the loading functions in polynomial-space statements.

\begin{proposition}
\label{sm:prop:higher_order_hermite_characterization}
Suppose that the variance recursion above preserves $\Rplus$ and that $\psi_{j+1}\in\Pol\left(\Rplus\right)$ for every $j\in\mathcal{E}_p$.

\begin{enumerate}[label=\textnormal{(\roman*)}]
\item The process $h^{(n)}$ is polynomial preserving on $\Rplus$ if and only if
\begin{equation*}\psi_1\in\Pol_1\left(\Rplus\right), \qquad \psi_2^2\in\Pol_2\left(\Rplus\right), \qquad \psi_{j+1}\in\Pol_1\left(\Rplus\right)\ \text{for every }j\in\mathcal{E}_p.\end{equation*}

\item The process $\left(y^{(n)},h^{(n)}\right)$ is polynomial preserving on $\mathbb{R}\times\Rplus$ if
\begin{equation*}\psi_0,\psi_1\in\Pol_1\left(\Rplus\right), \qquad \psi_{j+1}\in\Pol_1\left(\Rplus\right)\ \text{for every }j\in\mathcal{E}_p, \qquad \psi_2\in\operatorname{span}\left\{x\mapsto\sqrt{x}\right\}.\end{equation*}
These conditions are also necessary when $\varrho>0$. When $\varrho=0$, joint preservation is equivalent to the conditions in part~\textnormal{(i)} together with $\psi_0\in\Pol_1\left(\Rplus\right)$.
\end{enumerate}
\end{proposition}

\begin{proof}
Let $y_x^+$ and $h_x^+$ denote the one-step updates from variance state $x$ with innovation pair $\left(\epsilon_1,\epsilon_2\right)$, and suppress $\tau$ in the loading functions and parameters. The first moment of $h_x^+$ is $\psi_1\left(x\right)$, and Hermite orthogonality gives
\begin{equation*}\Var^{\Pm^{(n)}}\left[h_x^+\right]=\kappa_p\psi_2\left(x\right)^2+\sum_{j\in\mathcal{E}_p}j!\,\psi_{j+1}\left(x\right)^2,\qquad \kappa_p:=\sum_{j\in\mathcal{O}_p}c_j^2j!\geq1.\end{equation*}
Under marginal polynomial preservation, $\psi_1$ is affine and this variance belongs to $\Pol_2\left(\Rplus\right)$. Each squared even-order loading is bounded by a multiple of that nonnegative quadratic. Since the even-order loadings are polynomial, they are affine, and subtraction gives $\psi_2^2\in\Pol_2\left(\Rplus\right)$. Conversely, the parity argument in Appendix~\ref{app:proof_garch_polynomial_characterization} applies with the odd polynomial $\sum_{j\in\mathcal{O}_p}c_jH_j$ and the even polynomials $H_j$, $j\in\mathcal{E}_p$: only even powers of $\psi_2$ survive in each variance moment, and each surviving term in the $q$th moment has degree at most $q$. This proves part~\textnormal{(i)}.

Joint preservation requires the conditions in part~\textnormal{(i)} and $\psi_0\in\Pol_1\left(\Rplus\right)$. For every $\varrho\in\left[0,1\right]$, Gaussian conditioning and Hermite orthogonality give
\begin{equation*}\E^{\Pm^{(n)}}\left[\epsilon_1H_j\left(\epsilon_2\right)\right]=\varrho\E^{\Pm^{(n)}}\left[\epsilon_2H_j\left(\epsilon_2\right)\right]=\varrho\mathbf{1}_{\left\{j=1\right\}},\qquad j\geq1.\end{equation*}
Since $c_1=1$, it follows that
\begin{equation*}\E^{\Pm^{(n)}}\left[y_x^+h_x^+\right]=\psi_0\left(x\right)\psi_1\left(x\right)+\varrho\sqrt{\tau x}\,\psi_2\left(x\right).\end{equation*}
When $\varrho>0$, joint preservation therefore implies $q\left(x\right):=\sqrt{x}\,\psi_2\left(x\right)\in\Pol_2\left(\Rplus\right)$. Since $q^2=x\psi_2^2\in\Pol_3\left(\Rplus\right)$ and $q\left(0\right)=0$, $q\left(x\right)=cx$ for some $c\in\mathbb{R}$. Hence $\psi_2\left(x\right)=c\sqrt{x}$ for $x>0$, and the polynomial identity $\psi_2^2\left(x\right)=c^2x$ gives the same conclusion at zero.

For joint sufficiency, the parity and degree argument in Appendix~\ref{app:proof_garch_polynomial_characterization} again applies. In the expansion of $\left(y_x^+\right)^a\left(h_x^+\right)^b$, a term with $r$ return innovations and $s$ odd Hermite factors vanishes when $r+s$ is odd. Otherwise, its state-dependent coefficient has degree at most $a+b-\left(r+s\right)/2\leq a+b$, because the remaining loadings are affine and $\psi_2\left(x\right)=c\sqrt{x}$. This proves sufficiency for every $\varrho\in\left[0,1\right]$.

Finally, when $\varrho=0$, Gaussianity implies independence, so the mixed moments factor into return and variance moments. Under part~\textnormal{(i)} and affine $\psi_0$, these have degrees at most $a$ and $b$, respectively. The converse follows from the marginal moments, completing part~\textnormal{(ii)}.
\end{proof}

\bigskip

\clearpage

\section{Two-component polynomial volatility specifications}
\label{sm:sec:two_component_extension}

\bigskip

We extend the two canonical specifications using the short- and long-run variance decomposition of \citet{christoffersen2008option,chris10b}, in the centred parametrization of \citet{badescu2024discrete}. Write $h^{(n)}_{k\tau}=s^{(n)}_{k\tau}+q^{(n)}_{k\tau}$, where $s^{(n)}$ captures transitory deviations from the more persistent component $q^{(n)}$. The short-run component reverts to zero and may be negative; it is not a separate nonnegative variance process.

Fix $\tau$ and suppress its argument in the parameters. Retain the filtration and innovation law of Section~\ref{subsec:polynomial_garch_models_p}. Under $\Pm^{(n)}$, the innovation pairs $(\epsilon^{(n)}_{1,(k+1)\tau},\epsilon^{(n)}_{2,(k+1)\tau})$ are i.i.d. over $k$ and independent of $\F^{(n)}_{k\tau}$. A generic pair satisfies $(\epsilon_1,\epsilon_2)\overset{\mathcal D}{=}(-\epsilon_1,-\epsilon_2)$, has zero-mean, unit-variance marginals with finite absolute moments of all orders, and obeys $\E^{\Pm^{(n)}}[\epsilon_1\epsilon_2]=\varrho\in[0,1]$ and $\E^{\Pm^{(n)}}[\epsilon_2^4]>1$. Gaussianity is not required, and the marginal laws need not coincide. Take $(s^{(n)}_0,q^{(n)}_0)$ to be deterministic. For $\mathsf M\in\{\mathsf{VP},\mathsf{JP}\}$, consider
\begin{align}
y^{(n)}_{(k+1)\tau}&=r\tau+\lambda h^{(n)}_{k\tau}+\sqrt{\tau h^{(n)}_{k\tau}}\,\epsilon^{(n)}_{1,(k+1)\tau},\label{sm:eq:two_component_returns}\\
\Delta s^{(n)}_{(k+1)\tau}&=-(1-\phi_s)s^{(n)}_{k\tau}+\nu^{(n)}_{s,(k+1)\tau},\label{sm:eq:two_component_short}\\
\Delta q^{(n)}_{(k+1)\tau}&=(1-\phi_q)\left(\sigma^2-q^{(n)}_{k\tau}\right)+\nu^{(n)}_{q,(k+1)\tau},\label{sm:eq:two_component_long}
\end{align}
where $\Delta u^{(n)}_{(k+1)\tau}=u^{(n)}_{(k+1)\tau}-u^{(n)}_{k\tau}$ for $u\in\{s,q\}$, $0<\phi_s<\phi_q<1$, $\sigma^2>0$, and
\begin{equation}\label{sm:eq:two_component_noise}\nu^{(n)}_{u,(k+1)\tau}=\eta_u\sqrt{g_{\mathsf M}\!\left(h^{(n)}_{k\tau}\right)}\,\epsilon^{(n)}_{2,(k+1)\tau}+\left(b_{0,u}+b_{1,u}h^{(n)}_{k\tau}\right)\left[\left(\epsilon^{(n)}_{2,(k+1)\tau}\right)^2-1\right],\qquad u\in\{s,q\}.\end{equation}
Here $g_{\mathsf{VP}}(x)=a_0+a_1x+x^2$ and $g_{\mathsf{JP}}(x)=x$. Both components share the function $g_{\mathsf M}$ and the variance innovation, with component-specific coefficients. The component shocks $\nu^{(n)}_{s,(k+1)\tau}$ and $\nu^{(n)}_{q,(k+1)\tau}$ have conditional mean zero but need not be Gaussian or symmetric, even when the underlying innovation pair is Gaussian. The Markov variance state is $(s^{(n)},q^{(n)})$; total variance alone need not be Markov.

As in the Hermite extension, we distinguish polynomial preservation from state-space admissibility. We consider any Borel state space $E\subseteq\{(s,q)\in\mathbb R^2:q>0,\ s+q>0\}$ preserved by the component recursions, with $g_{\mathsf M}(s+q)\geq0$ on $E$. We do not impose separate nonnegativity on $s$ or derive coefficient restrictions ensuring this invariance. Positivity of $q$ alone does not ensure positivity of $h=s+q$.

With jointly Gaussian innovations and $\varrho=1$, the Heston--Nandi component variance dynamics of \citet{christoffersen2008option} are recovered under $\mathsf M=\mathsf{JP}$ by setting $b_{0,u}=\alpha_u$, $b_{1,u}=0$, and $\eta_u=-2\alpha_u\gamma_u$ for $u\in\{s,q\}$. The NGARCH component variance dynamics of \citet{chris10b} are recovered under $\mathsf M=\mathsf{VP}$ by setting $a_0=a_1=0$, $b_{0,u}=0$, $b_{1,u}=\alpha_u$, and $\eta_u=-2\alpha_u\gamma_u$. These reductions concern the variance recursions; the return mean remains the affine specification in Equation~\eqref{sm:eq:two_component_returns}.

\begin{proposition}
\label{sm:prop:two_component_preservation}
Under the preceding assumptions, $(s^{(n)},q^{(n)})$ is polynomial preserving on $E$ for both specifications. Under $\mathsf M=\mathsf{JP}$, $(y^{(n)},s^{(n)},q^{(n)})$ is polynomial preserving on $\mathbb R\times E$. The latter property also holds under $\mathsf M=\mathsf{VP}$ when the return and variance innovations are independent.
\end{proposition}

\begin{proof}
Fix $(s,q)\in E$, write $h=s+q$, and denote one-step updates by $s^+$, $q^+$, and $y^+$. The conditional means of $s^+$ and $q^+$ and the centred-quadratic innovation loadings are affine in $(s,q)$. In the expansion of $(s^+)^{p_s}(q^+)^{p_q}$, a term containing $j$ linear innovation factors has a factor $g_{\mathsf M}(h)^{j/2}\epsilon_2^j$, with the other innovation factors even in $\epsilon_2$. Symmetry eliminates odd $j$. For even $j$, the coefficient is a product of $g_{\mathsf M}(h)^{j/2}$ and at most $p_s+p_q-j$ affine factors. Since $\deg g_{\mathsf M}\leq2$, its degree is at most $p_s+p_q$. The innovation moment assumptions give finite absolute moments at every state, proving variance-state preservation.

For joint moments under $\mathsf M=\mathsf{JP}$, expand $(y^+)^a(s^+)^{p_s}(q^+)^{p_q}$. A term with $r$ return innovation factors and $j$ linear variance innovation factors vanishes when $r+j$ is odd, by joint symmetry. Otherwise, its square-root factors combine into $h^{(r+j)/2}$, giving degree at most $a+p_s+p_q-(r+j)/2$. Under independent innovations, the return and variance-state moments instead factor, with degrees at most $a$ and $p_s+p_q$, respectively. This also proves joint preservation for $\mathsf M=\mathsf{VP}$ in that case.
\end{proof}

Repeated conditioning in Equations~\eqref{sm:eq:two_component_short}--\eqref{sm:eq:two_component_long} gives
\begin{align*}
\E^{\Pm^{(n)}}\!\left[s^{(n)}_{(k+\ell)\tau}\mid\F^{(n)}_{k\tau}\right]&=\phi_s^\ell s^{(n)}_{k\tau},\\
\E^{\Pm^{(n)}}\!\left[q^{(n)}_{(k+\ell)\tau}\mid\F^{(n)}_{k\tau}\right]&=\sigma^2+\phi_q^\ell\left(q^{(n)}_{k\tau}-\sigma^2\right),\\
\E^{\Pm^{(n)}}\!\left[h^{(n)}_{(k+\ell)\tau}\mid\F^{(n)}_{k\tau}\right]&=\sigma^2+\phi_s^\ell s^{(n)}_{k\tau}+\phi_q^\ell\left(q^{(n)}_{k\tau}-\sigma^2\right),
\end{align*}
for $\ell\in\mathbb N_0$ with $k+\ell\leq nL$. Thus, the expected-variance term structure has two persistence rates. If the same chain is extended to all $k\geq0$, the limiting means of $s^{(n)}$, $q^{(n)}$, and $h^{(n)}$ are $0$, $\sigma^2$, and $\sigma^2$, respectively. Initializing $s^{(n)}_0=0$ and $q^{(n)}_0=\sigma^2$ gives these means at every date. The matrix formulas in Section~\ref{subsec:general_framework} apply to the enlarged state, including moments of cumulative variance and, under joint preservation, cumulative returns. Preservation under a pricing measure requires separate conditions.

\bigskip

\section{Polynomial preservation in the risk-neutral variance state}
\label{sm:q_preservation_hstar}

\bigskip

This section characterizes polynomial preservation directly in the risk-neutral variance state $h^{*\,(n)}$ under jointly Gaussian innovations with correlation $\varrho\in\left[0,1\right]$. We retain the positivity assumptions of Appendix~\ref{subsec:q_polynomial_preservation}, but allow $R_2\left(x;\tau\right)$ to diverge at zero. Suppose that $\xi_{2,k\tau}\left(\tau\right)=\xi_2\left(h^{(n)}_{k\tau};\tau\right)$. The risk-neutral variance state is $h^{*\,(n)}_{k\tau}=\Gamma\left(h^{(n)}_{k\tau};\tau\right)$, where
\begin{equation}\Gamma\left(x;\tau\right)=xR_1\left(x;\tau\right)=x\left[1-\varrho^2+\frac{\varrho^2}{1-2\xi_2\left(x;\tau\right)\psi_3\left(x;\tau\right)}\right],\qquad x>0.\label{eq:risk_neutral_variance_transformation}\end{equation}
Whenever $\Gamma\left(\cdot;\tau\right)$ is one-to-one, $h^{*\,(n)}$ remains a time-homogeneous Markov state, and the return dynamics take the form
\begin{equation*}y^{(n)}_{(k+1)\tau}=r\tau-\frac{\tau}{2}h^{*\,(n)}_{k\tau}+\sqrt{\tau h^{*\,(n)}_{k\tau}}\,\epsilon^{*\,(n)}_{1,(k+1)\tau}.\end{equation*}
To retain the innovation-loading form of the variance recursion, we consider the affine case
\begin{equation*}\Gamma\left(x;\tau\right)=c_0\left(\tau\right)+c_1\left(\tau\right)x,\qquad c_0\left(\tau\right)\geq0,\quad c_1\left(\tau\right)>0.\end{equation*}
For $\varrho>0$, set $a=1-\varrho^2$ and $\delta\left(\tau\right)=c_1\left(\tau\right)-a$. Admissibility requires $\delta\left(\tau\right)\geq0$, with strict inequality when $c_0\left(\tau\right)=0$. Equation~\eqref{eq:risk_neutral_variance_transformation} then gives
\begin{equation}\xi_2\left(x;\tau\right)=\frac{c_0\left(\tau\right)+\left[c_1\left(\tau\right)-1\right]x}{2\psi_3\left(x;\tau\right)\left[c_0\left(\tau\right)+\delta\left(\tau\right)x\right]},\qquad x>0.\label{sm:eq:affine_risk_neutral_variance_xi2}\end{equation}
In particular,
\begin{equation*}R_1\left(x;\tau\right)=c_1\left(\tau\right)+\frac{c_0\left(\tau\right)}{x},\qquad R_2\left(x;\tau\right)=\frac{c_0\left(\tau\right)+\delta\left(\tau\right)x}{\varrho^2x}>0.\end{equation*}
If $c_0\left(\tau\right)=0$, both ratios are constant and Proposition~\ref{prop:q_polynomial_preservation} applies with $R_1\left(\tau\right)=c_1\left(\tau\right)$ and $R_2\left(\tau\right)=\delta\left(\tau\right)/\varrho^2$. The following proposition characterizes the additional cases with $c_0\left(\tau\right)>0$.

\begin{proposition}
\label{sm:prop:q_polynomial_preservation_hstar}
Suppose that $0<\varrho\leq1$, that the pricing kernel in Equation~\eqref{eq:pricing_kernel} satisfies the no-arbitrage conditions in Equation~\eqref{eq:pricing_kernel_no_arbitrage}, and that $h^{(n)}$ takes values in $\left(0,\infty\right)$. Let the variance risk-preference parameter be given by Equation~\eqref{sm:eq:affine_risk_neutral_variance_xi2}, with $c_0\left(\tau\right)>0$, $c_1\left(\tau\right)>0$, and $\delta\left(\tau\right)\geq0$. The risk-neutral variance state space is $E_\tau^*=\left(c_0\left(\tau\right),\infty\right)$.
\begin{enumerate}[label=\textnormal{(\roman*)}]
\item For VPGARCH-SV, $h^{*\,(n)}$ is polynomial preserving on $E_\tau^*$ under $\Qm^{(n)}$ if and only if
\begin{equation*}b_0\left(\tau\right)=\eta\left(\tau\right)=0,\qquad \lambda\left(\tau\right)=-\frac{\tau}{2}\delta\left(\tau\right),\qquad a\delta\left(\tau\right)=0.\end{equation*}
\item For JPGARCH-SV, $\left(y^{(n)},h^{*\,(n)}\right)$ is jointly polynomial preserving on $\mathbb{R}\times E_\tau^*$ under $\Qm^{(n)}$ if and only if $b_0\left(\tau\right)=0$ and
\begin{equation*}\delta\left(\tau\right)\left[\lambda\left(\tau\right)+\frac{\tau}{2}c_1\left(\tau\right)\right]=0,\end{equation*}
\begin{equation*}a\eta\left(\tau\right)+\frac{2\varrho b_1\left(\tau\right)c_0\left(\tau\right)}{\sqrt{\tau}}\left[\lambda\left(\tau\right)+\frac{\tau}{2}\delta\left(\tau\right)\right]=0.\end{equation*}
\end{enumerate}
\end{proposition}

\begin{proof}
Fix $\tau$ and suppress it in the state functions and parameters. Write $v=c_0+c_1x$, $w=c_0+\delta x$, $s\left(x\right)=b_0+b_1x$, and $u\left(x\right)=\psi_2\left(x\right)$. The invertible affine transformation $x\mapsto v$ preserves polynomial degree. The Gaussian martingale restriction gives
\begin{equation*}P_1\left(x\right)=-d\sqrt{x}-\frac{q}{\sqrt{x}},\qquad d=\frac{\lambda}{\sqrt{\tau}}+\frac{\sqrt{\tau}}{2}c_1,\qquad q=\frac{\sqrt{\tau}}{2}c_0>0.\end{equation*}
The formulas in Supplementary Material~\ref{app:proof_q_polynomial_preservation} apply with $R=R_2=w/\left(\varrho^2x\right)$ and $K=R_1=v/x$. In either model class, polynomial preservation requires $\psi_1^*$ and $\psi_3^*=R_2s$ to be affine and $\left(\psi_2^*\right)^2$ to have degree at most two. Since $c_0>0$, affineness of $R_2s$ forces $b_0=0$, and the maintained restrictions then give $b_1>0$.

For VPGARCH-SV, put $g\left(x\right)=a_0+a_1x+x^2$. If $\eta\neq0$, the only possibly nonrational term in $\psi_1^*$ is a nonzero rational multiple of $\sqrt{xg\left(x\right)}$. Its coefficient is proportional to $d+q/x$, which cannot vanish identically since $q>0$. The square root of a cubic polynomial cannot agree with a rational function on an open interval, so $\eta=0$ is necessary. The conditional mean then becomes
\begin{equation*}\psi_1^*\left(x\right)=\psi_1\left(x\right)+\frac{b_1}{\varrho^2}w-b_1x+\frac{b_1}{\varrho^2}\left[\frac{w\left(q+dx\right)}{v}\right]^2.\end{equation*}
A rational square that is affine must be constant. Evaluating at zero and comparing coefficients therefore gives $w\left(q+dx\right)=qv$, or equivalently $d=\sqrt{\tau}a/2$ and $a\delta=0$. These are precisely the restrictions on $\lambda$ and $\delta$ in part~\textnormal{(i)}. Conversely, under these restrictions,
\begin{equation*}\psi_1^*=\psi_1+\frac{b_1}{\varrho^2}w-b_1x+\frac{b_1q^2}{\varrho^2},\qquad \left(\psi_2^*\right)^2=\frac{4b_1^2q^2}{\varrho^4}w,\qquad \psi_3^*=\frac{b_1}{\varrho^2}w.\end{equation*}
The Gaussian marginal characterization in Proposition~\ref{prop:garch_polynomial_characterization} proves sufficiency.

For JPGARCH-SV, define $H\left(x\right)=\eta-2\varrho b_1q-2\varrho b_1dx$. The transformed coefficients satisfy
\begin{equation*}\psi_1^*\left(x\right)=\psi_1\left(x\right)+\frac{b_1}{\varrho^2}w-b_1x+\frac{1}{4b_1}\left[\frac{w^2H\left(x\right)^2}{\varrho^4v^2}-\eta^2\right].\end{equation*}
Thus, affineness of the conditional mean forces $wH/v$ to be a constant $C$. Comparing coefficients in $wH=Cv$ gives
\begin{equation*}\delta d=0,\qquad -2\varrho b_1c_0d=a\left(\eta-2\varrho b_1q\right),\end{equation*}
which are equivalent to the restrictions in part~\textnormal{(ii)}.

To prove joint sufficiency without assuming a constant risk-neutral correlation, write $U=\sqrt{\tau v}\,\epsilon^{*\,(n)}_{1,(k+1)\tau}$ and $Z=\sqrt{w}\,\epsilon^{*\,(n)}_{2,(k+1)\tau}$ for the centred conditional Gaussian shocks at state $x$. Their covariance entries are
\begin{equation*}\Var^{\Qm^{(n)}}\left[U\mid x\right]=\tau v,\qquad \Cov^{\Qm^{(n)}}\left[U,Z\mid x\right]=\sqrt{\tau}\,w,\qquad \Var^{\Qm^{(n)}}\left[Z\mid x\right]=w.\end{equation*}
Moreover,
\begin{equation*}y^{(n)}_{(k+1)\tau}=r\tau-\frac{\tau}{2}v+U,\qquad h^{(n)}_{(k+1)\tau}=\psi_1^*\left(x\right)+\frac{C}{\varrho^3}Z+\frac{b_1}{\varrho^2}\left(Z^2-w\right).\end{equation*}
The conditional means and Gaussian covariance entries are affine in $x$. Gaussian moment expansion therefore makes each conditional mixed moment of total degree $m$ a polynomial of degree at most $m$ in $x$, and hence in $v$. The affine transformation of the next variance state preserves the same degree bound, proving joint polynomial preservation.
\end{proof}

When $\varrho=1$, the proposition recovers the Gaussian GARCH restrictions: $b_0\left(\tau\right)=0$ and $\lambda\left(\tau\right)=-\tau c_1\left(\tau\right)/2$ in both classes, with $\eta\left(\tau\right)=0$ additionally required for VPGARCH. For $0<\varrho<1$, VPGARCH-SV permits $c_0\left(\tau\right)>0$ only when $b_0\left(\tau\right)=\eta\left(\tau\right)=\lambda\left(\tau\right)=0$ and $c_1\left(\tau\right)=1-\varrho^2$. In JPGARCH-SV with $0<\varrho<1$, if $\delta\left(\tau\right)>0$, the conditions reduce to $\lambda\left(\tau\right)=-\tau c_1\left(\tau\right)/2$ and $\eta\left(\tau\right)=\varrho b_1\left(\tau\right)\sqrt{\tau}\,c_0\left(\tau\right)$. The boundary case $\delta\left(\tau\right)=0$ instead allows
\begin{equation*}\lambda\left(\tau\right)=-\frac{\left(1-\varrho^2\right)\eta\left(\tau\right)\sqrt{\tau}}{2\varrho b_1\left(\tau\right)c_0\left(\tau\right)},\end{equation*}
with $b_0\left(\tau\right)=0$ in both cases. The standardized risk-neutral correlation can depend on the state in these additional cases; the proof works directly with the conditional Gaussian covariance entries.

When $\varrho=0$, $R_1\left(x;\tau\right)=1$, so $\Gamma\left(x;\tau\right)=x$ and necessarily $c_0\left(\tau\right)=0$, $c_1\left(\tau\right)=1$. The transformation does not determine $R_2$. If $\eta\left(\tau\right)\neq0$, the constant-ratio characterization for independent innovations in Appendix~\ref{subsec:q_polynomial_preservation} remains necessary and sufficient. If $\eta\left(\tau\right)=0$, the corresponding polynomial property holds if and only if $\psi_3\left(x;\tau\right)R_2\left(x;\tau\right)$ is affine. These assertions follow from the independence argument in Supplementary Material~\ref{app:proof_q_polynomial_preservation}, which does not require a finite limit of $R_2$ at zero. In both cases $\lambda\left(\tau\right)$ is unrestricted, subject to the maintained admissibility conditions.

The additional cases above arise from allowing the innovation variance ratio to diverge at zero. The change of coordinates does not restore polynomiality to a nonpolynomial transition.

\bigskip

\section{Supporting results and proofs}
\label{sm:sec:Additional_proofs}

\bigskip

\subsection{A Gaussian loading lemma}
\label{app:hermite_loading_lemma}

The following lemma gives the stronger loading characterization available under Gaussian innovations. It is used to derive the affine restriction on the centred quadratic loading in the Gaussian specialization of Proposition~\ref{prop:garch_polynomial_characterization} and in the necessity part of Proposition~\ref{prop:q_polynomial_preservation}.

\begin{lemma}
\label{lem:hermite_loading_polynomiality}
Let $I\subseteq\mathbb{R}$ be an interval with nonempty interior and let $\mu,\sigma_1,\sigma_2:I\to\mathbb{R}$. For $x\in I$, define
\begin{equation*}Z_x=\mu\left(x\right)+\sigma_1\left(x\right)\epsilon+\sigma_2\left(x\right)\left(\epsilon^2-1\right), \qquad \epsilon\sim N\left(0,1\right).\end{equation*}
If $x\mapsto\E\left[Z_x^q\right]\in\Pol_q\left(I\right)$ for $q=1,2,3,4$, then
\begin{equation*}\mu\in\Pol_1\left(I\right), \qquad \sigma_2\in\Pol_1\left(I\right), \qquad \sigma_1^2\in\Pol_2\left(I\right).\end{equation*}
\end{lemma}

\begin{proof}
We use twice the following observation. If, for some integer $e\geq0$, $f:I\to\mathbb{R}$ satisfies $f^2=P$ for some $P\in\Pol_{2e}\left(I\right)$ and agrees with a rational function on $\{P\neq0\}$, then $f\in\Pol_e\left(I\right)$. The conclusion is immediate when $P\equiv0$. Otherwise, writing the rational function in lowest terms as $p/q$ gives $p^2=Pq^2$, so $q$ is constant. Hence, $f$ agrees on $\{P\neq0\}$ with a polynomial $R$ satisfying $R^2=P$. Since both $f$ and $R$ vanish at the zeros of $P$, they agree on $I$, and $\deg\left(R\right)\leq e$.

Since $\E\left[\epsilon\right]=\E\left[\epsilon^2-1\right]=0$, one has $\E\left[Z_x\right]=\mu\left(x\right)$ and therefore $\mu\in\Pol_1\left(I\right)$. Let $V$, $T$, and $K$ denote the second, third, and fourth cumulants of $Z_x$, respectively. The assumed degree restrictions on the first four moments imply
\begin{equation*}V\in\Pol_2\left(I\right), \qquad T\in\Pol_3\left(I\right), \qquad K\in\Pol_4\left(I\right).\end{equation*}
For notational simplicity, set $u=\sigma_1^2$ and $d=\sigma_2^2$. For $z$ in a neighborhood of zero, the Gaussian cumulant generating function
\begin{equation*}\mathcal{C}_{Z_x}\left(z\right)=z\mu-z\sigma_2-\frac{1}{2}\log\left(1-2z\sigma_2\right)+\frac{z^2u}{2\left(1-2z\sigma_2\right)}\end{equation*}
gives
\begin{equation*}V=u+2d, \qquad T=\sigma_2\left(6u+8d\right), \qquad K=48d\left(u+d\right).\end{equation*}
Consequently, with $S:=V^2-K/12=u^2$,
\begin{equation*}T^2=d\left(6u+8d\right)^2=2\left(V-u\right)\left(u+2V\right)^2,\end{equation*}
and hence
\begin{equation}\label{eq:lemma_u_identity}2uS=8V^3-T^2-6VS.\end{equation}

If $S\equiv0$, then $u\equiv0$. Otherwise, Equation~\eqref{eq:lemma_u_identity} shows that $u$ agrees with a rational function on $\{S\neq0\}$. Since $u^2=S\in\Pol_4\left(I\right)$, the preceding observation yields $u\in\Pol_2\left(I\right)$, and therefore $\sigma_1^2\in\Pol_2\left(I\right)$.

It follows that $d=\left(V-u\right)/2\in\Pol_2\left(I\right)$. If $d\equiv0$, then $\sigma_2\equiv0$. Otherwise, on $\{d\neq0\}$,
\begin{equation*}\sigma_2=\frac{T}{6u+8d},\end{equation*}
where the denominator is strictly positive. Since $\sigma_2^2=d$, the preceding observation gives $\sigma_2\in\Pol_1\left(I\right)$.
\end{proof}

\subsection{Proof of Proposition~\ref{prop:finite_dimensional_representation}}
\label{sm:proof_finite_dimensional_representation}

\bigskip

\begin{proof}

Let $p\in\Pol_m\left(E\right)$ have coordinate vector $\bm{p}$, and set $\tilde{p}_E:=1+p^2\in\Pol_{2m}\left(E\right)$. This nonnegative polynomial dominates $|p|$ on $E$. Since $\bm{X}^{(n)}$ is polynomial preserving, $\trans^{(n)}$ is positive and preserves $\Pol_{2m}\left(E\right)$, so $\left(\trans^{(n)}\right)^s\tilde{p}_E$ is nonnegative and finite on $E$ for every admissible $s$. By the transition-kernel representation in Equation~\eqref{eq:transition_operator} and repeated application of the Markov and tower properties to nonnegative random variables,
\begin{equation*}\E^{\Pm^{(n)}}\left[\left|p\left(\bm{X}^{(n)}_{(k+s)\tau}\right)\right|\,\middle|\,\F^{(n)}_{k\tau}\right]\leq\left(\trans^{(n)}\right)^s\tilde{p}_E\left(\bm{X}^{(n)}_{k\tau}\right)<\infty,\qquad 0\leq s\leq nL-k.\end{equation*}
Thus, the conditional expectations below are well defined in the generalized sense specified in Section~\ref{subsec:general_framework}. Localizing on the $\F^{(n)}_{k\tau}$-measurable sets where this bound is at most an integer justifies the subsequent applications of the tower property.

For every admissible $k,s$, the Markov property, time homogeneity, and iteration of Equation~\eqref{eq:basis_vector_transition} give
\begin{equation*}\E^{\Pm^{(n)}}\left[p\left(\bm{X}^{(n)}_{(k+s)\tau}\right)\,\middle|\,\F^{(n)}_{k\tau}\right]=\left(\trans^{(n)}\right)^s p\left(\bm{X}^{(n)}_{k\tau}\right)=\bm{\HT}_m\left(\bm{X}^{(n)}_{k\tau}\right)^\top\left(\bm{Q}^{(n)}_m\right)^s\bm{p},\end{equation*}
including $s=0$ under the identity-operator convention.
\end{proof}

\bigskip

\subsection{Proof of Corollary~\ref{cor:polynomial_cross_moments}}
\label{sm:proof_polynomial_cross_moments}

\bigskip

\begin{proof}
Write $t_j=k+s_1+\cdots+s_j$ and define the backward recursion
\begin{equation*}f_J=p_J,\qquad f_j=p_j\left(\trans^{(n)}\right)^{s_{j+1}}f_{j+1},\qquad j=J-1,\ldots,1.\end{equation*}
Polynomial preservation gives $f_j\in\Pol_{\mu_j}\left(E\right)$, and Equation~\eqref{eq:polynomial_multiplication_matrix} identifies its coordinate vector as $\bm{v}_j$. Applying the same recursion to the nonnegative polynomials $1+p_j^2$ gives a finite conditional bound for $\prod_{j=1}^J\left|p_j\left(\bm{X}^{(n)}_{t_j\tau}\right)\right|$. The same localization argument justifies backward conditioning and yields
\begin{equation*}\E^{\Pm^{(n)}}\left[\prod_{j=1}^Jp_j\left(\bm{X}^{(n)}_{t_j\tau}\right)\,\middle|\,\F^{(n)}_{k\tau}\right]=\left(\trans^{(n)}\right)^{s_1}f_1\left(\bm{X}^{(n)}_{k\tau}\right).\end{equation*}
Proposition~\ref{prop:finite_dimensional_representation} now gives Equation~\eqref{eq:conditional_polynomial_product}.
\end{proof}

\bigskip

\subsection{Proof of Proposition~\ref{prop:variance_polynomial_positivity}}
\label{sm:proof_variance_polynomial_positivity}

\bigskip

\begin{proof}
Fix $\tau$ and suppress it in the parameters. The restrictions on $a_0,a_1$ ensure $g\left(x\right):=a_0+a_1x+x^2\geq0$ on $\Rplus$. The variance recursion gives
\begin{equation*}h_x^+=\left(b_0+b_1x\right)\epsilon_2^2+\eta\sqrt{g\left(x\right)}\,\epsilon_2+\omega-b_0+\left(\phi-b_1\right)x.\end{equation*}
For $x>0$, its leading coefficient is strictly positive and its discriminant is
\begin{equation*}\Delta_\tau\left(x\right)=\eta^2g\left(x\right)-4\left(b_0+b_1x\right)\left[\omega-b_0+\left(\phi-b_1\right)x\right]=\delta_2x^2+\delta_1x+\delta_0.\end{equation*}
Thus, $h_x^+\geq0$ for every real $\epsilon_2$ and every $x>0$ if and only if $\Delta_\tau\left(x\right)\leq0$ for every $x>0$. This requires $\delta_2\leq0$ and $\delta_0\leq0$, by considering $x\to\infty$ and $x\downarrow0$, respectively. Under these two inequalities, the condition is equivalent to
\begin{equation*}\delta_1\leq\inf_{x>0}\left\{\left(-\delta_2\right)x+\frac{-\delta_0}{x}\right\}=2\sqrt{\delta_2\delta_0},\end{equation*}
where the identity also holds when either coefficient is zero. Hence Equation~\eqref{eq:positivity_delta_restrictions} is necessary and sufficient for $h_x^+\geq0$ for every $x>0$ and every real $\epsilon_2$.

At $x=0$, if $b_0>0$, the nonpositive discriminant $\Delta_\tau\left(0\right)=\delta_0$ gives $h_0^+\geq0$. If $b_0=0$, the restriction $\delta_0=\eta^2a_0\leq0$ forces $\eta\sqrt{a_0}=0$, so $h_0^+=\omega\geq0$. Hence the recursion preserves $\Rplus$.
\end{proof}

\bigskip

\subsection{Proof of Proposition~\ref{prop:joint_polynomial_positivity}}
\label{sm:proof_joint_polynomial_positivity}

\bigskip

\begin{proof}
For each $x\in\Rplus$, denote by $h_x^+$ the next conditional variance given current variance $x$. By Equations~\eqref{eq:general_variance_dynamics_p}, \eqref{eq:common_mean_parametrizations}, and~\eqref{eq:joint_polynomial_loading_functions},
\begin{equation*}
h_x^+=\left[b_0\left(\tau\right)+b_1\left(\tau\right)x\right]\epsilon_2^2+\eta\left(\tau\right)\sqrt{x}\,\epsilon_2+\omega\left(\tau\right)+\phi\left(\tau\right)x-b_0\left(\tau\right)-b_1\left(\tau\right)x,
\end{equation*}
where $\epsilon_2\overset{\Pm^{(n)}}{\sim}D_2\left(0,1\right)$.

Nonnegativity requires the coefficient of $\epsilon_2^2$ to be nonnegative on $\Rplus$, giving $b_0\left(\tau\right)\geq0$ and $b_1\left(\tau\right)\geq0$. Setting $\epsilon_2=0$ similarly gives $b_0\left(\tau\right)\leq\omega\left(\tau\right)$ and $b_1\left(\tau\right)\leq\phi\left(\tau\right)$. Thus, Equation~\eqref{eq:joint_positivity_loading_restrictions} is necessary.

Suppose first that $x>0$. By Equation~\eqref{eq:joint_positivity_loading_restrictions} and $b_0\left(\tau\right)+b_1\left(\tau\right)>0$, $b_0\left(\tau\right)+b_1\left(\tau\right)x>0$, so $h_x^+$ is a strictly convex quadratic function of $\epsilon_2$. Its discriminant is
\begin{equation*}
\Delta_\tau\left(x\right)=\eta\left(\tau\right)^2x-4\left[b_0\left(\tau\right)+b_1\left(\tau\right)x\right]\left\{\omega\left(\tau\right)-b_0\left(\tau\right)+\left[\phi\left(\tau\right)-b_1\left(\tau\right)\right]x\right\}.
\end{equation*}
Hence $h_x^+\geq0$ for every $\epsilon_2\in\mathbb{R}$ and every $x>0$ if and only if
\begin{align*}
\eta\left(\tau\right)^2&\leq4\inf_{x>0}\frac{\left[b_0\left(\tau\right)+b_1\left(\tau\right)x\right]\left\{\omega\left(\tau\right)-b_0\left(\tau\right)+\left[\phi\left(\tau\right)-b_1\left(\tau\right)\right]x\right\}}{x}\\
&=4\left\{\sqrt{b_0\left(\tau\right)\left[\phi\left(\tau\right)-b_1\left(\tau\right)\right]}+\sqrt{b_1\left(\tau\right)\left[\omega\left(\tau\right)-b_0\left(\tau\right)\right]}\right\}^2.
\end{align*}
The equality follows from $\inf_{x>0}\left(a/x+bx\right)=2\sqrt{ab}$ for $a,b\geq0$, including cases where either coefficient is zero. This gives Equation~\eqref{eq:joint_positivity_leverage_restriction}.

At $x=0$, Equation~\eqref{eq:joint_positivity_loading_restrictions} gives
\begin{equation*}
h_0^+=b_0\left(\tau\right)\epsilon_2^2+\omega\left(\tau\right)-b_0\left(\tau\right)\geq0.
\end{equation*}
Thus, the JPGARCH-SV recursion maps $\Rplus$ into itself.
\end{proof}

\bigskip

\subsection{Proof of Proposition~\ref{prop:garch_moment_existence}}
\label{sm:proof_garch_moment_existence}

\begin{proof}
By the univariate specialization in Supplementary Material~\ref{sm:matryoshka_details}, $q^{(n)}_{jj}$ is the coefficient of $x^j$ in $\trans^{(n)}x^j=\E^{\Pm^{(n)}}\left[\left(h_x^+\right)^j\right]$, where $h_x^+$ is the one-step update defined in the proof of Proposition~\ref{prop:garch_polynomial_characterization}. The case $j=0$ is immediate. For $j\geq1$ and fixed $\tau$, the model loadings give $h_x^+/x\to A_{\mathsf M}$ as $x\to\infty$ and
\begin{equation*}\left|h_x^+/x\right|^j\leq C_j\left(1+\epsilon_2^2\right)^j,\qquad x\geq1,\end{equation*}
for a finite constant $C_j$ independent of $x$ and $\epsilon_2$. The maintained innovation moments make this bound integrable. Polynomial preservation and dominated convergence therefore give
\begin{equation*}q^{(n)}_{jj}=\lim_{x\to\infty}\E^{\Pm^{(n)}}\left[\left(h_x^+/x\right)^j\right]=\E^{\Pm^{(n)}}\left[A_{\mathsf M}^{\,j}\right].\end{equation*}

Writing $A_{\mathsf{VP}}=b_1\left(\tau\right)\epsilon_2^2+\eta\left(\tau\right)\epsilon_2+\phi\left(\tau\right)-b_1\left(\tau\right)$, Proposition~\ref{prop:variance_polynomial_positivity} gives $b_1\left(\tau\right)>0$ and $\delta_2\left(\tau\right)\leq0$, hence $A_{\mathsf{VP}}\geq0$. Similarly, $A_{\mathsf{JP}}=b_1\left(\tau\right)\epsilon_2^2+\phi\left(\tau\right)-b_1\left(\tau\right)\geq0$ by Proposition~\ref{prop:joint_polynomial_positivity}. Therefore, Condition~\eqref{sm:eq:matryoshka_spectral_condition} is equivalent to $\E^{\Pm^{(n)}}\left[A_{\mathsf M}^{\,j}\right]<1$ for $j=1,\ldots,m$. Necessity of Equation~\eqref{eq:garch_moment_condition} follows by taking $j=m$, while sufficiency follows from Lyapunov's inequality. Proposition~\ref{sm:prop:matryoshka_stationary_moments} then gives the stated convergence of moments.
\end{proof}

\bigskip

\subsection{Proof of Corollary~\ref{cor:polynomial_temporal_aggregates}}
\label{sm:proof_polynomial_temporal_aggregates}

\begin{proof}
For each $x\in\Rplus$, let $y_x^+$ and $h_x^+$ denote the one-step updates defined in the proof of Proposition~\ref{prop:garch_polynomial_characterization}, and set
\begin{equation*}\widetilde y_x^+:=y_x^+-\psi_0\left(x;\tau\right)=\sqrt{\tau x}\,\epsilon_1, \qquad \epsilon_1\overset{\Pm^{(n)}}{\sim}D_1\left(0,1\right).\end{equation*}
In each case below, the Markov property follows directly from the one-step recursion and the independence of the next innovation pair from the current filtration. By linearity, it suffices to consider monomials.

For $\bm U^{(n,k)}$, conditional on the current state $\left(v,x\right)^\top$, the next state has the same distribution as $\left(v+\tau x,h_x^+\right)^\top$. With $g_b\left(x;\tau\right):=\E^{\Pm^{(n)}}\left[\left(h_x^+\right)^b\right]\in\Pol_b\left(\Rplus\right)$, the monomial $f\left(v,x\right)=v^ax^b$ satisfies
\begin{equation*}\E^{\Pm^{(n)}}\left[\left(v+\tau x\right)^a\left(h_x^+\right)^b\right]=\sum_{u=0}^{a}\binom{a}{u}v^{a-u}\left(\tau x\right)^u g_b\left(x;\tau\right),\end{equation*}
which has total degree at most $a+b$.

For $\bm V^{(n,k)}$, the next state has the same distribution as $\left(\tau x\epsilon_1^2,h_x^+\right)^\top$. Expanding $\left(h_x^+\right)^b$, every term is, up to a constant independent of $x$, of the form $\psi_1\left(x;\tau\right)^i\psi_2\left(x;\tau\right)^r\psi_3\left(x;\tau\right)^\ell\epsilon_2^r\left(\epsilon_2^2-1\right)^\ell$ with $i+r+\ell=b$. After multiplication by $\left(\tau x\epsilon_1^2\right)^a$, joint central symmetry gives $\E^{\Pm^{(n)}}\left[\epsilon_1^{2a}\epsilon_2^r\left(\epsilon_2^2-1\right)^\ell\right]=0$ for odd $r$, while for even $r$,
\begin{equation*}\deg\left(x^a\psi_1^i\left(\psi_2^2\right)^{r/2}\psi_3^\ell\right)\leq a+i+r+\ell=a+b.\end{equation*}
This proves part~\textnormal{(i)}.

Under JPGARCH-SV, or under VPGARCH-SV with independent innovations, conditional on $\bm Z^{(n,k)}_{j\tau}=\left(z,x\right)^\top$, the next state has the same distribution as $\left(z+y_x^+,h_x^+\right)^\top$. Let $g_{u,b}\left(x;\tau\right):=\E^{\Pm^{(n)}}\left[\left(y_x^+\right)^u\left(h_x^+\right)^b\right]\in\Pol_{u+b}\left(\Rplus\right)$ for $u=0,\ldots,a$, by Proposition~\ref{prop:garch_polynomial_characterization}. Then
\begin{equation*}\E^{\Pm^{(n)}}\left[\left(z+y_x^+\right)^a\left(h_x^+\right)^b\right]=\sum_{u=0}^{a}\binom{a}{u}z^{a-u}g_{u,b}\left(x;\tau\right),\end{equation*}
whose terms have total degree at most $a+b$. This proves part~\textnormal{(ii)}.
\end{proof}

\bigskip

\subsection{Proof of Proposition~\ref{prop:general_q_dynamics}}
\label{app:proof_general_q_dynamics}

\begin{proof}
The normalization in Equation~\eqref{eq:pricing_kernel_no_arbitrage} makes the density process in Appendix~\ref{subsec:pricing_kernel_q_dynamics} a strictly positive unit-mean $\Pm^{(n)}$-martingale. For this proof, abbreviate $A_i=A_{i,k\tau}\left(\tau\right)$, $B=B_{k\tau}\left(\tau\right)$, $P_i=P_{i,k\tau}\left(\tau\right)$, and $R_i=R_{i,k\tau}\left(\tau\right)$. Applying the conditional change-of-measure formula gives, wherever finite,
\begin{equation*}\mathcal{C}^{\Qm^{(n)}}_{\epsilon^{(n)}_{1,(k+1)\tau},\epsilon^{(n)}_{2,(k+1)\tau}}\left(z_1,z_2\right)=\mathcal{C}^{\Pm^{(n)}}\left(A_1+z_1,A_2+z_2,B\right)-\mathcal{C}^{\Pm^{(n)}}\left(A_1,A_2,B\right).\end{equation*}
Differentiation at $\left(z_1,z_2\right)=\left(0,0\right)$, justified by the assumed local finiteness in the first two arguments, gives the conditional means and variances in Equation~\eqref{eq:general_variance_risk_ratio}. Standardization yields
\begin{equation*}\begin{aligned}\mathcal{C}^{\Qm^{(n)}}_{\epsilon^{*\,(n)}_{1,(k+1)\tau},\epsilon^{*\,(n)}_{2,(k+1)\tau}}\left(z_1,z_2\right)={}&\mathcal{C}^{\Pm^{(n)}}\left(A_1+\frac{z_1}{\sqrt{R_1}},A_2+\frac{z_2}{\sqrt{R_2}},B\right)-\mathcal{C}^{\Pm^{(n)}}\left(A_1,A_2,B\right)-\sum_{i=1}^2\frac{z_iP_i}{\sqrt{R_i}}.\end{aligned}\end{equation*}

Substituting $\epsilon^{(n)}_{i,(k+1)\tau}=P_i+\sqrt{R_i}\,\epsilon^{*\,(n)}_{i,(k+1)\tau}$, for $i=1,2$, into Equations~\eqref{eq:general_return_dynamics_p}--\eqref{eq:general_variance_dynamics_p} and collecting constant, linear, and centred quadratic terms gives Equations~\eqref{eq:general_return_dynamics_q}--\eqref{eq:general_variance_dynamics_q} and the stated coefficients. The risky-asset pricing restriction gives the equivalent conditional-cgf expression for $\psi_0^*$.

For the Gaussian specialization in Remark~\ref{rem:jointly_gaussian_q_innovations} of the main paper, completing the square gives
\begin{equation*}\mathcal{C}^{\Pm^{(n)}}\left(z_1,z_2,z_3\right)=\frac{1-\varrho^2}{2}z_1^2-\frac{1}{2}\log\left(1-2z_3\right)+\frac{\left(z_2+\varrho z_1\right)^2}{2\left(1-2z_3\right)}, \qquad z_3<\frac{1}{2}.\end{equation*}
Its first and second derivatives give the displayed conditional means and variances and the conditional covariance $\varrho R_{2,k\tau}\left(\tau\right)$, hence the stated standardized correlation.
\end{proof}

\subsection{Proof of Proposition~\ref{prop:q_polynomial_preservation}}
\label{app:proof_q_polynomial_preservation}

\begin{proof}
Under the maintained positivity restrictions, the variance update from any $x>0$ is a nonnegative quadratic in the Gaussian variance innovation with strictly positive quadratic coefficient. It can vanish only at finitely many innovation values, a null event. Thus, $h_0>0$ implies strictly positive variance at every trading date under both equivalent measures.

Fix $\tau$ and suppress it in the state functions and parameters. Set $a=1-\varrho^2$, $s\left(x\right)=\psi_3\left(x\right)=b_0+b_1x$, $u\left(x\right)=\psi_2\left(x\right)$, $R\left(x\right)=R_2\left(x\right)$, and $K\left(x\right)=R_1\left(x\right)=a+\varrho^2R\left(x\right)$. The Gaussian formulas in Appendix~\ref{subsec:pricing_kernel_q_dynamics} and the martingale restriction give
\begin{equation*}P_1\left(x\right)=-\sqrt{x}\,d\left(x\right), \qquad d\left(x\right)=\frac{\lambda}{\sqrt{\tau}}+\frac{\sqrt{\tau}}{2}K\left(x\right),\end{equation*}
and
\begin{equation*}P_2\left(x\right)=\frac{a\left[R\left(x\right)-1\right]u\left(x\right)/(2s\left(x\right))-\varrho R\left(x\right)d\left(x\right)\sqrt{x}}{K\left(x\right)}.\end{equation*}
Consequently,
\begin{equation*}\psi_2^*\left(x\right)=\frac{R\left(x\right)^{3/2}}{K\left(x\right)}\left[u\left(x\right)-2\varrho s\left(x\right)d\left(x\right)\sqrt{x}\right], \qquad \psi_3^*\left(x\right)=R\left(x\right)s\left(x\right),\end{equation*}
and completion of the square yields
\begin{equation*}\psi_1^*\left(x\right)=\psi_1\left(x\right)+\left[R\left(x\right)-1\right]s\left(x\right)+\frac{R\left(x\right)^2\left[u\left(x\right)-2\varrho s\left(x\right)d\left(x\right)\sqrt{x}\right]^2/K\left(x\right)^2-u\left(x\right)^2}{4s\left(x\right)}.\end{equation*}
By Lemma~\ref{lem:hermite_loading_polynomiality}, variance polynomial preservation requires $\psi_1^*$ and $\ell:=\psi_3^*=Rs$ to be affine and $\left(\psi_2^*\right)^2$ to have degree at most two. These conditions also suffice, by the Gaussian marginal characterization in Proposition~\ref{prop:garch_polynomial_characterization}.

We first consider VPGARCH-SV with $\varrho>0$. Here $u=\eta\sqrt{g}$, where $g\left(x\right)=a_0+a_1x+x^2$, and $b_1>0$. Since $\ell$ is affine, $R$, $K$, and $d$ are rational. If $\eta\neq0$, the only possibly nonrational term in $\psi_1^*$ is
\begin{equation*}-\frac{\varrho\eta R\left(x\right)^2d\left(x\right)}{K\left(x\right)^2}\sqrt{xg\left(x\right)}.\end{equation*}
The square root of the cubic polynomial $xg\left(x\right)$ cannot agree with a rational function on an open interval. Affineness of $\psi_1^*$ therefore forces $d=0$.

If $\eta=0$, put $T=as+\varrho^2\ell$ and
\begin{equation*}W=\frac{\ell\left[\lambda s/\sqrt{\tau}+\sqrt{\tau}T/2\right]}{T}.\end{equation*}
Then $W$ is rational and $\psi_1^*-\psi_1-\ell+s=\varrho^2xW^2/s$ is affine. When $b_0>0$, an identity $xW^2/s=M$ with $M$ affine forces $M=0$: otherwise the multiplicity at $x=0$ first forces $M$ to be proportional to $x$, after which $W^2$ would be proportional to the nonconstant affine polynomial $s$, which is impossible for a rational square. When $b_0=0$, the finite-limit assumption on $R=\ell/s$ forces $\ell$, and hence $T$ and $W$, to be proportional to $x$; affineness of $xW^2/s$ again requires $W=0$. Since $\ell$ and $T$ are positive on $\left(0,\infty\right)$, this also gives $d=0$.

Thus $K$, and consequently $R$, are constant, with $\lambda=-\tau K/2$. The transformed variance coefficients reduce to
\begin{equation*}\psi_1^*=\psi_1+\left(R-1\right)s+\frac{\eta^2}{4}\left(\frac{R^2}{K^2}-1\right)\frac{g}{s},\end{equation*}
\begin{equation*}\psi_2^*=\frac{R^{3/2}}{K}\eta\sqrt{g}, \qquad \psi_3^*=Rs.\end{equation*}
The last two coefficients have the required polynomial degrees. Since $R-K=a\left(R-1\right)$, the first is affine if and only if $\eta a\left(R-1\right)=0$ or $s$ divides $g$. The latter condition is $b_1^2a_0-b_0b_1a_1+b_0^2=0$. Finally, constancy of $R=\left(1-2\xi_2s\right)^{-1}$ is equivalent to $\xi_2\left(x\right)=\xi_2\left(\tau\right)/s\left(x\right)$, with $\xi_2\left(\tau\right)<1/2$, and $R=1$ is equivalent to $\xi_2\left(\tau\right)=0$. This proves necessity and sufficiency in part~\textnormal{(i)}.

We next consider JPGARCH-SV with $\varrho>0$, for which $u\left(x\right)=\eta\sqrt{x}$. Joint polynomial preservation implies that the conditional return mean $r\tau-\tau xK\left(x\right)/2$ is affine. The finite-limit assumption forces $K$, and hence $R$, to be constant. The standardized Gaussian pair then has constant positive correlation $\varrho\sqrt{R/K}$. Rescaling the return coordinate by $K^{-1/2}$ allows application of the joint characterization in Proposition~\ref{prop:garch_polynomial_characterization}. In particular,
\begin{equation*}\psi_2^*\left(x\right)=\frac{R^{3/2}}{K}\sqrt{x}\left[\eta-2\varrho d\left(b_0+b_1x\right)\right]\end{equation*}
must be proportional to $\sqrt{x}$, so $b_1d=0$. If $b_1=0$, all transformed coefficients have the required degrees without further restrictions. If $b_1>0$, then $d=0$ and
\begin{equation*}\psi_1^*\left(x\right)=\psi_1\left(x\right)+\left(R-1\right)s\left(x\right)+\frac{\eta^2}{4}\left(\frac{R^2}{K^2}-1\right)\frac{x}{s\left(x\right)}.\end{equation*}
This is affine if and only if $b_0=0$ or $\eta a\left(R-1\right)=0$. Together with $b_1d=0$, these are precisely the conditions in part~\textnormal{(ii)}. Conversely, they make $\psi_0^*,\psi_1^*,\psi_3^*$ affine and $\psi_2^*$ proportional to $\sqrt{x}$; the same joint characterization proves sufficiency.

For completeness, let $\varrho=0$. The tilted innovations are independent, $K=1$, and the return mean is $r\tau-\tau x/2$. The corresponding variance coefficients are
\begin{equation*}\psi_1^*=\psi_1+\ell-s+\frac{\left(R^2-1\right)u^2}{4s}, \qquad \left(\psi_2^*\right)^2=R^3u^2, \qquad \psi_3^*=\ell.\end{equation*}
If $\eta=0$, the variance characterization is therefore equivalent to affineness of $\ell=Rs$. If $\eta\neq0$ and $s$ is nonconstant, affineness of $\ell$ and polynomiality of $\ell^3u^2/s^3$ force $\ell$ to be proportional to $s$: otherwise $s^3$ would divide the polynomial $u^2$, whose degree is at most two. If $s$ is constant, only the JPGARCH-SV case is possible, and $\ell^3u^2/s^3$ has degree at most two only when $\ell$ is constant. Thus $R$ is constant whenever $\eta\neq0$. The displayed drift then gives the product restrictions in the proposition, without any restriction on $\lambda$. Independence makes the conditional mixed moments factor, so the variance characterization also suffices for the joint property of JPGARCH-SV. This proves the statements for $\varrho=0$.
\end{proof}

\bigskip

\section{Additional tables and figures}
\label{sm:sec:additional_results}

\bigskip

\begin{figure}[!htbp]
	\centering
	\includegraphics{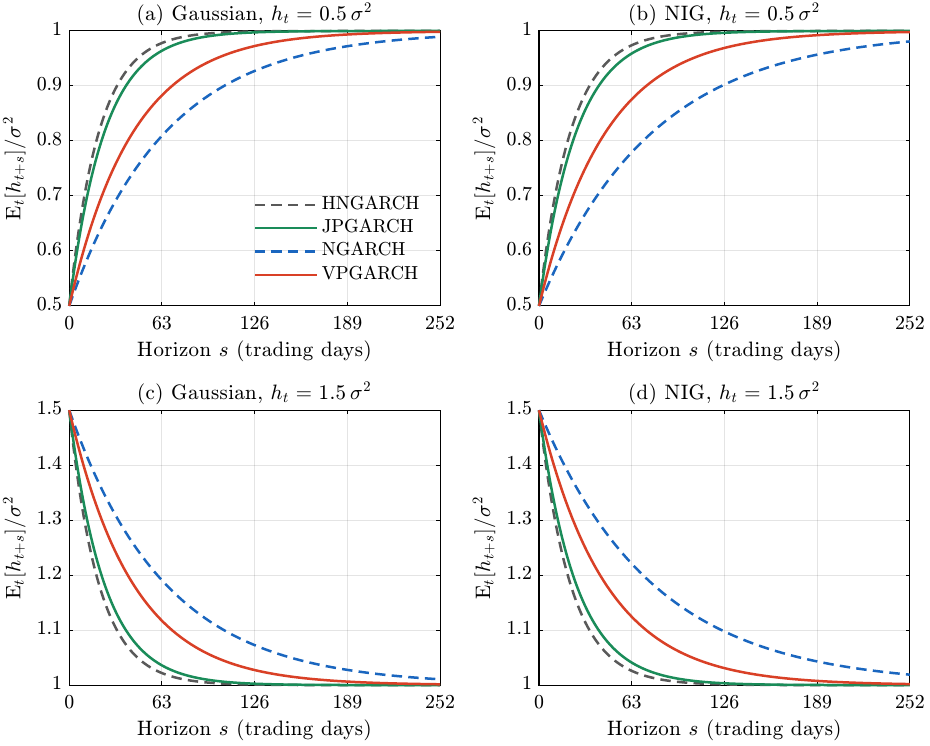}
	\caption{\textbf{Term Structure of the Expected Variance}\newline
		\footnotesize
		\emph{Notes:} The figure shows $\E_t[h_{t+s}]/\sigma^2$ for horizons $s$ up to one year, starting from $h_t=0.5\sigma^2$ (top panels) and $h_t=1.5\sigma^2$ (bottom panels), where $\sigma^2$ is the long-run variance of each model. The left panels use the Gaussian estimates and the right panels the NIG estimates in Table~\ref{tab:return_mle}.}
	\label{fig:variance_term_structure}
\end{figure}

\begin{figure}[!htbp]
	\centering
	\includegraphics{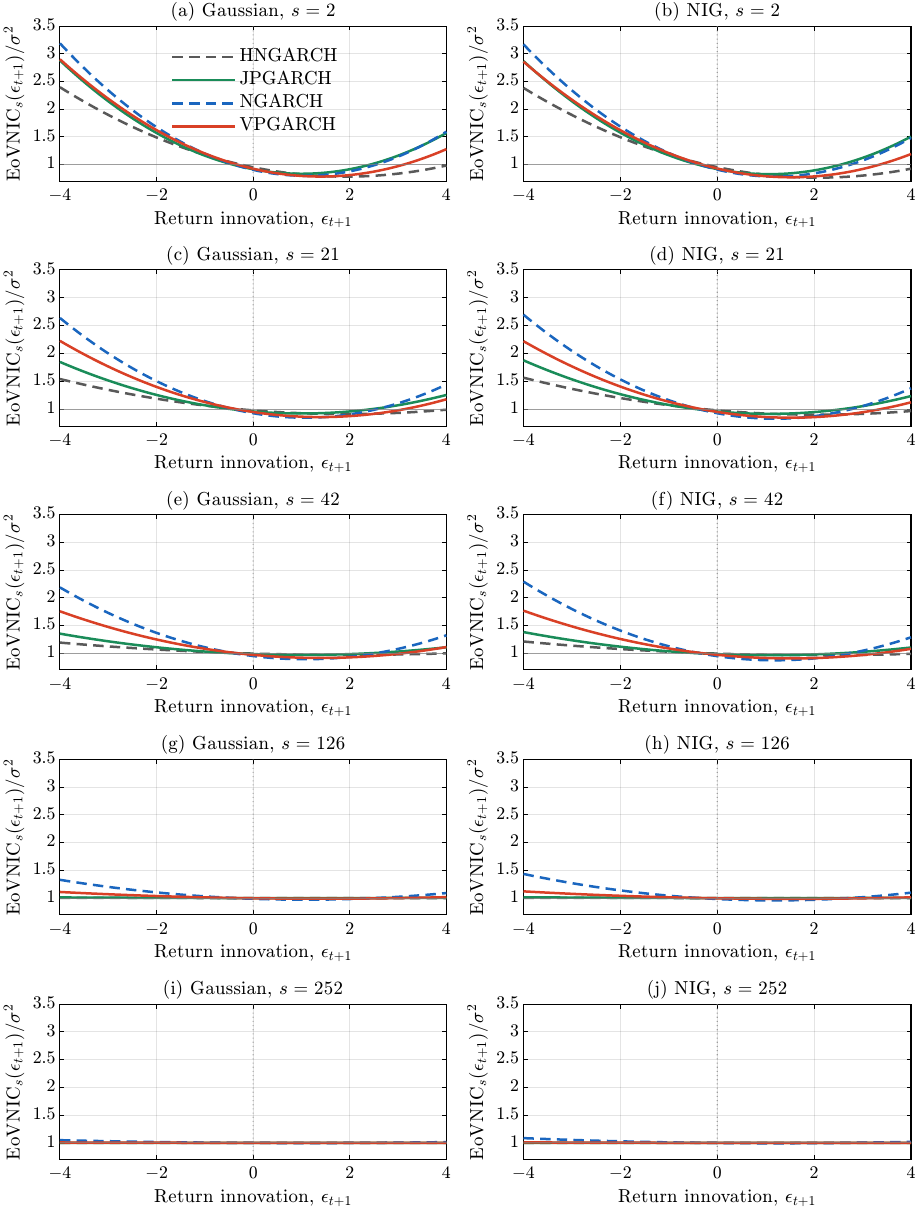}
	\caption{\textbf{Expected-Variance News Impact Curves}\newline
		\footnotesize
		\emph{Notes:} Each panel shows $\mathrm{EoVNIC}_s(\epsilon_{t+1})/\sigma^2$ for the horizon $s$ indicated, where $\mathrm{EoVNIC}_s(\epsilon)=\E_{t+1}[h_{t+s}]$ is evaluated at $h_t=\sigma^2$ and $\epsilon_{t+1}=\epsilon$. The left panels use the Gaussian estimates and the right panels the NIG estimates in Table~\ref{tab:return_mle}. The horizontal line marks the long-run variance.}
	\label{fig:eovnic}
\end{figure}

Table~\ref{tab:option_data} summarizes the option sample. Figures~\ref{fig:variance_term_structure} and~\ref{fig:eovnic} report the expected-variance term structures and news impact curves discussed in Section~\ref{subsec:return_estimation} of the main paper.

\begin{sidewaystable}
	\caption{\textbf{S\&P 500 Index Option Data by Moneyness and Maturity}\label{tab:option_data}}
	\scriptsize
	\renewcommand{\arraystretch}{1.25}
		\newcommand{\mc}[1]{\multicolumn{1}{c}{#1}}
	\newsavebox{\NumCellBox}
	\newcolumntype{N}[1]{>{\setbox\NumCellBox=\hbox\bgroup}c<{\egroup\phantom{#1}\llap{\box\NumCellBox}}}
	\centering
	\begin{tabular*}{\linewidth}{@{\extracolsep{\fill}} cl N{000,000} *{7}{N{00,000}} N{000,000}}
		\toprule
		Maturity & & \multicolumn{9}{c}{Moneyness} \\
		\cmidrule(lr){1-1}\cmidrule(lr){3-11}
		& & \mc{$[0,0.900)$} & \mc{$[0.900,0.950)$} & \mc{$[0.950,0.975)$} & \mc{$[0.975,1.000)$} & \mc{$[1.000,1.025)$} & \mc{$[1.025,1.050)$} & \mc{$[1.050,1.100)$} & \mc{$[1.100,\infty)$} & \mc{Overall} \\
		\cmidrule(lr){3-3}\cmidrule(lr){4-4}\cmidrule(lr){5-5}\cmidrule(lr){6-6}\cmidrule(lr){7-7}\cmidrule(lr){8-8}\cmidrule(lr){9-9}\cmidrule(lr){10-10}\cmidrule(lr){11-11}
		\multirow{2}{*}{$[7,30)$}    & Number of contracts    & 24,258  & 22,604 & 14,446 & 14,791 & 15,295 & 11,251 & 8,063  & 2,347  & 113,055 \\
		& Implied volatility (\%) & 41.0    & 25.5   & 20.5   & 17.5   & 15.1   & 15.4   & 19.5   & 29.2   & 24.4 \\[1.5ex]
		\multirow{2}{*}{$[30,90)$}   & Number of contracts    & 112,041 & 49,611 & 27,251 & 28,785 & 28,255 & 26,260 & 33,461 & 17,429 & 323,093 \\
		& Implied volatility (\%) & 34.9    & 22.6   & 19.6   & 17.7   & 15.7   & 14.3   & 15.1   & 21.1   & 24.0 \\[1.5ex]
		\multirow{2}{*}{$[90,180)$}  & Number of contracts    & 58,179  & 16,856 & 8,839  & 9,629  & 9,276  & 8,413  & 12,884 & 13,486 & 137,562 \\
		& Implied volatility (\%) & 33.7    & 22.5   & 20.5   & 19.0   & 17.4   & 16.1   & 15.4   & 17.4   & 24.9 \\[1.5ex]
		\multirow{2}{*}{$[180,270)$} & Number of contracts    & 25,954  & 5,367  & 2,758  & 2,755  & 2,546  & 2,259  & 4,334  & 9,069  & 55,042 \\
		& Implied volatility (\%) & 32.8    & 22.0   & 20.4   & 19.3   & 18.2   & 17.2   & 16.0   & 16.4   & 25.1 \\[1.5ex]
		\multirow{2}{*}{$[270,365]$} & Number of contracts    & 15,155  & 3,554  & 1,823  & 1,916  & 2,030  & 1,644  & 2,943  & 6,775  & 35,840 \\
		& Implied volatility (\%) & 30.6    & 21.5   & 20.1   & 19.3   & 18.4   & 17.6   & 16.4   & 16.1   & 23.4 \\[1.5ex]
		\multirow{2}{*}{Overall}     & Number of contracts    & 235,587 & 97,992 & 55,117 & 57,876 & 57,402 & 49,827 & 61,685 & 49,106 & 664,592 \\
		& Implied volatility (\%) & 34.7    & 23.2   & 20.1   & 18.0   & 16.0   & 15.1   & 15.9   & 18.9   & 24.3 \\
		\bottomrule
	\end{tabular*}

	\begin{minipage}{\linewidth}
		\footnotesize
		\emph{Notes:} This table reports the number of S\&P 500 index options and their average Black--Scholes implied volatility, in percent, by moneyness and maturity for the 1,443 weekly dates from January 1996 to August 2023. Moneyness is defined as the strike-to-forward ratio $K/F$, where $F=S\exp\left((r-q)(T-t)\right)$ is the forward price and $T-t$ is the maturity in years. Options with $K/F<1$ are out-of-the-money puts and options with $K/F\geq1$ are out-of-the-money calls. Maturity is measured in calendar days.
	\end{minipage}
\end{sidewaystable}

\clearpage

Table~\ref{tab:vov_coefficients} reports the coefficients of the one-step conditional variance of the variance, which enter the variance-of-variance news impact curves in Section~\ref{subsec:implied_variance_dynamics} of the main paper. Tables~\ref{tab:option_parameters_gaussian} and~\ref{tab:option_parameters_nig} report quantiles of the parameters calibrated to the weekly cross-sections of options in Section~\ref{subsec:option_pricing} of the main paper, under Gaussian and NIG innovations, respectively.

Under Gaussian innovations, the risk premium parameter $\lambda$ of HNGARCH is not identified from option prices, because the risk-neutral dynamics depend on $\lambda$ and $\gamma$ only through $\gamma^*=\gamma+\lambda+\frac{1}{2}$ \citep{heston2000closed}.\footnote{With the parameter mapping in Equation~\eqref{eq:hngarch_parameter_mapping} of the main paper, the risk-neutral HNGARCH dynamics implied by the pricing kernel with $\xi_{2,t}=0$ are again of the Heston--Nandi form, with $\gamma$ replaced by $\gamma^*$. Option prices are therefore unchanged when $\lambda$ and $\gamma$ vary with $\gamma^*$ and the other Heston--Nandi parameters held fixed.} The values of $\lambda$, $\phi$, $\eta$ and $\sigma^2$ reported for HNGARCH in Table~\ref{tab:option_parameters_gaussian} therefore represent one of many parameterizations that imply the same option prices. This argument relies on Gaussian innovations, for which the change of measure merely shifts the innovation; under NIG innovations, $\lambda$ also changes the shape of the conditionally transformed innovation distribution at admissible states, so the argument does not carry over. Finally, the parameters $a_0$ and $a_1$ of VPGARCH enter only through the lower-order terms of the leverage loading $\psi_2(x)=\eta\sqrt{a_0+a_1x+x^2}$, so that their relative contribution is largest at low variance levels. The estimates of $a_0$ are below $10^{-12}$, near its lower bound of zero, in more than 35\% of the weeks.

\begin{figure}[!htbp]
	\centering
	\includegraphics{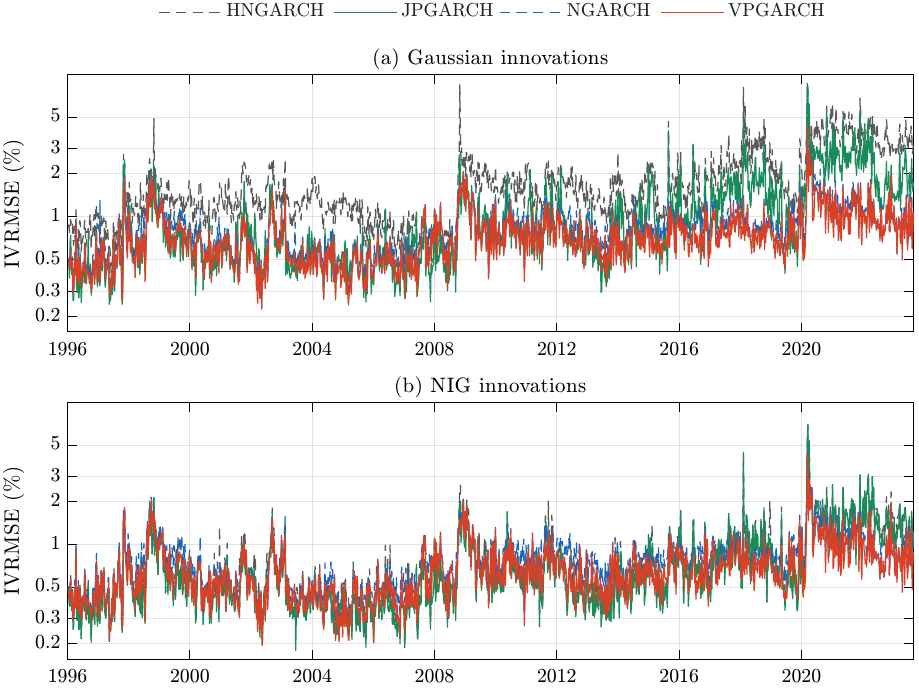}
	\caption{\textbf{Weekly Implied Volatility Root Mean Squared Errors by Innovation Distribution}\newline
		\footnotesize
		\emph{Notes:} Panels~(a) and~(b) show the weekly IVRMSEs, in percent and on a logarithmic scale, of the four models under Gaussian and NIG innovations, respectively. Each model is calibrated separately on each of the 1,443 weekly dates from January 1996 to August 2023 with $\xi_{2,t}=0$. Figure~\ref{fig:weekly_ivrmse_distribution} of the main paper shows exponentially weighted moving averages of these series.}
	\label{fig:sm_weekly_ivrmse_distribution}
\end{figure}

Figures~\ref{fig:sm_weekly_ivrmse_distribution} and~\ref{fig:sm_weekly_ivrmse_model} show the raw weekly IVRMSEs whose exponentially weighted moving averages are plotted in Figure~\ref{fig:weekly_ivrmse_distribution} of the main paper and Figure~\ref{fig:weekly_ivrmse_model} below. Under Gaussian innovations, each model attains its largest weekly IVRMSE in the week of March~11, 2020, ranging from 4.42\% for VPGARCH to 8.63\% for HNGARCH and JPGARCH. NIG innovations lower the weekly IVRMSE of HNGARCH in every week.

\begin{table}
	\caption{\textbf{One-Step Conditional Variance of the Variance}}\label{tab:vov_coefficients}
	\scriptsize
	\renewcommand{\arraystretch}{1.25}
	\setlength{\tabcolsep}{10pt}
	\begin{tabular*}{\textwidth}{@{\extracolsep{\fill}} l *{4}{c}}
		\toprule
		\multicolumn{5}{l}{\textbf{Panel A: Coefficients}} \\
		\midrule
		& HNGARCH & JPGARCH & NGARCH & VPGARCH \\
		\cmidrule(l){2-2} \cmidrule(l){3-3} \cmidrule(l){4-4} \cmidrule(l){5-5}
		$d_0$ & $(m_4-1)b_0^2$ & $(m_4-1)b_0^2$ & $0$ & $\eta^2a_0+(m_4-1)b_0^2$ \\
		$d_1$ & $\eta^2$ & $\eta^2+2(m_4-1)b_0b_1$ & $0$ & $\eta^2a_1+2(m_4-1)b_0b_1$ \\
		$d_2$ & $0$ & $(m_4-1)b_1^2$ & $\eta^2+(m_4-1)b_1^2$ & $\eta^2+(m_4-1)b_1^2$ \\
		\toprule
		\multicolumn{5}{l}{\textbf{Panel B: Values at the Gaussian Estimates}} \\
		\midrule
		& HNGARCH & JPGARCH & NGARCH & VPGARCH \\
		\cmidrule(l){2-2} \cmidrule(l){3-3} \cmidrule(l){4-4} \cmidrule(l){5-5}
		$d_0/\sigma^4$ & 0.0047 & 0.0011 & 0 & 0.0004 \\
		$d_1/\sigma^2$ & 0.0344 & 0.0349 & 0 & 0.0049 \\
		$d_2$ & 0 & 0.0076 & 0.0587 & 0.0488 \\
		\toprule
		\multicolumn{5}{l}{\textbf{Panel C: Values at the NIG Estimates}} \\
		\midrule
		& HNGARCH & JPGARCH & NGARCH & VPGARCH \\
		\cmidrule(l){2-2} \cmidrule(l){3-3} \cmidrule(l){4-4} \cmidrule(l){5-5}
		$d_0/\sigma^4$ & 0.0073 & 0.0016 & 0 & 0.0006 \\
		$d_1/\sigma^2$ & 0.0365 & 0.0400 & 0 & 0.0082 \\
		$d_2$ & 0 & 0.0118 & 0.0711 & 0.0520 \\
		\bottomrule
	\end{tabular*}
	\footnotesize
	\emph{Notes:} The one-step conditional variance of the variance is $\Var_t\left[h_{t+1}\right]=d(h_t)=d_0+d_1h_t+d_2h_t^2$, where $d(x)=\psi_2(x)^2+(m_4-1)\psi_3(x)^2$ for symmetric innovations with fourth moment $m_4$. Panel~A reports the coefficients; see Sections~\ref{subsec:variance_polynomial_dynamics}, \ref{subsec:joint_polynomial_dynamics}, and~\ref{subsec:moment_implications} of the main paper. The HNGARCH and NGARCH columns follow from the JPGARCH and VPGARCH columns with $b_1=0$ and with $a_0=a_1=b_0=0$, respectively. Panels~B and~C report the coefficients at the Gaussian estimates, for which $m_4=3$, and at the NIG estimates, for which $m_4=3+3\nu$, in Table~\ref{tab:return_mle} of the main paper, with $a_0=0$ for VPGARCH. The values are scaled so that $d(u\sigma^2)/\sigma^4=d_0/\sigma^4+(d_1/\sigma^2)\,u+d_2u^2$.
\end{table}

\begin{table}
	\caption{\textbf{Calibrated Parameters under Gaussian Innovations}}\label{tab:option_parameters_gaussian}
	\scriptsize
	\renewcommand{\arraystretch}{1.25}
	\setlength{\tabcolsep}{10pt}
	\begin{tabular*}{\textwidth}{@{\extracolsep{\fill}} l *{5}{c}}
		\toprule
		& \multicolumn{5}{c}{Quantiles} \\
		\cmidrule(lr){2-6}
		& 5th & 25th & 50th & 75th & 95th \\
		\cmidrule(l){2-2} \cmidrule(l){3-3} \cmidrule(l){4-4} \cmidrule(l){5-5} \cmidrule(l){6-6}
		\textit{HNGARCH} &&&&& \\
		$\lambda$ & 3.99E-08 & 2.38E-05 & 0.3251 & 1.1959 & 3.5549 \\
		$\sigma^2\times10^{4}$ & 0.7949 & 1.1974 & 1.7397 & 2.4889 & 4.0699 \\
		$\phi$ & 0.8578 & 0.9277 & 0.9533 & 0.9683 & 0.9841 \\
		$\eta\times10^{3}$ & -9.2280 & -5.7184 & -4.2735 & -3.2962 & -2.1705 \\
		$b_0\times10^{6}$ & 2.6229 & 4.7658 & 7.5235 & 14.2039 & 32.4493 \\
		$h_0\times10^{4}$ & 0.2453 & 0.5799 & 1.0947 & 1.9561 & 5.4310 \\
		\midrule
		\textit{JPGARCH} &&&&& \\
		$\lambda$ & 6.56E-08 & 0.0404 & 0.4617 & 1.0636 & 2.4394 \\
		$\sigma^2\times10^{4}$ & 0.8524 & 1.3081 & 1.9286 & 2.8420 & 5.3330 \\
		$\phi$ & 0.9123 & 0.9523 & 0.9683 & 0.9789 & 0.9910 \\
		$\eta\times10^{3}$ & -7.4889 & -4.9096 & -3.8205 & -2.9881 & -2.0744 \\
		$b_0\times10^{6}$ & 0.7103 & 2.7110 & 4.7213 & 7.6732 & 17.4189 \\
		$b_1$ & 0.0523 & 0.1105 & 0.1470 & 0.1960 & 0.3722 \\
		$h_0\times10^{4}$ & 0.3048 & 0.6014 & 1.1202 & 1.9218 & 4.8973 \\
		\midrule
		\textit{NGARCH} &&&&& \\
		$\lambda$ & 1.07E-08 & 5.29E-07 & 5.53E-06 & 0.0804 & 0.7766 \\
		$\sigma^2\times10^{4}$ & 0.9180 & 1.4109 & 2.0779 & 3.1704 & 6.7607 \\
		$\phi$ & 0.9077 & 0.9493 & 0.9660 & 0.9774 & 0.9896 \\
		$\eta$ & -0.7312 & -0.4477 & -0.3304 & -0.2545 & -0.1716 \\
		$b_1$ & 0.0197 & 0.0668 & 0.1191 & 0.1801 & 0.3267 \\
		$h_0\times10^{4}$ & 0.2139 & 0.5196 & 1.0203 & 1.8297 & 4.7466 \\
		\midrule
		\textit{VPGARCH} &&&&& \\
		$\lambda$ & 2.13E-08 & 6.42E-07 & 6.71E-06 & 0.0689 & 0.8780 \\
		$\sigma^2\times10^{4}$ & 0.8877 & 1.3426 & 2.0026 & 2.9942 & 5.9243 \\
		$\phi$ & 0.8996 & 0.9440 & 0.9626 & 0.9749 & 0.9874 \\
		$\eta$ & -0.6486 & -0.3852 & -0.2747 & -0.2091 & -0.1365 \\
		$a_0\times10^{10}$ & 1.00E-08 & 2.20E-04 & 0.2441 & 2.7399 & 10.0171 \\
		$a_1\times10^{5}$ & 0 & 0.9863 & 4.0113 & 9.6055 & 56.9202 \\
		$b_0\times10^{6}$ & 2.44E-05 & 0.7465 & 4.4238 & 7.7459 & 15.9720 \\
		$b_1$ & 0.0204 & 0.0664 & 0.1127 & 0.1704 & 0.3251 \\
		$h_0\times10^{4}$ & 0.1839 & 0.4986 & 1.0134 & 1.8321 & 4.7945 \\
		\bottomrule
	\end{tabular*}
	\footnotesize
	\emph{Notes:} This table reports selected quantiles of the parameters obtained by calibrating each model separately to the cross-section of options on each of the 1,443 weekly dates from January 1996 to August 2023, with $\xi_2=0$ and Gaussian innovations. Only the parameters calibrated for each model are reported; the intercept is implied by $\omega=(1-\phi)\sigma^2$, and $h_0$ is the initial conditional variance on each date.
\end{table}

\begin{table}
	\caption{\textbf{Calibrated Parameters under NIG Innovations}}\label{tab:option_parameters_nig}
	\scriptsize
	\renewcommand{\arraystretch}{1.25}
	\setlength{\tabcolsep}{10pt}
	\begin{tabular*}{\textwidth}{@{\extracolsep{\fill}} l *{5}{c}}
		\toprule
		& \multicolumn{5}{c}{Quantiles} \\
		\cmidrule(lr){2-6}
		& 5th & 25th & 50th & 75th & 95th \\
		\cmidrule(l){2-2} \cmidrule(l){3-3} \cmidrule(l){4-4} \cmidrule(l){5-5} \cmidrule(l){6-6}
		\textit{HNGARCH} &&&&& \\
		$\lambda$ & 1.37E-05 & 0.4290 & 1.1197 & 2.2038 & 4.2045 \\
		$\sigma^2\times10^{4}$ & 0.7739 & 1.1344 & 1.6180 & 2.3472 & 3.7937 \\
		$\phi$ & 0.8967 & 0.9479 & 0.9650 & 0.9753 & 0.9857 \\
		$\eta\times10^{3}$ & -8.2177 & -4.7699 & -3.6768 & -2.6527 & -1.8909 \\
		$b_0\times10^{6}$ & 1.3349 & 3.0797 & 4.7906 & 8.1332 & 21.0482 \\
		$h_0\times10^{4}$ & 0.2781 & 0.5922 & 1.1152 & 1.8807 & 4.9729 \\
		$\nu$ & 1.5802 & 3.6502 & 6.7482 & 11.8139 & 28.7209 \\
		\midrule
		\textit{JPGARCH} &&&&& \\
		$\lambda$ & 8.67E-06 & 0.4004 & 1.0184 & 2.0922 & 4.0522 \\
		$\sigma^2\times10^{4}$ & 0.7555 & 1.1383 & 1.6408 & 2.3989 & 3.8576 \\
		$\phi$ & 0.9168 & 0.9515 & 0.9671 & 0.9771 & 0.9865 \\
		$\eta\times10^{3}$ & -7.5585 & -4.6436 & -3.6011 & -2.7138 & -1.8822 \\
		$b_0\times10^{6}$ & 0.9503 & 2.2752 & 3.9657 & 6.3541 & 15.2686 \\
		$b_1$ & 0 & 8.63E-06 & 0.0043 & 0.0257 & 0.1319 \\
		$h_0\times10^{4}$ & 0.2973 & 0.6058 & 1.1204 & 1.9092 & 4.7866 \\
		$\nu$ & 0.0027 & 2.5207 & 5.7611 & 10.6671 & 28.3782 \\
		\midrule
		\textit{NGARCH} &&&&& \\
		$\lambda$ & 2.00E-08 & 7.57E-07 & 1.26E-05 & 0.1709 & 0.9230 \\
		$\sigma^2\times10^{4}$ & 0.9290 & 1.4123 & 2.0889 & 3.1580 & 6.6600 \\
		$\phi$ & 0.9132 & 0.9525 & 0.9673 & 0.9783 & 0.9898 \\
		$\eta$ & -0.7355 & -0.4360 & -0.3284 & -0.2490 & -0.1688 \\
		$b_1$ & 0.0131 & 0.0306 & 0.0625 & 0.1256 & 0.2876 \\
		$h_0\times10^{4}$ & 0.2111 & 0.5245 & 1.0202 & 1.8126 & 4.7329 \\
		$\nu$ & 0 & 0.0329 & 0.4503 & 1.6649 & 4.8046 \\
		\midrule
		\textit{VPGARCH} &&&&& \\
		$\lambda$ & 5.10E-08 & 1.84E-06 & 5.82E-05 & 0.1798 & 1.1202 \\
		$\sigma^2\times10^{4}$ & 0.8970 & 1.3382 & 1.9870 & 2.9669 & 5.8286 \\
		$\phi$ & 0.9043 & 0.9469 & 0.9640 & 0.9760 & 0.9879 \\
		$\eta$ & -0.6213 & -0.3615 & -0.2633 & -0.1943 & -0.1103 \\
		$a_0\times10^{10}$ & 1.00E-08 & 2.40E-04 & 0.2485 & 3.3289 & 19.9975 \\
		$a_1\times10^{5}$ & 4.21E-07 & 1.0048 & 5.7994 & 16.1639 & 105.1894 \\
		$b_0\times10^{6}$ & 2.44E-05 & 0.5060 & 3.8126 & 7.0989 & 15.3679 \\
		$b_1$ & 0.0085 & 0.0262 & 0.0597 & 0.1248 & 0.3003 \\
		$h_0\times10^{4}$ & 0.1915 & 0.5047 & 1.0177 & 1.8441 & 4.7962 \\
		$\nu$ & 0 & 0.0200 & 0.3666 & 1.4991 & 4.8719 \\
		\bottomrule
	\end{tabular*}
	\footnotesize
	\emph{Notes:} This table reports the same quantiles as Table~\ref{tab:option_parameters_gaussian} for the calibrations with NIG innovations, with the same scaling of the parameters. The shape of the NIG distribution is reported through $\nu=1/\alpha$, where $\alpha$ is the shape parameter in Equation~\eqref{eq:nig_mixture} of the main paper, so that the fourth moment of the innovations is $m_4=3+3\nu$ and $\nu=0$ corresponds to Gaussian innovations.
\end{table}

\clearpage

\begin{figure}[!htbp]
	\centering
	\includegraphics{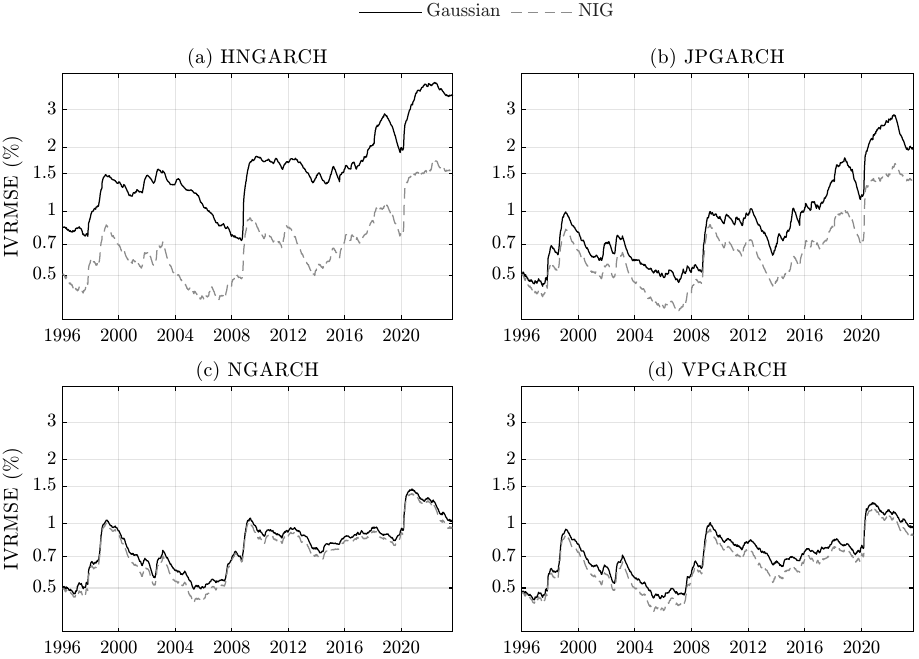}
	\caption{\textbf{Smoothed Weekly Implied Volatility Root Mean Squared Errors by Model}\newline
		\footnotesize
		\emph{Notes:} Smoothed weekly IVRMSEs under Gaussian (solid) and NIG (dashed) innovations, in percent and on a logarithmic scale. Smoothing follows Figure~\ref{fig:weekly_ivrmse_distribution} of the main paper; the raw series appear in Figure~\ref{fig:sm_weekly_ivrmse_model}.}
	\label{fig:weekly_ivrmse_model}
\end{figure}

\clearpage

\begin{figure}[!htbp]
	\centering
	\includegraphics{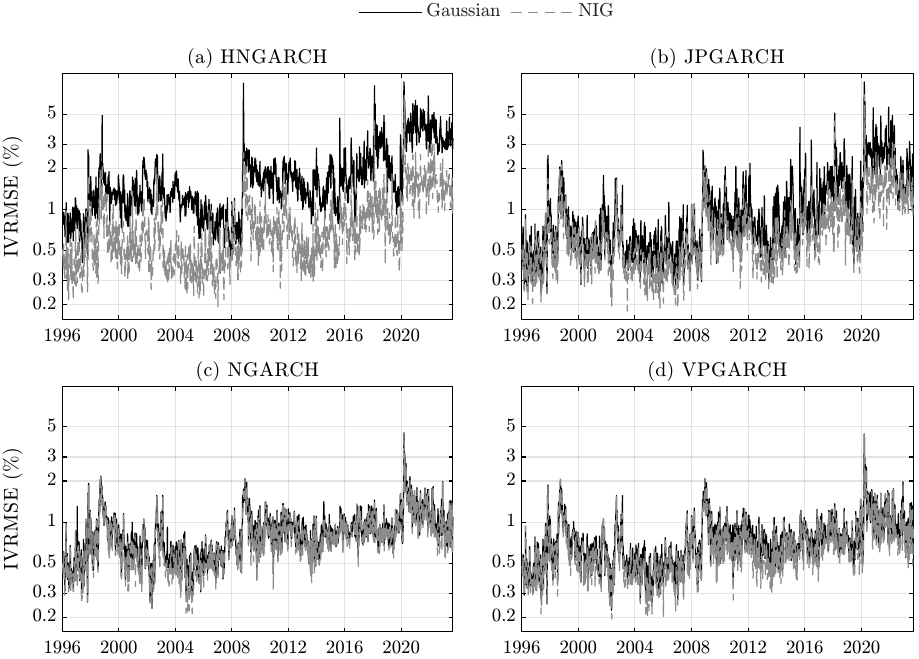}
	\caption{\textbf{Weekly Implied Volatility Root Mean Squared Errors by Model}\newline
		\footnotesize
		\emph{Notes:} Each panel compares, for one model, the weekly IVRMSEs obtained under Gaussian innovations (solid line) and NIG innovations (dashed line), in percent and on a logarithmic scale. The calibrations are those of Figure~\ref{fig:sm_weekly_ivrmse_distribution}. Figure~\ref{fig:weekly_ivrmse_model} shows exponentially weighted moving averages of these series.}
	\label{fig:sm_weekly_ivrmse_model}
\end{figure}

\clearpage
\setlength{\bibsep}{3pt}
\titlespacing*{\section}{0pt}{6pt}{12pt}
\singlespacing
\begin{spacing}{0.0}
	\newcommand{\SortNoop}[1]{}

\end{spacing}

\ArxivBibScope{}
\end{document}